# A WAVELET ANALYSIS OF INTER-DEPENDENCE, CONTAGION AND LONG MEMORY AMONG GLOBAL EQUITY MARKETS

A Thesis Submitted to the University of Hyderabad in
Partial Fulfilment of the Requirements for the Award of

DOCTOR OF PHILOSOPHY
IN
ECONOMICS

BY

**AVISHEK BHANDARI**

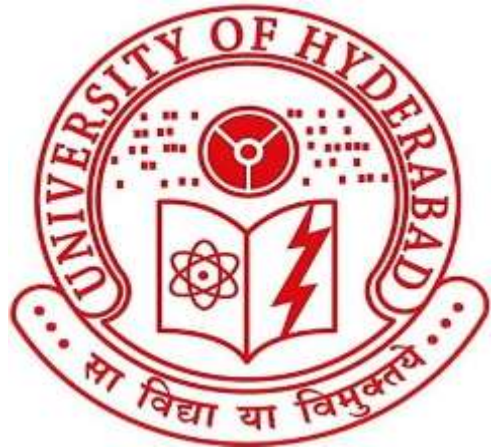

**School of Economics
University of Hyderabad
(P.O) Central University, Gachibowli
Hyderabad-500 046
April 2017**

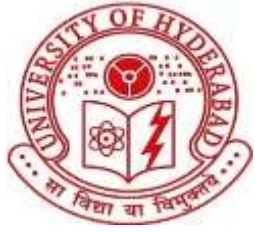

# CERTIFICATE

This is to certify that "**A Wavelet Analysis of Inter-dependence, Contagion and Long memory among Global Equity Markets**" submitted by **Avishek Bhandari** bearing Registration Number **12SEPH10** in partial fulfilment of the requirements for award of **Doctor of Philosoph**y in the School of Economics is a bonafide work carried out by him under my supervision and guidance.

This thesis is free from plagiarism and has not been submitted previously in part or in full to this or any other university or institution for award of any degree or diploma.

Papers related to this thesis have been:

A. Published in the following publication

1. Bhandari, A., Bandi, K. On the Dynamics of Inflation-Stock Returns in India. J. Quant. Econ. 16, 89–99 (2018). https://doi.org/10.1007/s40953-017-0075-6

B. Presented in the following conferences:

1. IV International conference on Applied Econometrics, IBS, Hyderabad, (March 2014).
2. 51st annual conference of the Indian Econometric society (TIES), Panjabi University, Patiala, (December 2015).
3. 53$^{rd}$ Annual Conference of The Indian Econometric Society, NISER, Bhubaneswar, (December 2016)

Further, the student has passed the following coursework requirement for Ph.D. / was exempted from doing coursework (recommended by Doctoral Committee) on the basis of the following courses passed during his M.Phil. Program and the M.Phil. degree was awarded:

| S. No. | Course Code | Name | Credits | Pass/Fail |
|---|---|---|---|---|
| 1. | SE-701 | Advanced Economic Theory | 4.00 | PASS |
| 2. | SE-702 | Social Accounting and Data Base | 4.00 | PASS |
| 3. | SE-703 | Research Methodology | 4.00 | PASS |
| 4. | SE-751 | Study Area | 4.00 | PASS |
| 5. | SE-752 | Dissertation | 16.00 | PASS |

**Dean**                                                                                      **(Prof. B. Kamaiah)**

**School of Economics**                                                             **Research Supervisor**

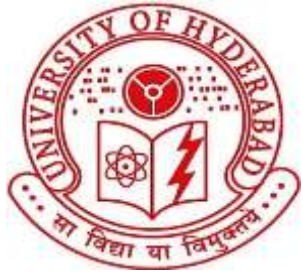

# DECLARATION

I, Avishek Bhandari, hereby declare that the work embodied in the present dissertation entitled, "**A Wavelet Analysis of Inter-dependence, Contagion and Long memory among Global Equity Markets**" submitted by me under the guidance and supervision of Prof. Bandi Kamaiah is a bonafide research work for the award of Doctor of Philosophy in Economics from the University of Hyderabad. I also declare to the best of my knowledge that it has not been submitted previously in part or in full to this University or any other university or institution for the award of any degree or diploma. I hereby agree that my thesis can be deposited in Shodganga/INFLIBNET.

Date:                                                                                 Avishek Bhandari

Place: Hyderabad                                                  Reg. No. 12SEPH10

# CONTENTS

# LIST OF TABLES

# LIST OF FIGURES

# LIST OF ABBREVIATIONS

| ADF | Augmented Dickey Fuller |
|---|---|
| CEE | Central and Eastern European countries COI |
| CWT | Continuous Wavelet Transform |
| DCC | Dynamic conditional correlation |
| DCC-GARCH | Dynamic Conditional Correlation- Generalized Autoregressive Conditional Heteroskedasticity |
| DWT | Discrete Wavelet Transform |
| EMH | Efficient Market Hypothesis |
| EMU | European Monetary Union |
| GARCH | Generalized Autoregressive Conditional Heteroskedasticity |
| GFC | Global Financial Crisis |
| KPSS | Kwiatkowski–Phillips–Schmidt–Shin |
| LA (8) | Least-Asymmetric (8) Filter |
| LIBOR | London interbank offered rate |
| LRD | Long-range Dependence |
| MODWT | Maximal Overlap Discrete Wavelet Transform |
| MRA | Multi-Resolution Analysis |
| OECD | Organisation for Economic Co-operation and Development |
| R/S | Rescaled Range |
| WMC | Wavelet Multiple Correlation |
| WMCC | Wavelet Multiple Cross-Correlation |

*Acknowledgment*

# Chapter 1

# Introduction, Objectives and Contributions of the Study

## 1.1 Introduction and Framework

The present study attempts to investigate into the structure and features of global equity markets from a time-frequency perspective. An analysis grounded on this framework allows one to capture information from a different dimension, as opposed to the traditional time domain analyses, where multiscale structures of financial markets are clearly extracted. In financial time series, multiscale features manifest themselves due to presence of multiple time horizons. The existence of multiple time horizons necessitates a careful investigation of each time horizon separately as market structures are not homogenous across different time horizons. The presence of multiple time horizons, with varying levels of complexity, requires one to investigate financial time series from a heterogeneous market perspective where market players are said to operate at different investment horizons.

The existence of investment heterogeneity is first explored in Muller et al. (1997) where the theory of heterogeneous market hypothesis is expounded. This hypothesis is motivated by the presence of multiple scales, or fractals, in financial time series, which is argued to be induced by the behaviour of a group of market participants or investors. These groups are not homogenous with regard to their investment decisions, inasmuch as market participants differ from one another based on their investment holding period. Therefore, markets can be broken down, particularly owing to the diversity of participants' investment holding periods, into several investment horizons, trading horizons or timescales. A particular investment horizon or timescale has a group of investors operating on it who share similar time perspective. For example, investors who operate on shorter timescale or investment horizon of one or few days are primarily interested in speculative activity as opposed to investors with longer time horizons, say agents indulged in investment decision making of large institutions. This inherent diversity of market players and their investment decisions, which is a function of the respective timescales, leads to the formation of multiple layers of investment time-horizons, ranging from seconds to years. This dissemination of information at dissimilar timescales, which traditional time domain econometric methods cannot capture, calls for an alternative method which can accurately capture information from multiple investment horizons.

Wavelet methods of the time-frequency class, for instance, provide powerful tools that can disentangle information from multiple timescales (see for eg. Percival and Walden, 2000; Gencay et al., 2002). It is this property of wavelets that allows one to carefully investigate global equity markets within the theoretical framework of heterogeneous market hypothesis.

This thesis extends the application of time-frequency based wavelet techniques to: i) analyse the interdependence of global equity markets from a heterogeneous investor perspective with a special focus on the Indian stock market, ii) investigate the contagion effect, if any, of financial crises on Indian stock market, and iii) to study fractality and scaling properties of global equity markets and analyse the efficiency of Indian stock markets using wavelet based long memory methods.

## 1.2 Review of Spectral and Wavelet Methods

This thesis primarily uses methods from the wavelet domain in analysing the relationship among global equity markets. However, basic concepts from Fourier based spectral methods are reviewed in order to provide a glimpse into the frequency domain counterpart of time series analysis. Moreover, since wavelet methods are based on theories from Fourier analysis (Mallat, 1999), a brief review of spectral analysis is provided. This section briefly reviews Fourier based methods and then moves on to the wavelet based concepts and justifies need for the use of wavelet methods in this thesis.

*1.2.1 Spectral methods*

A financial time-series can be decomposed into its periodic or regular components to model the repetitive and oscillatory behaviour of the underlying time series. This is done by expressing the time series using combinations of periodic functions like sines and cosines with different frequencies. Unlike time series analysis, spectral analysis focuses on identifying the dominant frequencies present in the signal. However, time domain information is completely lost when the time series is analysed in the frequency space. A time series can be viewed from a frequency lens by transforming the time signal, say $x(t)$, into the frequency space by means of Fourier series approximation given by:

$$x(t) = \frac{q_0}{2} + \sum_{\omega=1}^{n} q_\omega \cos \omega t + r_\omega \sin \omega t) \qquad (1.1)$$

where the signal $x(t)$ is a polynomial of order $n$ and contains $(2n+1)$ Fourier coefficients given by $q_0, q_1 \ldots q_n$, $r_1, r_2 \ldots r_n$. The signal $x(t)$ is a linear combination of periodic sines and cosines with period $2\pi$. By defining $c_\omega = (q_\omega + ir_\omega)/2$, $c_{-\omega} = (q_\omega - ir_\omega)/2$ and $c_0 = a_0/2$, the above equation can be represented as

$$x(t) = \sum_{\omega=-n}^{n} c_\omega e^{-i\omega t} \qquad (1.2)$$

The main idea underlying the transformation of time signal to its spectral representation is to re express $x(t)$ as a new sequence $f(\omega)$ which describes the significance of each frequency component $\omega$ in the dynamics of new series (Masset, 2008). Accordingly

$$f(\omega) = \sum_{t=-\infty}^{\infty} x(t) e^{-i\omega t} \qquad (1.3)$$

The above equation, which is the discrete Fourier transform of $x(t)$, projects the time signal $x(t)$ onto a set of sinusoidal functions, where each component correspond to a unique frequency. Moreover, the original signal $x(t)$ can be obtained from the frequency domain signal by the inverse transform given by

$$x(t) = \frac{1}{2\pi} \int_{-\pi}^{\pi} f(\omega) e^{-i\omega t} d\omega \qquad (1.4)$$

Using the aforementioned concepts of Fourier analysis, one can analyse a covariance stationary time series from a frequency domain perspective by transforming the time series process, say $\{X_t\}$, into the spectral domain. Formally, the autocovariance function of the time series is transformed into the Fourier domain, which gives the spectral density function of the process $\{X_t\}$, and is given by

$$f_X(\omega) = \frac{1}{2\pi} \sum_{j=-\infty}^{\infty} \gamma_j e^{-i\omega j} \qquad -\pi < \omega < \pi \qquad (1.5)$$

The plot of the above spectral density against frequency gives the power spectrum and is useful in detecting the presence of dominant frequencies, thereby aiding in understanding the dominant cyclical components in the time series. The estimator of the spectral density is known as the periodogram which is an inconsistent estimator. Therefore, windowed version of the periodogram estimator, where consistency is maintained, is used in practical applications. A detailed exposition of time series in spectral domain is provided in Priestley (1981). Moreover, Nachane (2006) highlights some important procedures, like i) Aliasing, ii) Filtering, iii) Tapering, iv) Window closing, and v) Fast Fourier Transform, that needs to be checked and implemented while using spectral techniques. However, methods from spectral analysis fail to capture the time information present in the time-series as Fourier transformation completely eliminates information from time domain. Nevertheless, simultaneous information from both spectral and time domain can be obtained by the use of short-term Fourier transform, also known as Gabor transform, where the signal is portioned into several blocks and Fourier transform is employed for each and every block. However, the use of fixed sized windows in Gabor method entails significant loss of information while trying to simultaneously obtain information from both time and frequency. This drawback of Gabor method is mitigated by wavelet analysis where wavelet windows allows flexible alteration of its size and are very useful in obtaining good resolutions in both time and frequency.

*1.2.2 Wavelet analysis*

The drawbacks of both spectral and Gabor methods can be mitigated by the use of wavelet techniques. Wavelet analysis does not include the assumption of covariance stationarity and decompositions based on wavelet transforms does not inherit the limitations of spectral methods. They provide a relatively better decomposition of time series which is localized concurrently in time and frequency. A wavelet window, which can be altered according to specific needs, is applied on a time signal to extract finer and detailed information of the signal from both time and frequency domains.

Initially, low-frequency components of time series giving poor time resolution are extracted using a broad scaling window. The scaling window is then subsequently shortened in length to extract out higher frequencies that give better time resolution. Windows that capture high frequency components, via the use of short length scaling window, gives good time localization (thus, poor frequency information), and vice versa. The information obtained out of each scale of resolution, which is dependent on scaling (altering the window size) and translating (shifting the wavelet window across the time signal) the window, is taken out and analysed until the whole signal is properly captured. Therefore, by continuously implementing this process of information extraction until the desired signal decomposition is achieved, excellent time and frequency resolution of a given signal can be obtained. This is the central principle of multiresolution analysis that make wavelets very useful in practical applications.

A wavelet is a wave which is limited in size and has the property of compact support. A function is said to be compactly supported if it is finite and is zero outside a certain interval. Hence, wavelets are smaller waves with different shapes and sizes based on the type of wavelet function. Therefore, a wavelet is a function $\psi(\cdot)$ defined on $\mathbb{R}$ such that $\int_{\mathbb{R}} \psi(t)\, dt = 0$ and $\int_{-\infty}^{\infty} |\psi(t)|^2\, dt = 1$. A reference wavelet $\psi_{b,s}(t)$, known as the mother wavelet, is chosen to perform wavelet analysis and is defined as

$$\psi_{b,s}(t) = \frac{1}{\sqrt{s}} \psi\left(\frac{t-b}{s}\right) \quad (1.6)$$

where $s \neq 0$ and $b$ are real constants. The parameter $s$ is the *scaling parameter* (used to determine window widths), whereas the parameter $b$ denotes the *translation parameter* (which controls the location of the window). The following diagram describes the implementation of mother wavelets in extracting detailed and finer information from a time signal.

Figure 1.1 Diagrammatic representation of scaling and translation

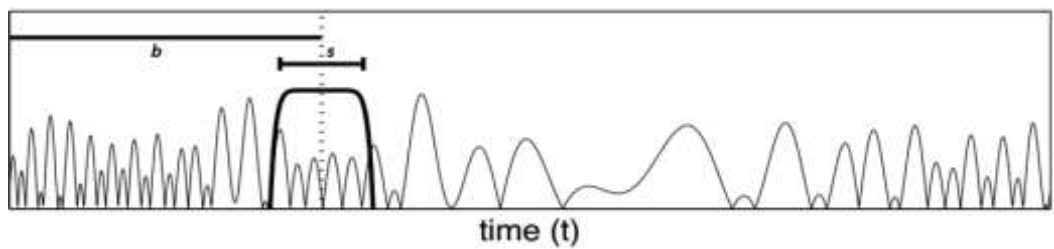

time (t)

The mother wavelet with width $s$ (known as dilation parameter) and located at $b$ (known as translation parameter), given by the compactly supported bold wave in the above figure, is placed in the time signal and slid throughout the signal to extract information pertaining to a certain wavelet scale. In order to extract finer time localisation, the window can be shortened in length by a factor of two. For example, the window in the above figure can be halved in length to $\frac{s}{2}$, thereby enabling the shrunken window to capture finer details encompassing higher frequencies from the signal. Similarly, the wavelet window can be doubled in size to enable it capture lower frequencies thereby providing good frequency localisation. This process can be continued until detailed information from both time and frequency encompassing multiple resolutions are captured.

The mother wavelet $\psi_{b,s}(t)$, which is governed by the dilation parameter *s* and the translation parameter *b*, is used to define the "*continuous wavelet transform*" (CWT) of a time signal $x(t)$ and is given by

$$W^X(b,s) = \int_{-\infty}^{\infty} x(t)\, \overline{\psi_{b,s}(t)}\, dt \qquad (1.7)$$

provided the following *admissibility condition* is satisfied

$$C_\psi = \int_{-\infty}^{\infty} \frac{|\Psi(\omega)|^2}{|\omega|} d\omega < \infty \qquad (1.8)$$

where $\Psi(\omega)$ is the Fourier transform of the mother wavelet $\psi_{a,b}(t)$ which is given by $\Psi(\omega) = \int_{-\infty}^{\infty} \psi(t)\, e^{-i\omega t} dt$. The admissibility condition given above is a theoretical requirement that allows the reconstruction of $x(t)$ from the CWT. The use of wavelet transforms allows one to obtain information about the time series at different resolution levels, thereby enabling one to extract details from the data made possible by the working of multiresolution algorithm. Figure 1.2 helps in understanding the working of multiresolution analysis (MRA).

Plot 1 in the Figure 1.2 depicts time series analysis where there is good localisation with respect to time. On the contrary, Plot 2 gives the Fourier transformed version of the time series depicted in Plot 1 where only frequency information is present. Plot 3 depicts the Gabor method where information from both time and frequency, as given in Plot 1 and Plot 2, is present. However, due to the inherent information trade-off between some amount of accuracy in both time and frequency is lost. Finally, Plot 4 depicts the wavelet method which describes the working of multiresolution algorithm. The frequencies are read from bottom to top of the box where frequencies increases as we move to the top. In the first step, a broader wavelet window is employed to extract the entire frequency information giving no time localisation.

Figure 1.2 Box diagram representing time-frequency methods

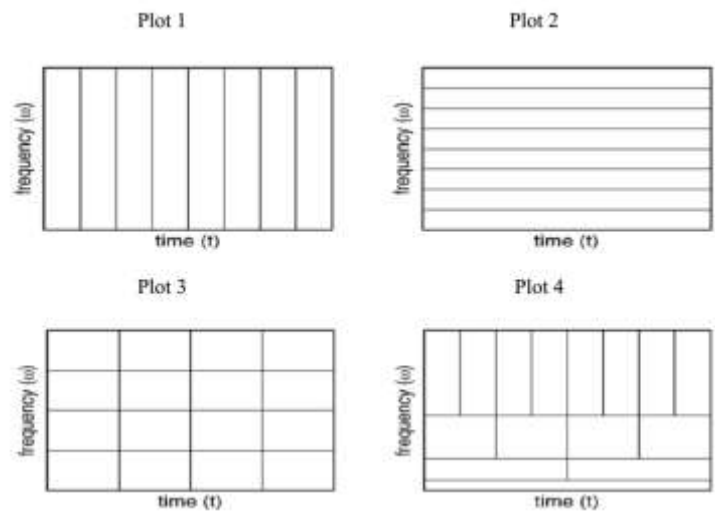

The wavelet window is then shrunken to half of its original length in step two, by reducing the length of the dilation parameter *s,* thereby capturing some time resolution by trading off some frequency localisation. This process is implemented until the required information of both time and frequency space is attained. Therefore, the figure above is very crucial in understanding the process of multiresolution algorithm.

The analysis of a one-dimensional time series using wavelet transform results in a two-dimensional output as the output of continuous wavelet methods constitute a plane. Therefore, the transform results in redundancy as information from a one dimensional signal is represented in a two dimensional space. This implies that common information is shared by some neighbouring coefficients in the time-scale plane. However, this issue is resolved by sampling the time-scale plane, mathematically made possible by the theory of MRA, thereby keeping only discrete coefficients from the continuous wavelet space sufficient enough to describe the signal without any loss of information. Therefore, in most of the practical applications, the "discrete version of the wavelet transform" is implemented. Multiresolution analysis, with the help of discrete wavelet based pyramid algorithm given in Mallat (1989), allows partitioning of time series into several *crystals* containing wavelet coefficients corresponding to different layers of resolutions and frequencies. The original time series is "decomposed into several details" (or crystals), each containing corresponding high or low frequency coefficients that are also known as wavelet *atoms*. Formally, MRA constitutes a group of wavelet subspaces $\{W_s\}_{s \in \mathbb{Z}}$ that are nested (Daubechies, 1992) and satisfies the following properties:

1. $\bigcap_{\_\_} \quad \bigcup_{\_}$ e in $L^2(R)$

2. $W_s \subset W_{s-1}$

3. $x(t) \in W_s \Leftrightarrow x(2^s t) \in W_0$

4. $W_0$ is the space of approximation coefficients that contains a scaling function, or father wavelet, denoted by $\phi_0(t)$. The collection of scaling functions $\{\phi_0(t-b), b \in \mathbb{Z}\}$ forms a Riesz basis for $W_0$. Finally, it follows from the above that the collection of dilated and shifted scaling functions $\{\phi_{b,s}(t) = 2^{-s/2} \phi_0(2^{-s} t - b), b \in \mathbb{Z}\}$ forms the basis for the wavelet space $W_j$. The same is true for the mother wavelet $\psi_{b,s}(t)$. Since $W_s \subset W_{s-1}$, the coarser coefficients are contained in $W_s$ whereas $W_{s-1}$ contains approximations that are less coarser than $W_s$, and so on. In practice, the value of $s$ is finite such that $s = 0, ..., S$, thereby $W_S \subset W_{S-1} \subset ........ \subset W_0$.

Therefore, the basic idea behind MRA is to examine coarser approximations by removing high frequency details from the signal or time-series. On the other hand, the detail coefficients are obtained while moving from one coarser crystal to another. The detail coefficient of the time series $X_t$ at the scaling level $s$ and time position $b$ is denoted as $d_X(s,b)$, and the wavelet details at level $s$ and time position $b$ is given by the inner product of $d_X(s,b)$ and the mother wavelet $\psi_{b,s}(t)$. Similarly, the approximation coefficient at scaling level $S$ and time position $b$ is given by $a_X(S,b)$, and the level $S$ approximation is obtained by taking the inner product of $a_X(S,b)$ and the father wavelet $\phi_{b,s}(t)$. Therefore, the time series $X_t$ can be broken down into its approximation (low frequency trend) and details (high frequencies) as:

$$X_t = \sum_b a_X(S,b)\phi_{b,s}(t) + \sum_{s=1}^{S}\sum_b d_X(s,b)\ \psi_{b,s}(t) \qquad (1.9)$$

Given the father wavelet or scaling function $\phi_{b,s}(t)$ and the mother wavelet $\psi_{b,s}(t)$, the time series $X_t$ can be transformed via the discrete wavelet transform, by performing inner product of $X_t$ with both $\phi_{b,s}(t)$ and $\psi_{b,s}(t)$, thereby generating $a_X(s,b)$ and $d_X(s,b)$, respectively. More generally, the discrete wavelet transformed version of the time series $X_t$ comprises the collection of coefficients $\{\{a_X(S,b), b \in \square \quad \square\ .$

Since approximations $\{a_X(S,b), b \in \square$ describe the long-run trend or smooth behaviour of time series by providing information at the coarsest resolution, the scaling function $\phi_{b,s}(t)$ acts like a low pass filter. On the other hand, the details $\{d_X(s,b), b \in \square \quad S\}$ are arrived at by subtracting the neighbouring approximations, and storing the resulting output. Hence the mother wavelet $\psi_{b,s}(t)$ behaves like a bandpass filter, and thus resembles a short wave, or *a wavelet*. The number of resolutions that a time signal can be disintegrated into is given by the scaling parameter. Formally, a time series of length $N$ can be decomposed into $s = \log_2(N)$ levels, where the level or wavelet scale $s$ is also known as an *octave*. The pyramid algorithm of Mallat makes it possible to compute the details and wavelet approximations. This is achieved by convolving the discrete wavelet and scaling filters, $g_1$ and $h_1$ derived from mother and father wavelets, with the approximation crystal $a_X(s-1,b)$ at level $s$-1. This outputs the approximation $a_X(s,b)$ and the detail $d_X(s,b)$. The process is continued until the highest level of approximation $a_X(S,b)$ is attained. For the purpose of this thesis, six-eight levels of wavelet decomposition is carried out for empirical analysis giving six-eight details and one long-run approximation. However, only wavelet details are given importance as varying levels of time-horizons are captured making it suitable for analyses based on this thesis. The reader is referred to Mallat (1989), Daubechies (1992), Strang (1996) and Mallat (2006) for a deeper understanding of multiresolution algorithm and wavelet theory.

## 1.3 Wavelet Multiresolution Applications in Finance and Economics

The shortcomings of spectral methods, particularly the requirement of stationarity, can be relatively mitigated by the use of time-scale based wavelet analysis, as wavelet analysis does not enforce the assumption of stationary time-series, thereby making it suitable for the study of financial data. Moreover, wavelets can uniquely isolate the dynamics in a time-series over different scales and horizons. A time-series signal, at first observation, might look stationary but a deeper analysis of the signal with excellent time localization, made possible by the use of windowed Fourier transforms or wavelet filters, might help detect the presence of discontinuities. Nonetheless, at a finer and detailed level of signal analysis, presence of non-stationarity could be detected, Capobianco (2004). Therefore, wavelet analysis, which by allowing us to analyze the data at different scales of resolution, is definitely a good choice for economic and financial time-series analysis as it gives us the information about both time and frequency-varying components of the signal. The information extracted using highly time-localized wavelet windows, from non-stationary financial time-series, can be very useful due to the importance of the information available from minute details of the signal. Furthermore, wavelet based techniques are suitable for detection of regime shifts, financial market shocks and discontinuities present in financial data of any frequency, Ramsey and Zhang (1997).

According to Crowley (2005), the rising interest among economic researchers in experimenting with various data decomposition techniques, forecasting methods, density estimation and other aspects of data mining have led to the introduction of several wavelet based algorithms suitable for implementation in the vast area of financial and economic research. The introduction of the "*maximal overlap discrete wavelet transform*" (MODWT, hereafter) by Percival and Walden (2000) marked an important development in the analysis of financial time-series with non-dyadic length. MODWT is an upgraded version of the "*discrete wavelet transform*" (DWT) introduced by Nason and Silverman (1994) for statistical analysis. However, MODWT loses orthogonality but still is very useful in the analysis of financial time-series as the length of time-series is not always dyadic. Furthermore, the introduction of continuous wavelet based coherence analysis by Grinsted et al. (2004) made possible the analysis of correlation in the wavelet space. Similarly, several discrete wavelet based correlation methods which are highly appropriate for financial data are presented in Percival and Walden (2000) and Gencay et al. (2001).

Since the induction of wavelets into the field of statistical data analysis starting mid-1990s, there has been significant amount of wavelet based applications in the analysis of financial and economic data. Ramsey and Zhang (1997), for example, use waveform dictionaries and evidences the existence of high energy at higher wavelet scales encompassing low frequencies. Moreover, the presence of noisy bursts of energy at higher frequencies was recorded, which were distributed across higher frequencies throughout the year. Ramsey and Lampart (1998), using a hybridisation of Granger causality and MRA find evidence of scale dependent causal association between income and money supply. The order of integration for Canadian and U.S. interest rates was estimated, by Tkacz (2000), using the wavelet long memory estimator of Jensen (2000). With respect to the analysis of a similar long-memory type fractal parameter, a wavelet version of the estimator of effective Holder exponent is employed by Struzik (2001) to uncover the correlation characteristics of the S&P 500 index and to unearth its local spectral contents.

Capobianco (2004), on performing the MRA of Nikkei equity returns unearthed hidden periodic elements. Similarly, Crowley and Lee (2005) while investigating business cycles of some European economies detected unrelated frequency components among weakly integrated markets. Moreover, the use of MODWT in analyzing scaling of financial markets, timescale volatility decomposition, and systematic risk is presented in Gencay et al. (2001) and Gencay et al. (2005). Similarly, Gallegati (2008) using a MODWT based correlation technique investigates associations between the industrial production of the U.S. and equity returns.

Wavelet analysis is used by Lee (2004) to study the phenomena of global transmission mechanism of stock market dynamics. Moreover, Gabor transform and wavelet analysis were used by Raihan et al. (2005) to study the behavior of US business cycle. Wavelet analysis was found to be superior to Gabor transform based analysis. Rua and Nunes (2009), on the other hand, using continuous wavelet analysis unearthed the existence of time-horizon dependent interdependence among some developed markets. Similarly, Barunik et al. (2011) using continuous wavelet technique find scale dependent comovement between some European markets. A partial least square regression technique in the wavelet domain was developed by Huang (2011) by combining wavelet analysis to kernel regressions. As compared with the other existing time-domain models, the wavelet based model was found to be more parsimonious, generating very accurate forecasts. More recently, the study of correlation structure between S&P 500 and other international markets is investigated by Benhmad (2013) using wavelet analysis. S&P 500 and European stock markets were found to exhibit strong interdependencies, which changed according to changes in time-scale. Studies based on wavelets, hybridization of MRA and other time-series techniques, continuous wavelet analysis of financial data, fractal estimators using wavelet methods, have gained prominence in recent times. Other important works using wavelets in analyzing financial data will be reviewed in the subsequent three chapters.

## 1.4 Objectives and Structure of the Study

The first chapter introduces the methodology of wavelets and reviews some important applications of wavelets in economics and finance. A thorough review of some seminal papers, relevant to financial and economic studies, is given along with latest contributions and applications. The theoretical framework of heterogeneous market hypothesis, as discussed before, is used to analyse the research objectives that this thesis addresses. The complex structure of financial market linkages, within the theoretical framework of heterogeneous market hypothesis, and lack of studies incorporating this framework in the Indian context to analyse interdependence, contagion and long memory, calls for an extensive study to bridge this existing research gap.

The second chapter empirically investigates the interdependence among global equity markets using novel methods from the discrete wavelet class. A survey of relevant methods, latest contributions and applications in studying global market interdependence are presented.

A thorough analysis of the structure of global equity market interdependence with a special focus on Indian investors and the subsequent empirical evidences, generated using a battery of classical and advanced wavelet correlation techniques, attempts to delineate the effects of heterogeneity in investment horizons on international portfolio diversification, using Indian market as a case study. As a result, the following are addressed, namely, (i) should Indian investor invest in developed or emerging markets to gain benefits from international portfolio diversification? , and (ii) how will international portfolio diversification change investor stock holding period?

The third chapter surveys the literature on contagion and analyses, using both continuous and discrete wavelet methods, the effects of major financial crises on Indian markets. Accordingly, the third objective of the thesis is to investigate the contagious effects of financial crises on Indian markets and access its implication for international portfolio diversification.

The fourth chapter investigates the long memory behaviour of global equity markets and tries to empirically justify the multifractal nature of financial markets. The dynamic evolution of developed and emerging markets, in terms of efficiency, is analysed using time varying long memory estimates. Since the presence of long memory will have serious implications on empirical evidences from previous chapters, the fourth objective of the thesis is to investigate the efficiency (or inefficiency) and multifractality of Indian stock markets using wavelet based long memory estimators.

The final chapter summarizes the contributions and highlights the usefulness of the results for policymakers, which is then followed by limitations and scope for further studies.

## 1.5 Contributions

Wavelet based studies, underscoring the implications of multiscale investment horizons on international portfolio diversification, are very few. Nevertheless, none among them focuses on the study of linkages between Indian and global equity markets from a wavelet based multiscale perspective. This dearth of information, on the multiscale nature of equity market interdependence between Indian and global markets, motivates this study of interdependence, and expounds the benefits of the resultant multiscale information for Indian investors. The study on contagion implements a multi horizon comovement approach to identify contagion particularly in the Indian context, which can effectively identify the evolution of correlation across markets in both time and frequencies. Lack of studies investigating contagion from a multiscale viewpoint, with a special focus on Indian equity markets, justify the need for this study.

The chapter on long memory particularly delves upon the advantages of applying wavelet based long memory estimators in analysing long memory behaviour of global equity returns in general and Indian equity returns in particular. Dynamic evolution of long memory parameter, using a rolling wavelet estimator to generate the time varying Hurst series, is studied for the Indian case to distinguish between phases of efficiency and inefficiency. Lack of studies based on time varying long memory of Indian equity markets justify the need for this study. Moreover, the chapter on long memory contributes to the literature by implementing a latest multivariate wavelet based long memory method, which to the best of the authors' knowledge is the first application in finance and economics.

# Chapter 2

## Interdependence among Global Equity Markets

### 2.1 Introduction

The strength of interdependence among global markets, which in the literature[1] concerning global market integration is measured using a plethora of methods, can act as a proxy for determining the nature of integration between markets. As risk mitigating portfolio amalgamations has been linked with imperfect correlations among assets in the portfolio, the strength of correlation among equity markets helps international investors in gauging the nature of risk befalling their portfolios. Information about correlation structure of equity markets allow investors in making optimal portfolio strategies by formulating risk minimising portfolio combinations.

This chapter investigates the nature and structure of interdependence among global equity markets with special focus on the Indian market. The structure of correlation and cross-correlation among select pairs of global markets is inspected in the time-frequency space via a wavelet lens. The interdependence between Indian and global markets is examined by analysing the correlation structure between Indian and select markets at varying time-horizons, enabling one to efficiently capture investment risks befalling on non-homogenous market participants. Furthermore, the heterogeneity of market participants' space of operation is effectively captured in a wavelet framework allowing diverse investors, with variegated risks and investment preferences, to access risks concomitant with disparate investment periods.

Therefore, the strength of correlation between Indian and select global markets is analysed at different resolutions or investment horizons, facilitating the enunciation and assessment of market linkages at various regions encompassing the time-frequency space. Moreover, the inhomogeneity of cross-market correlations, among Indian and global markets at varying time horizons, facilitates Indian investors in accessing risks associated at different investment periods, thereby allowing them to carefully formulate investment decisions. The prime motivation that drives this chapter lies in the need to understand the opportunities facing Indian investors with regard to their investment periods in which they operate. Since wavelet based methods can effectively capture diverse investment horizons, the analyses carried out in this chapter can assist heterogeneous Indian investors in making strategic investment choices based on their time-horizon of investment.

---
[1] Kearney and Lucey (2004) review major studies on interdependence.

The following section surveys some important literature crucial to the analyses based on this chapter. In doing so, some seminal work on market interdependence and integration is reviewed.

## 2.2 Literature Review

Since the pivotal work on portfolio diversification by Grubel (1968), where diversification is demonstrated to reduce risks, there has been a colossal amount of literature concerning global market interrelations. Portfolios that are strategically spread out carries less risk compared to those that comprise of less diverse combinations (Dajcman, 2012). This branching out of portfolios, among diverse stocks from different global markets, can be advantageous only if correlations among the selected global markets are lower (Grubel and Fadner, 1971). Thus it logically follows that high degree of comovement among global markets curtails any benefit arising from branched out assortment of portfolios (Ling and Dhesi, 2010). However, as theoretically demonstrated by French and Poterba (1991), most investors are engulfed by home-bias when composing their portfolios as they expect returns in home market to be higher than markets abroad.

The empirical literature presents mixed evidences regarding the benefits of diversifying the portfolios, with some demonstrating favourable investment scenarios as opposed to others that report less gains from diversified portfolios due to significant correlation between markets. Moreover, divergent results regarding the economic linkages, that drives interdependence and synchronicity between markets, are present in literature. For example, a variety of factors[2], ranging from trade and regional proximity to financial market similarities, determine the strength of interdependence among markets (see Roll, 1992; Flavin et al., 2002, etc.).

Agmon (1972) finds evidence of significant interrelation between markets of the U.S., U.K., Germany and Japan. Strong interdependence between these countries is evidenced as shocks in the equity market of the U.S. immediately impacts other three markets. Moreover, market leadership of the American market is demonstrated as price changes in other markets follow changes in the market of the U.S. On the other hand, Lessard (1973) espouses the formation of an investment union to reap benefits of strategic investment decisions. Moreover, benefits of portfolio diversification is demonstrated for developing countries from a particular regional block. Similarly, Solnik (1974) demonstrates the existence of larger benefits, mainly in terms of reduced portfolio risks, arising out of internationally diversified portfolios as opposed to domestically composed portfolios. Moreover, Jorion and Schwartz (1986) find less evidence of market integration between Canadian and global markets using a maximum likelihood approach. Existence of market segmentation is established thereby evidencing opportunities for risk diversification.

Bertero and Mayer (1990), while investigating market interdependence during the 1987 crisis, find evidence of strong interrelations between the studied markets. The degree of interdependence, as determined by the correlation structure between markets, was more pronounced after the market crash thereby diminishing any benefits from diversified portfolios. On the other hand, Harvey (1995) finds less correlation among developed markets and twenty emerging markets and espouses the inclusion of assets from emerging markets in portfolio combinations.

Eun and Shim (1989) find evidence of strong interdependence among developed stock markets with the U.S. market leading all others in the sample. Moreover, Becker et al. (1990) demonstrate the leading behaviour of the U.S. market over the Japanese market with strong correlation between the two markets. In a similar vein, Hamao et al. (1990) find strong short-run interdependence among markets of the U.S., U.K. and Japan with unidirectional volatility spillover from U.S. to Japan and U.K.

---

[2] These differences in factors explaining market linkages, as argued by Beine and Candelon (2011), exist due to the heterogeneous nature of markets.

The existence of long-run interdependence among developed markets is evidenced by Kasa (1992) using a cointegration method, thereby negating diversification benefits for investors with long-run investment horizons. Similarly, Arshanapalli and Doukas (1993) using a cointegration technique finds evidence in support of increasing co-movement among markets of the U.S. and those of France, Germany and U.K. with the U.S. as the market leader. However, no evidence of interdependence with the Japanese market was found. On the contrary, Aggarwal and Park (1994) find strong evidence of integration between the markets of the U.S. and Japan. Nevertheless, Smith et al. (1993) using a rolling Granger causality approach find no evidence of strong causal relationship among the markets of the U.S., U.K., Japan and Germany, thereby suggesting positive benefits from portfolio diversification.

Phylaktis and Ravazzolo (2002) identified trade among the countries of pacific basin region with Japan and the U.S. as the driving factor behind strong market integration among these economies. However, Phylaktis (2005) after incorporating portfolio and foreign exchange restrictions, find evidence of diversification benefits if equities from emerging economies of pacific basin regions are included in the portfolio. Moreover, short-run diversification benefits are found to be more pronounced than the long-run ones. Gilmore and McManus (2002), on the other hand, find low short-run correlation between the markets of the U.S. and central Europe. Furthermore, after the application of the Johansen cointegrating technique, no long-run relationship among these markets was evinced, indicating diversification benefits for the U.S. investors holding assets from central European markets.

Goetzmann et al. (2001) examine the correlation structure of major world markets for a period of about 150 years and find an increase in market interdependence over the years. Moreover, evidence of time-varying benefits of diversification is found with assets from emerging markets being the pivotal force behind diversification opportunities. Butler and Joaquin (2002), while investigating the benefits of globally diversified portfolios, find significant upsurge in correlations during bear market phases, thereby nullifying any benefits from diversified portfolios. Similarly, Li et al. (2003), after imposing short-selling constraint on G7 economies, find significant diminution of payoffs from diversified investment holdings. On the contrary, Fletcher and Marshall (2005) finds evidence supporting diversification benefits for investors from the U.K. operating in developed markets, even after imposing short-selling constraints.

On the other hand, Pretorius (2002) while investigating the determinants of interdependence, finds trade and industrial growth to be the driving factor behind increased cross-country interrelations. Moreover, markets within a specific geographical region are found to be more correlated than non-regional blocks. However, economic fundamentals were found to be the sole determinant of increasing interdependence among the emerging markets, thereby providing evidence of diversification opportunities while holding emerging markets' assets.

The investigation of market integration between eleven European markets and the U.S., via the use of cointegration methods, is carried out by Laopodis (2005). The existence of cointegration among these economies varied and was not consistent across these groups. Moreover, diversification opportunities for U.S. investors, with both short and long-run investment horizons, is documented. However, the author stresses upon the phenomena of increasing integration over time due to the rising correlation[3] trends among developed markets.

Click and Plummer (2005), in their study of market integration among the south East Asian markets, find evidence of some integration among these markets. Moreover, they find that long-run interdependence is more pronounced than short-run relations, implying reduced diversification benefits for investors with long-run investment horizons. However, they demonstrate that there still exist some degree of diversification opportunities when considering assets from East Asian economies. Nevertheless, the aspect of improved efficiency, as highlighted by the possibility of increasing equity market integration among these countries, is stressed upon as reinforced by the results from cointegration.

De Santis and Gerard (2006) attempt to trace out the determinants of financial market integration by investigating thirty global markets. The influence of the European monetary union (EMU) on investors' portfolio allocation is also documented. In the backdrop of increasing capital flows and rising proportion of savings allocated to global equity markets, the authors demonstrate evidences supporting diminution of home-bias among some European markets. Moreover, the creation of the EMU strengthened the interdependence among Euro area markets.

---

[3] The dynamic evolution of correlation structure across global equity markets is theoretically explained in Bracker and Koch (1999).

Driessen and Laeven (2007) investigate the benefits of portfolio reallocation in both developed and emerging markets. Diversification benefit for home investors investing abroad is documented even after imposing short selling constraints in emerging markets. However, they unveil market risk to be the prime determinant of diversification opportunities, with investors from riskier economies holding more benefits of diversifying their assets. Moreover, investors operating in developing markets are found to enjoy better diversification opportunities as these markets are not properly integrated with global markets.

Hatchondo (2008) develops a theoretical model to explain portfolio diversification mechanism and investors' behaviour. The influence of asymmetric information is demonstrated to induce home bias where investors from the home market invest in their own country's stock and are averse to investing in assets abroad. Moreover, investors in home market outperform their counterpart abroad in correctly detecting the rank of better investment prospects. However, Goetzmann and Kumar (2008), after studying the diversification behaviour of about sixty thousand investors of the U.S., find the dominance of under diversified portfolios. Furthermore, novice investors with lower skill set are found to hold undiversified portfolios, whereas advanced investors hold comparatively better investment bundles. Nevertheless, some erudite and well informed investors intentionally hold less diversified portfolios because of superior information.

The existence of diversification opportunities of local investors from East Asian and South American countries, in the presence of short selling and other investment constraints, is studied by Chiou (2008). Investor from these markets are found to benefit from both local and international diversification. Moreover, inclusion of assets from markets in North America and Europe significantly lowered risks for investors from other markets abroad. Similarly, Middleton et al. (2008) find evidence supporting benefits from diversification by including assets from markets of Eastern European economies. The rise, over the years, in integration and interdependence among global markets notwithstanding, investors from emerging economies still benefit from diversifying internationally. Furthermore, Chiou (2009) investigates the benefits of diversification, in the presence of some investment constraints, for investors from the U.S. and finds reduced opportunities from diversified asset combinations. However, benefits from diversifying internationally are not entirely eliminated.

Flavin and Panopoulu (2009) explore the stability of diversification opportunities over time for the G7 markets using various regime-switching models of volatility. Stability of market interrelations is evidenced whereas increasing interdependence between markets during times of financial turmoil, as reported in a vast majority of studies, is not established. Moreover, investors from the U.S. holding assets from foreign markets are found to possess lesser risks. Additionally, diversification benefits are robust to periods of varying volatility as benefits are not reduced for investors during these periods.

Interdependence among markets from several Asian economies, the U.S. and the U.K. is examined by Awokuse et al. (2009). Cointegrating method in a rolling window framework is applied to investigate the time dependent nature of cointegration. Significant increase in inter-market linkages and integration is reported after periods of financial liberalization, where markets from the U.S. and Japan are found to lead all other markets in the studied sample. Furthermore, the instable nature of changing cointegration relationships over time is conjectured to potentially limit investors from pursuing international diversification strategies.

The integration of equity markets of India and twelve Asian markets is studied by Mukherjee and Mishra (2010) using a GARCH based framework. Significant bi-directional transmission of return spillovers between India and major Asian markets is evidenced in the short-run, thereby reducing any short-run diversification opportunities for Indian investors.

Coeurdacier and Guibaud (2011) investigate inter market equity relationships to study local investors' portfolio allocation in the presence of home-bias and other endogenous preferences. Evidence of investors allocating their assets with markets possessing superior diversification benefits is documented. Nevertheless, the presence of investors' home-bias does not shield some informationally sophisticated investors from reaping the benefits arising out of international diversification of assets comprising their portfolios.

Liu (2013) examines the interdependence structure of developing and developed markets and attempts to trace various linkages that affect the nature and degree of interdependence among these markets. The linkages driving interdependence among developed markets are found to be different than linkages driving interdependence among emerging markets. The similarity of industrial organisation is found to be the major factor influencing interrelations among developed markets whereas financial linkages are found to influence the comovement among emerging markets.

Zhang and Li (2014) investigate interdependence between the markets of the U.S. and China using dynamic conditional correlation (DCC) and cointegration methodologies. No evidence of long run relationship between the two markets is found. However, correlation between these markets are found to be time dependent and rising over the years, thereby indicating the diminution of diversification benefits for both local and foreign investors. Moreover, short-term shocks from the U.S., especially during periods of financial turmoil, are demonstrated to have an impact on the Chinese market. More recently, Al Nasser and Hajilee (2016) investigate interdependence among five emerging markets and developed markets of the U.S., Germany and U.K. Evidence of short-run integration between the selected emerging and developed markets is documented.

The majority of studies on equity market integration and interdependence, however, use traditional time domain econometric and statistical methodologies to investigate the relationship between equity markets. However, analysis of long-run and short-run interrelationships are carried out using traditional methods fail to simultaneous capture investors' decision encompassing heterogeneous time horizons. Nevertheless, few recent studies attempt to uncover both short and long run dynamics using wavelets and related time-frequency techniques, thereby allowing to capture information about heterogeneous investors' choices and investment decisions.

Earlier studies using wavelet based time-frequency techniques find an increase in comovement between developed equity markets at lower frequencies associated with long-run investment holding periods, thereby diminishing diversification benefits for investors who operate in long-run investment horizons ( see Rua and Nunes, 2009, Ranta, 2010). Similarly, Dajcman et al. (2012), using wavelet methods, document the existence of scale dependent comovements among markets of select developed economies.

In a similar vein, Graham et al. (2012) investigate interdependence among twenty global equity markets using continuous wavelet methods, allowing them to identify, i) investors' diversification opportunities at varying time horizons, and ii) the dynamic evolution and time varying nature of equity market comovements. In their study of comovement between emerging markets and the U.S., lower comovement is documented at short-run investment horizons, thereby providing diversification opportunities for investors with short-run horizons. Similarly, Graham and Nikkinen (2011) study the market interrelations among Finland and select global markets and find potential diversification benefits for Finnish investors in the short-run.

The comovement between select Asian equity markets is examined by Tiwari et al. (2013) using wavelet multiple cross-correlation methods. The selected Asian markets are found to be strongly interdependent at lower time-horizons but not integrated at shorter time-horizons, thereby providing investors with short-run diversification opportunities. More recently, Tiwari et al. (2016), using both continuous and discrete wavelet methods, uncover interesting information on comovements among the European markets during the aftermath of the Eurozone crisis. Evidence of contagion between some European markets and the related implications for portfolio diversification are discussed.

Interestingly, Lehkonen and Heimonen (2014), use a hybridization of DCC-GARCH and wavelet methods to investigate comovements between markets from BRIC and other developed economies. The level of interdependence is found to be associated with geographical region. Moreover, some diversification benefits are demonstrated to exist when including assets from the BRIC markets. Furthermore, using a similar hybridisation algorithm, Najeeb et al. (2015) test for time-scale dependent interdependence between Malaysian and select developed and emerging markets. Diversification opportunities for Malaysian investors are evidenced to exist only for short-run investment horizons whereas in the long-run Malaysian equity market exhibit high correlation with other markets in the sample.

Alaoui et al. (2015), using both discrete and continuous wavelet methods, find significant impact of the London interbank offered rate (LIBOR) on Islamic emerging markets at certain investment holding periods, providing investors with useful information for strategizing purposes.

Nonetheless, all of the above aforementioned studies on market interdependence and integration does not explore the portfolio diversification implications for Indian investors. Moreover, the existence of heterogeneous Indian investors and their related investment holding periods, suitably captured by wavelet based multiresolution methods, necessitates an investigation on these lines where both heterogeneity and timescale dependency of inter market correlation structure are effectively reconnoitred. Furthermore, the lack of studies on these lines, investigating equity market integration with a special focus on India, calls for analyses predominantly focusing on portfolio diversification opportunities for Indian investors. The following section briefly describe the relevant methods and tools used in this chapter.

## 2.3 Methodology

A wavelet is a function $\psi(.)$ defined on $\mathbb{R}$ such that $\int_{\mathbb{R}} \psi(t)dt = 0$ and $\int_{-\infty}^{\infty} |\psi(t)|^2 dt = 1$. A signal can be decomposed into its finer detail and smoother components by projecting the signal onto mother and father wavelets given by $\psi$ and $\phi$ respectively. Dilation and translation operation is performed on both mother and father wavelets to form a basis for the space of squared integrable function, $L^2(\mathbb{R})$. Therefore, any function $x(t)$ in $L^2(\mathbb{R})$ can be represented as linear combinations of these basis functions. The dilated and translated versions of mother and father wavelets are denoted by $\psi_{b,s}(t)$ and $\phi_{b,s}(t)$ respectively, where

$$\psi_{b,s}(t) = \frac{1}{\sqrt{s}} \psi\left(\frac{t-b}{s}\right) \quad (2.1)$$

$$\phi_{b,s}(t) = \frac{1}{\sqrt{s}} \psi\left(\frac{t-b}{s}\right) \quad (2.2)$$

$s$ and $b$ represents the scaling (dilation) and translation parameter, respectively. Here $s=1,......S$ controls the number of multiresolution elements. Formally, a function $x(t)$ can be represented in the wavelet space as

$$x(t) = \sum_b a_{S,b} \phi_{S,b}(t) + \sum_b d_{S,b} \psi_{S,b}(t) + \sum_b d_{S-1,b} \psi_{S-1,b}(t) + ... + \sum_b d_{1,b} \psi_{1,b}(t) \quad (2.3)$$

where $a_{S,b}$ are coefficients describing coarser features of $x(t)$, and $d_{S,b}$ are detail coefficients that captures information from multiple resolutions or time-horizons.

*Wavelet based correlation and cross-correlation*

Let $X_t = (x_{1,t}, x_{2,t})$ be a "bivariate stochastic process with univariate spectra" (autospectra) $S_1(f)$ and $S_2(f)$ respectively, and let $W_{s,b} = (w_{1,s,b}, w_{2,s,b})$ be the scale $s$ wavelet coefficients computed from $X_t$. These wavelet coefficients are obtained by applying the wavelet transform to all elements of $X_t$. The obtained wavelet coefficient contains both $a_{S,b}$ (coarser approximations) and $d_{s,b}$ (wavelet details). For a given scale $s$, the wavelet covariance between $x_{1,t}$ and $x_{2,t}$ is given by

$$\gamma_X(s) = \frac{1}{2\pi} Cov(w_{1,s,b}, w_{2,s,b}) \quad (2.4)$$

The wavelet covariance "decomposes the covariance of a bivariate process on a scale-by-scale basis", i.e.

$$\sum_{s=1}^{\infty} \gamma_X(s) = Cov(x_{1,t}, x_{2,t}) \quad (2.5)$$

By introducing an integer lag $\tau$ between $w_{1,s,b}$ and $w_{2,s,b}$, the notion of wavelet cross-covariance can be introduced, and is given by

$$\gamma_{X,\tau}(s) = \frac{1}{2\pi} Cov(w_{1,s,b}, w_{2,s,b+\tau}) \quad (2.6)$$

In some situations it may be beneficial to normalize the wavelet covariance by wavelet variance, which gives us wavelet correlation

$$\rho_X(s) = \frac{\gamma_X(s)}{\sigma_1(s)\sigma_2(s)} \quad (2.7)$$

where $\sigma_1^2(s)$ and $\sigma_2^2(s)$ are the wavelet variances of $x_{1,t}$ and $x_{2,t}$ (at scale $s$), respectively. Just like the usual correlation coefficient between two random variables, $|\rho_X(s)| < 1$. However, wavelet correlation gives correlation among variables from a multiscale dimension Also, by allowing the two processes $x_{1,t}$ and $x_{2,t}$ to differ by an integer lag $\tau$, we can define wavelet cross-correlation, which gives us the lead-lag relationship between two processes, on a scale-by scale basis. The approximate confidence bands for the estimates of wavelet correlation and cross-correlation is given in Percival and Walden (2000) and Gencay et al. (2002). Moreover, the reader is referred to Fernandez-Macho (2012) for the technique of wavelet multiple correlation (WMC) and multiple cross-correlation (WMCC).

## 2.4 Empirical Analysis of Interdependence

2.4.1 *Empirical data*

The empirical data consists of twenty four major stock indices comprising both developed and emerging markets. The stock indices included are BSE 30 (India), Nasdaq (U.S.), S&P 500 (U.S.), DJIA (U.S.), FTSE 100 (Great Britain), CAC40 (France), DAX 30 (Germany), NIKKEI 225 (Japan), KOSPI (Korea), KLSE (Malaysia), JKSE (Indonesia), TAIEX (Taiwan), SSE (China), STI (Singapore), HSI (Hong Kong), BEL20 (Belgium), ATX (Austria), AEX (Netherlands), IBEX 35 (Spain), SMI (Switzerland), STOXX50 (Eurozone), ASX 200 (Australia), KSE100 (Pakistan), and IBOV (Brazil). The period of study ranges from 01-07-1997 to 20-01-2014 consisting of 4096 dyadic length observations making it suitable for various wavelet methods. Returns of all the stock indices are calculated by taking first order logarithmic differences.

The descriptive statistics of index returns is summarised in Table 2. The table also reports Shapiro-Wilk and Jarque-Bera tests of normality along with the p-values in parentheses. Normality is rejected for returns from all markets. Moreover, results from the ADF and KPSS unit root tests are also reported in the table. The null of non-stationarity is rejected by the ADF test whereas the p-values from the KPSS test fail to reject the null of stationarity. Therefore, results from both unit root tests show that returns for all markets are stationary. The following section proceeds with the classical analysis of wavelet correlation and wavelet cross-correlation for select market pairs.

2.4.2 *Results from classical wavelet correlation analysis.*

The empirical analysis begins with the classical wavelet correlation and cross-correlation analysis of select stock market pairs. The returns from all markets are decomposed using the MODWT method into six levels of resolution, corresponding to the first six details. The extracted MODWT detail coefficients *d1, d2, d3, d4, d5, d6* correspond to the time-scale, or investment-horizon, of one-two days, two-four days, four-eight days, eight-sixteen days, sixteen-thirty two days, and thirty two-sixty four days, respectively. The filter used in the wavelet multiresolution decomposition is the "*Daubechies least asymmetric*" with length eight (LA8). This filter is said to be the most appropriate filter in decomposing financial time-series (see Percival and Walden, 2000; Gencay et al., 2002, among others.). Moreover, edge effects are taken care by implementing the brick-wall[4] boundary condition on the decomposed MODWT series. In the next step, the estimator of wavelet correlation and cross-correlation is calculated from the MODWT decomposed returns. Figure 2.1 displays the plot of wavelet correlation, along with the associated lower and upper confidence bands, among the stock returns of BSE 30 and the developed markets of CAC40, DAX, FTSE, SMI, SP500 and STOXX50.

Table 2.1 Descriptive statistics of stock returns

---

[4] As finite length time-series causes boundary problems, the decomposed MODWT coefficients near the boundaries are replaced by null values during computation. This is known as the brick-wall method.

Table 2. Stock Returns Descriptive Statistics.

| Index return | BSE 30 | FTSE 100 | SP 500 | CAC 40 | DAX | DJIA | NASDAQ | NIKKEI | KOSPI | JKSE | KLSE | TAIEX |
|---|---|---|---|---|---|---|---|---|---|---|---|---|
| Mean | 0.0004 | 0.0001 | 0.0002 | 0.0001 | 0.0002 | 0.0002 | 0.0002 | -0.0001 | 0.0002 | 0.0005 | 0.0001 | 0.0000 |
| Median | 0.0010 | 0.0000 | 0.0006 | 0.0004 | 0.0008 | 0.0004 | 0.0011 | 0.0001 | 0.0008 | 0.0009 | 0.0003 | 0.0002 |
| Min | -0.1181 | -0.0926 | -0.0947 | -0.0947 | -0.0743 | -0.0820 | -0.1017 | -0.1211 | -0.1280 | -0.1273 | -0.2415 | -0.0994 |
| Max | 0.1599 | 0.0938 | 0.1096 | 0.1059 | 0.1080 | 0.1051 | 0.1325 | 0.1323 | 0.1128 | 0.1313 | 0.2082 | 0.0852 |
| Std.dev. | 0.0165 | 0.0126 | 0.0131 | 0.0154 | 0.0160 | 0.0123 | 0.0174 | 0.0158 | 0.0194 | 0.0172 | 0.0149 | 0.0153 |
| Skewness | -0.0901 | -0.1396 | -0.2077 | 0.0026 | -0.0399 | -0.1251 | -0.0339 | -0.3271 | -0.2072 | -0.1923 | 0.4339 | -0.1515 |
| Kurtosis | 5.5653 | 5.3788 | 7.1822 | 4.1753 | 3.5807 | 7.2239 | 4.5284 | 5.4478 | 4.3604 | 6.9235 | 53.2819 | 2.6866 |
| ADF Test (prob.) | -15.2644(0.01) | -16.1477(0.01) | -15.9515(0.01) | -15.5581(0.01) | -15.6685(0.01) | -16.1165(0.01) | -15.0395(0.01) | -16.3033(0.01) | -15.0912(0.01) | -13.8821(0.01) | -15.1110(0.01) | -15.3926(0.01) |
| KPSS Test (prob.) | 0.1186(0.10) | 0.0659(0.10) | 0.0886(0.10) | 0.1330(0.10) | 0.0736(0.10) | 0.0542(0.10) | 0.0950(0.10) | 0.1882(0.10) | 0.0664(0.10) | 0.2426(0.10) | 0.2067(0.10) | 0.1004(0.10) |
| Jarque-Bera (prob.) | 5,299.5(0.00) | 4,958.5(0.00) | 8,845.4(0.00) | 2,980.3(0.00) | 2,193.2(0.00) | 8,929.2(0.00) | 3,506.2(0.00) | 5,145.8(0.00) | 3,279.6(0.00) | 8,217.7(0.00) | 4,85,145.0(0.00) | 1,250.1(0.00) |
| Shapiro-Wilk (prob.) | 0.9518(0.00) | 0.9426(0.00) | 0.9251(0.00) | 0.9542(0.00) | 0.9574(0.00) | 0.9264(0.00) | 0.9464(0.00) | 0.9541(0.00) | 0.9393(0.00) | 0.9173(0.00) | 0.7084(0.00) | 0.9655(0.00) |
| Count | 4096 | 4096 | 4096 | 4096 | 4096 | 4096 | 4096 | 4096 | 4096 | 4096 | 4096 | 4096 |

| Index return | SSE | STI | HSI | BEL 20 | ATX | AEX | IBEX | KSE 100 | SMI | STOXX 50 | IBOVESPA | ASX |
|---|---|---|---|---|---|---|---|---|---|---|---|---|
| Mean | 0.0001 | 0.0001 | 0.0001 | 0.0000 | 0.0002 | 0.0000 | 0.0001 | 0.0007 | 0.0001 | 0.0000 | 0.0003 | 0.0002 |
| Median | 0.0000 | 0.0002 | 0.0003 | 0.0003 | 0.0007 | 0.0005 | 0.0007 | 0.0013 | 0.0005 | 0.0003 | 0.0009 | 0.0004 |
| Min | -0.0926 | -0.0915 | -0.1473 | -0.0832 | -0.1025 | -0.0959 | -0.0959 | -0.1321 | -0.0811 | -0.0821 | -0.1721 | -0.0870 |
| Max | 0.0940 | 0.1287 | 0.1725 | 0.0933 | 0.1202 | 0.1003 | 0.1348 | 0.1276 | 0.1079 | 0.1044 | 0.2883 | 0.0572 |
| Std.dev. | 0.0158 | 0.0139 | 0.0176 | 0.0132 | 0.0148 | 0.0155 | 0.0157 | 0.0161 | 0.0127 | 0.0156 | 0.0217 | 0.0103 |
| Skewness | -0.1170 | 0.0197 | 0.1180 | 0.0379 | -0.3738 | -0.0967 | 0.0301 | -0.3939 | -0.0390 | -0.0180 | 0.3375 | -0.4811 |
| Kurtosis | 4.5100 | 6.9774 | 9.4260 | 5.2511 | 6.7962 | 5.1402 | 4.4219 | 5.8745 | 5.1257 | 3.8713 | 13.0613 | 5.8776 |
| ADF Test (prob.) | -14.2153(0.01) | -14.8924(0.01) | -15.4102(0.01) | -15.8007(0.01) | -14.6775(0.01) | -15.2686(0.01) | -14.8537(0.01) | -14.2848(0.01) | -16.0505(0.01) | -15.5020(0.01) | -14.7384(0.01) | -16.5294(0.01) |
| KPSS Test (prob.) | 0.1217(0.10) | 0.0708(0.10) | 0.0768(0.10) | 0.0925(0.10) | 0.1481(0.10) | 0.0784(0.10) | 0.0949(0.10) | 0.1315(0.10) | 0.0656(0.10) | 0.1313(0.10) | 0.0941(0.10) | 0.0751(0.10) |
| Jarque-Bera (prob.) | 3,486.4(0.00) | 8,320.7(0.00) | 15,192.7(0.00) | 4,714.2(0.00) | 7,989.4(0.00) | 4,522.6(0.00) | 3,343.1(0.00) | 6,004.3(0.00) | 4,491.8(0.00) | 2,562.5(0.00) | 29,228.1(0.00) | 6,062.8(0.00) |
| Shapiro-Wilk (prob.) | 0.9397(0.00) | 0.9268(0.00) | 0.9140(0.00) | 0.9432(0.00) | 0.9212(0.00) | 0.9354(0.00) | 0.9574(0.00) | 0.9277(0.00) | 0.9407(0.00) | 0.9543(0.00) | 0.9193(0.00) | 0.9414(0.00) |
| Count | 4096 | 4096 | 4096 | 4096 | 4096 | 4096 | 4096 | 4096 | 4096 | 4096 | 4096 | 4096 |

Note: Results from ADF and KPSS unit root tests along with respective p-values, in parentheses, are reported.

The correlation between BSE 30 and all other five markets, at all six levels of wavelet details, is very low. Moreover, there exist no statistically significant wavelet correlations as the confidence interval span the zero axis in all six data pairs. Wavelet correlation between BSE30 and the markets of ATX, IBOV, HSI, KLSE, KOSPI and NIKKEI is shown in Figure 2.2. There is a rise in correlation from short-run timescales to long-run timescales. Strong wavelet correlation between BSE 30 and ATX, BSE30 and IBOV, BSE30 and HSI and BSE30 and KLSE, after 4-8 days investment horizon, can be evidenced from the correlation plots. This shows that Indian investors should be careful while including assets from these markets as diversification benefits are significantly reduced with rise in correlations.

Moreover, BSE30 seems to be significantly correlated with KOSPI at the time-horizon of 32-64 days. However wavelet correlation between the BSE30-KOSPI pair is not statistically significant up to 16-32 days investment horizons, thereby providing Indian investors with some short-run diversification benefits while including assets from the South Korean equity market.

Note: The wavelet correlation plots with the lower and upper confidence bands, indexed as L and U, is given along with six levels of wavelet decomposition[5]. The horizontal axis shows the level of decomposition whereas the vertical axis gives correlation values ranging from -1 to 1. All wavelet correlation computations are performed with the LA (8) wavelet filter after tackling the boundary effects using the brick-wall condition.

Figure 2.1 Wavelet correlation of BSE 30 with CAC40, DAX, FTSE, SMI, SP500 and STOXX

---

[5] The highest number of levels that a series of length $N$ can be decomposed into is given by $\log_2(N)$.

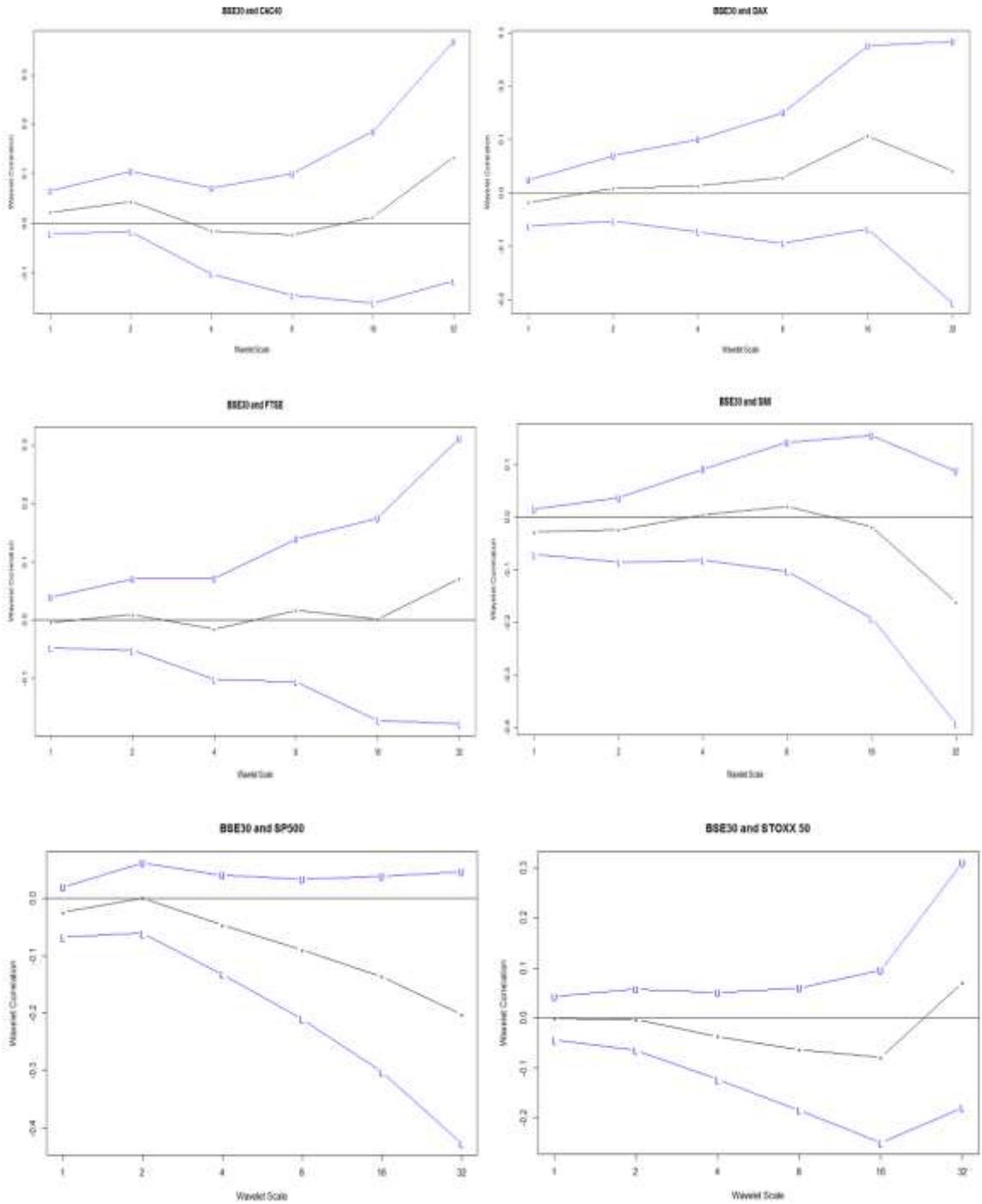

Figure 2.2 Wavelet correlation of BSE 30 with ATX, IBOV, HSI, KLSE, KOSPI and NIKKEI

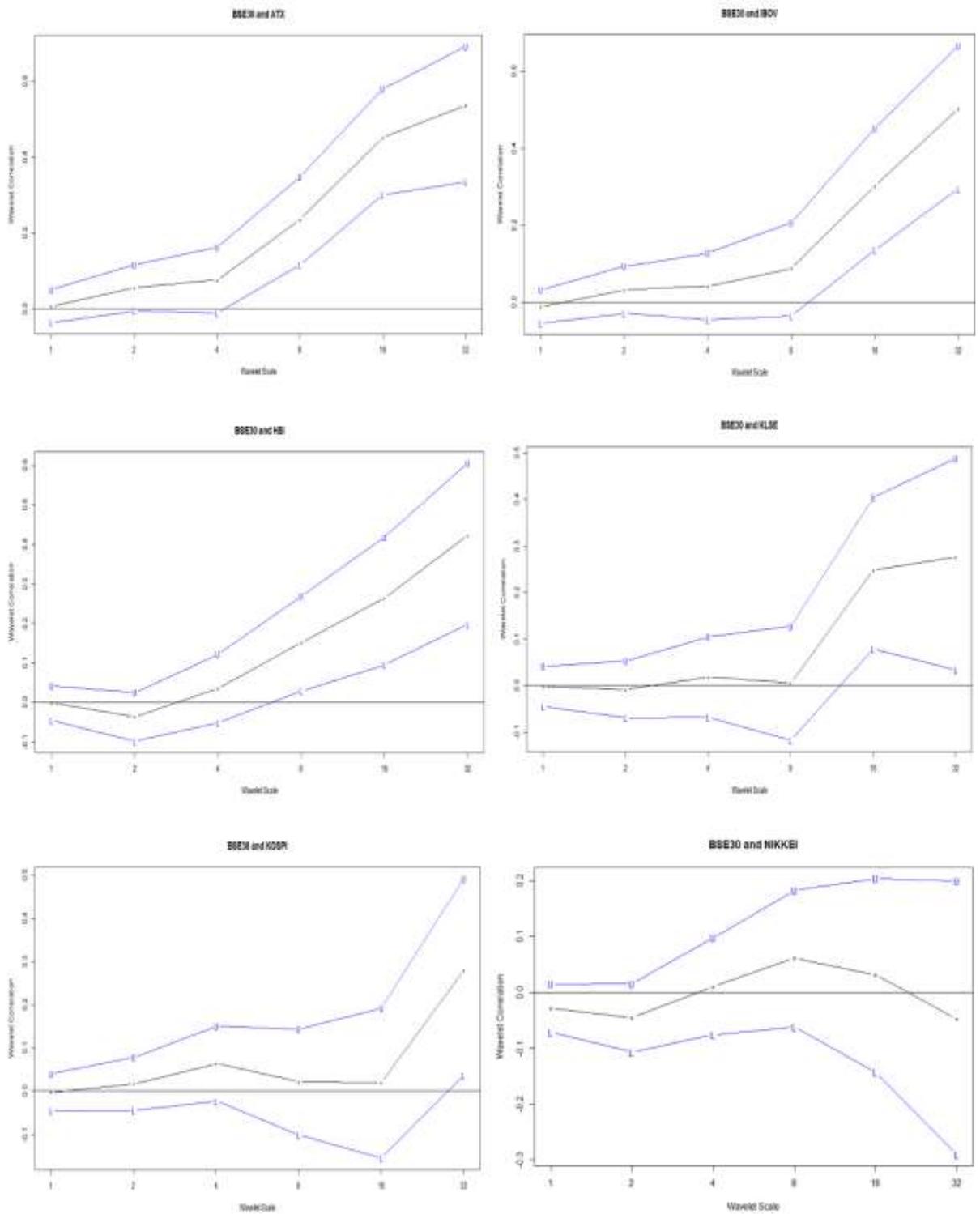

Furthermore, wavelet correlation between BSE 30 and NIKKEI is not statistically significant as the lower confidence band span the zero axis.

Figure 2.3 Wavelet cross-correlation of BSE30-ATX, BSE30-HSI and BSE30-KOSPI

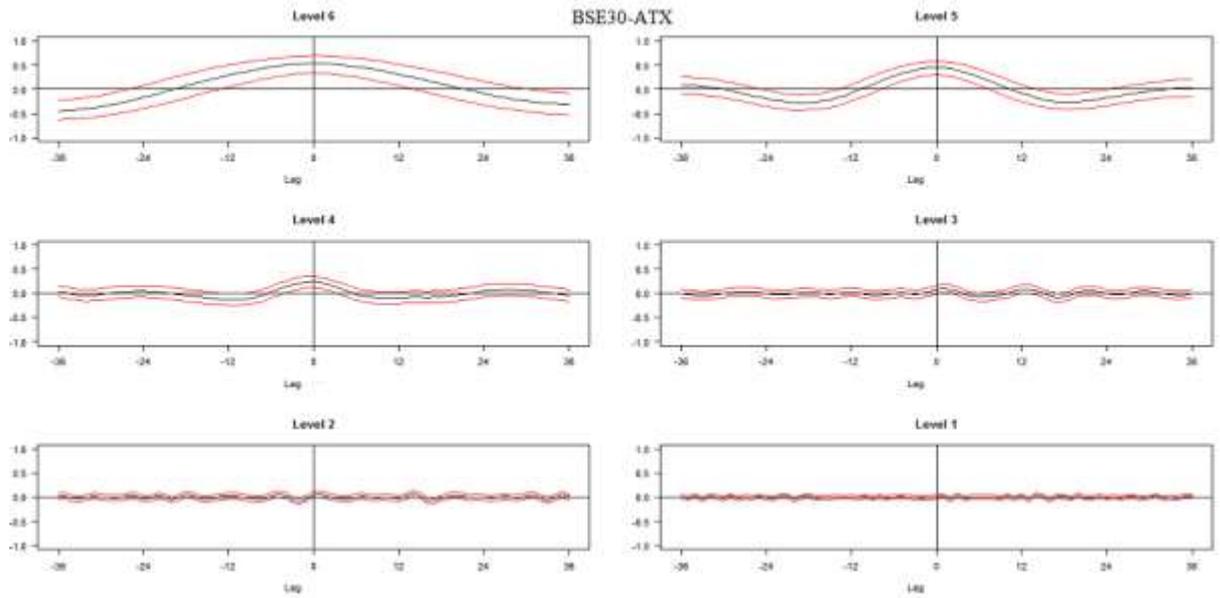

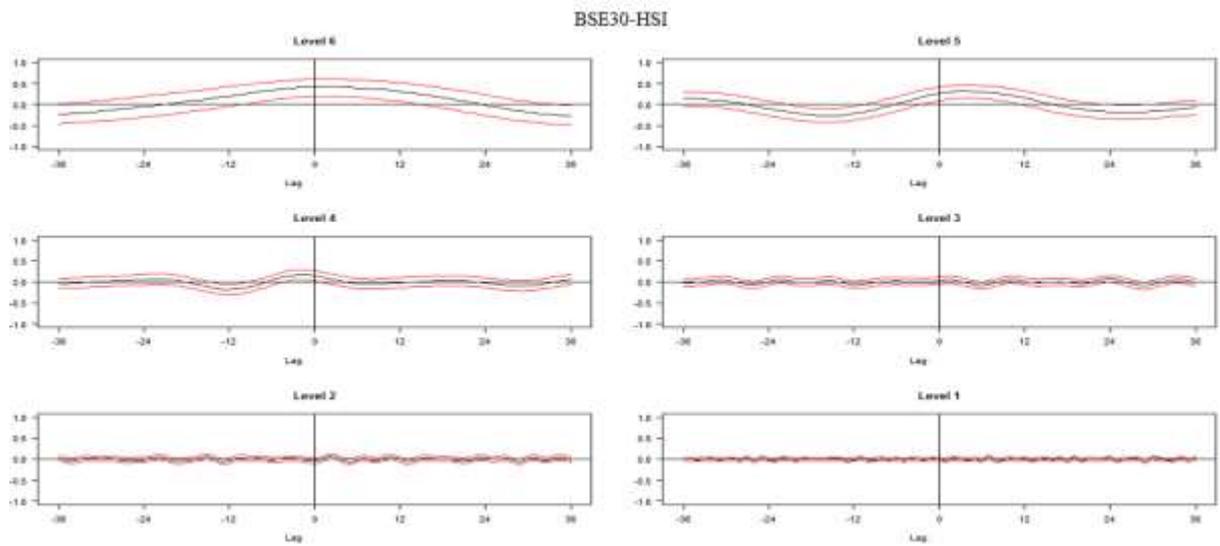

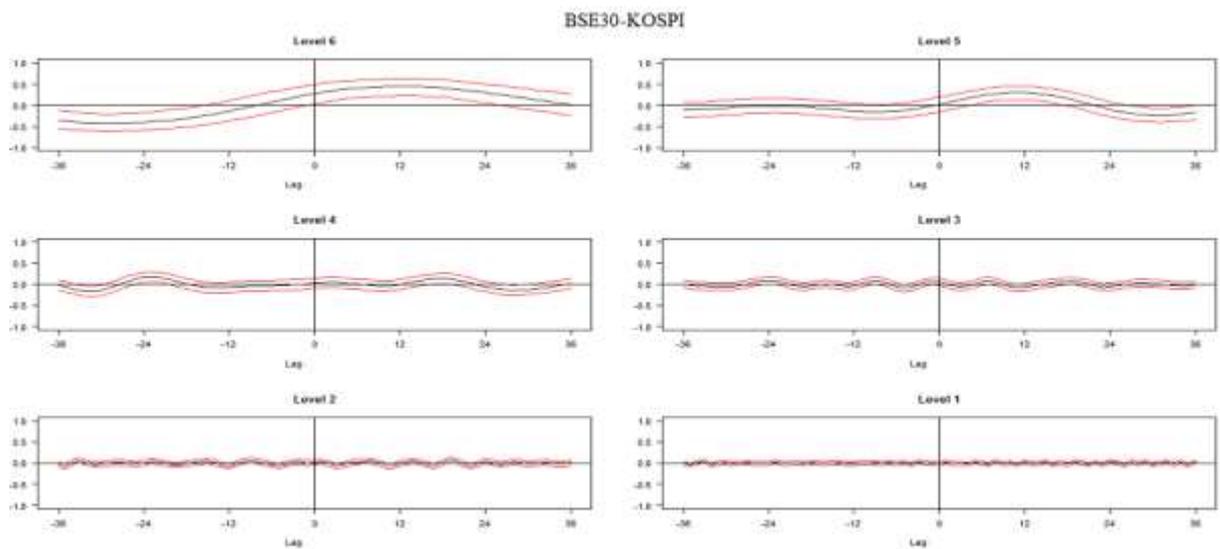

Figure 2.3 gives the wavelet cross-correlation plots of BSE30-ATX, BSE30-HSI and BSE30-KOSPI. The horizontal axis displays lags (in days) whereas in the vertical axis correlations are shown. Significant contemporaneous correlation between BSE30 and ATX at level 5, corresponding to the time-horizon of 16-32 days, or monthly time-scale, can be observed from the plot. The same is true for time-horizons of 8-16 and 32-64 days, corresponding to levels four and six respectively. Moreover, the cross-correlation plots at these levels seem to be slightly skewed towards the right, indicating the leading behaviour of ATX over BSE30. Similar results can be observed with BSE30-HSI and BSE-KOSPI pairs, where some significant correlation can be observed beyond monthly time horizon. Both KOSPI and HSI seem to lead BSE30 at the monthly time horizon and beyond. However, BSE30 leads HSI at level 4 corresponding to time horizon of 8-16 days. This means that changes in BSE30 is followed by changes in HSI eight to sixteen days later.

Figure 2.4 demonstrates wavelet cross-correlations for BSE30-IBOV and BSE30-KLSE pairs. The correlation seem to increase as the timescale increases. BSE30 and IBOV show signs of some cross-correlation at level 2, corresponding to 2-4 day timescale, at a lead (negative six lag) of around six day. This means that the present day returns of IBOV is related to the returns of BSE30 six days later. Moreover, some signs of left asymmetry shows that BSE30 leads IBOV at the time-horizon of 2-4 days, implying that changes in BSE30 are followed by changes in IBOV 2-4 days later. Furthermore, some strong correlations between BSE30 and IBOV beyond level 4 wavelet decomposition can be evidenced from the plot. At levels four and five, since the cross-correlation plot is skewed to the left, BSE30 leads IBOV at both fortnightly and monthly time-horizons. This skewness is not very pronounced at the sixth level, corresponding to the investment horizon of 32-64 days, thereby making it difficult to interpret the lead-lag behaviour at this level. However, the contemporaneous correlation at this timescale seem to be strong between the returns of BSE30 and IBOV.

Some signs of cross-correlation between BSE30 and KLSE is observed at 2-4 days timescale around 22 and 26 days lag. The leading behaviour of KLSE is also apparent owing to right skewness of the plot. However, no statistically significant cross-correlations at levels three and four, corresponding to the investment horizons of 4-8 and 8-16 days, can be observed as the lower confidence band is homogeneously distributed below the zero axis. However, both contemporaneous correlation and cross-correlations up to lag of six days are found to be statistically significant at the time-horizon of 16-32 days. The cross-correlation at level 6 up to six days lag is also significant where KLSE leads BSE30.

Figure 2.4 Wavelet cross-correlation of BSE30-IBOV and BSE30-KLSE

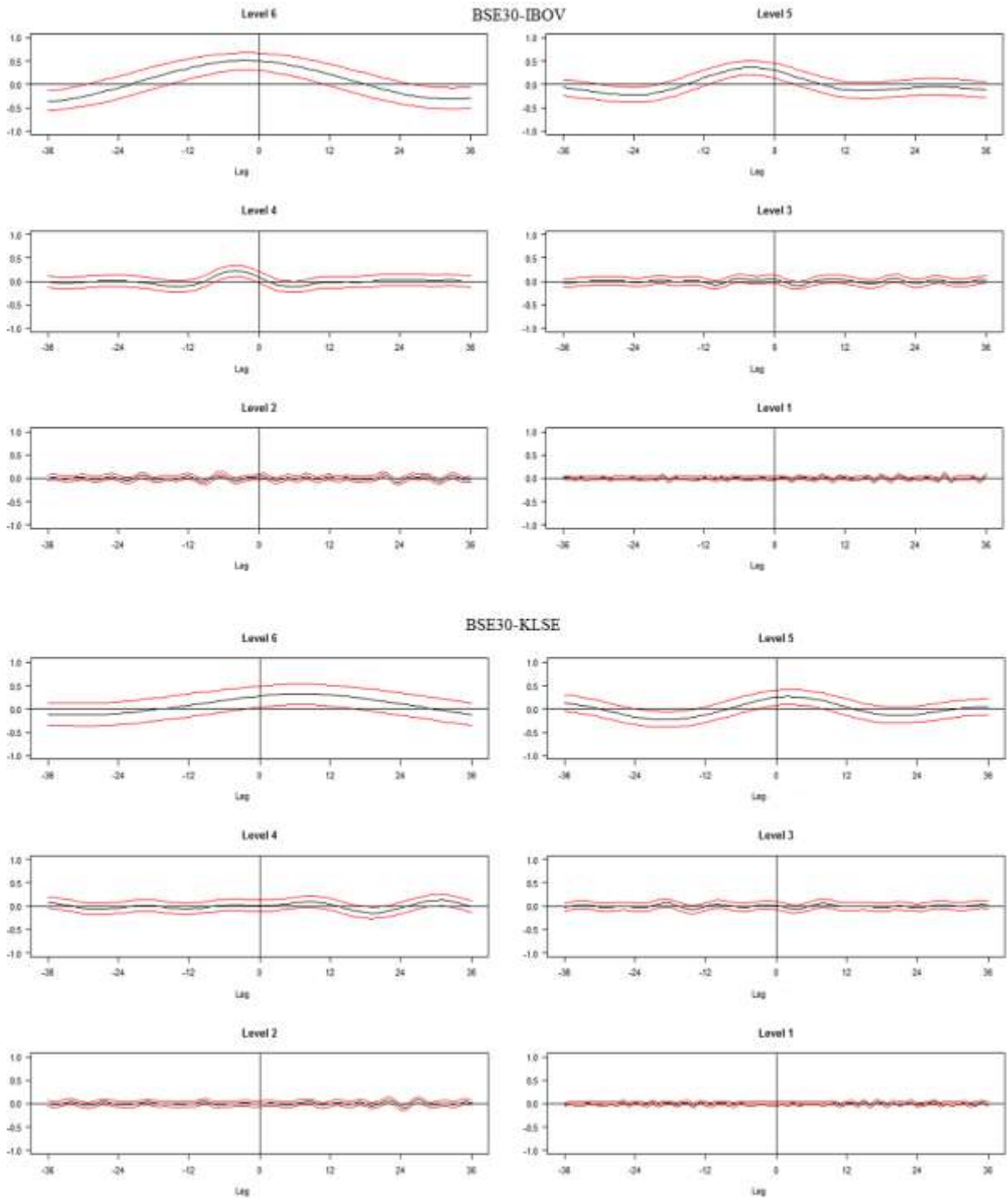

The leading behaviour of KLSE over BSE30 at time-horizons of 16-32 and 32-64 days makes it clear that changes in KLSE is followed by changes in BSE30 up to time-horizon of two months. Therefore, Indian investors who operate at these time-horizons need to be careful while considering Malaysian assets in their portfolios.

Figure 2.5 Wavelet cross-correlation of CAC40-DAX and CAC40-SP500

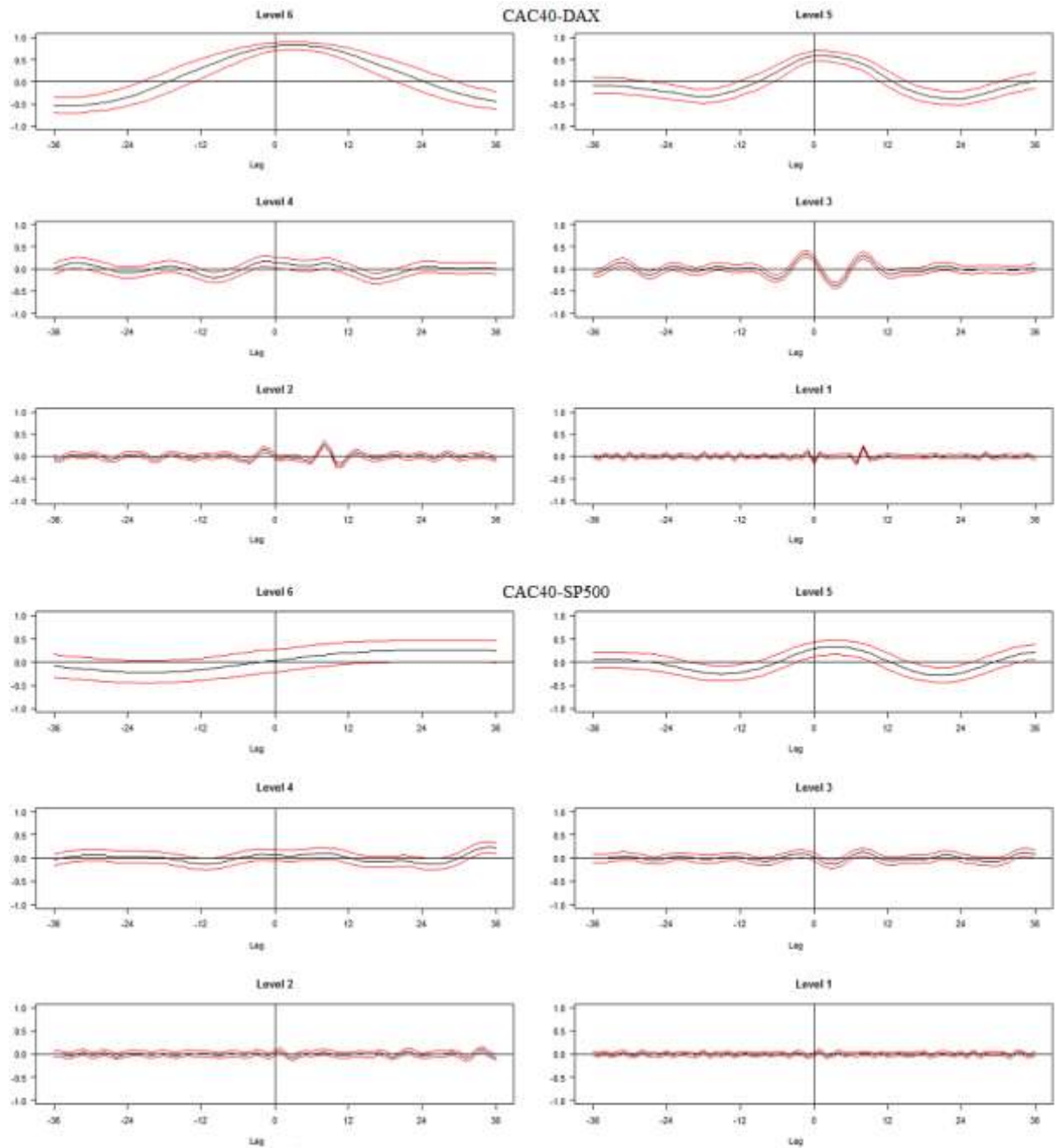

Cross-correlations among the developed markets of Germany, France and the U.S. is given in Figure 2.5. Strong market integration can be evidenced from the cross-correlation of CAC40-DAX pair with cross-correlations increasing as we move towards long-run time-horizons. DAX seems to leads CAC40 at levels one, two, three, and five corresponding to daily, intra-weekly, weekly and monthly time-horizons. Similarly, some significant cross-correlation can be observed between SP500 and CAC40 at levels three and five, corresponding to investment horizons.

Figure 2.6 Wavelet cross-correlation of DAX-IBEX and SP500-IBEX

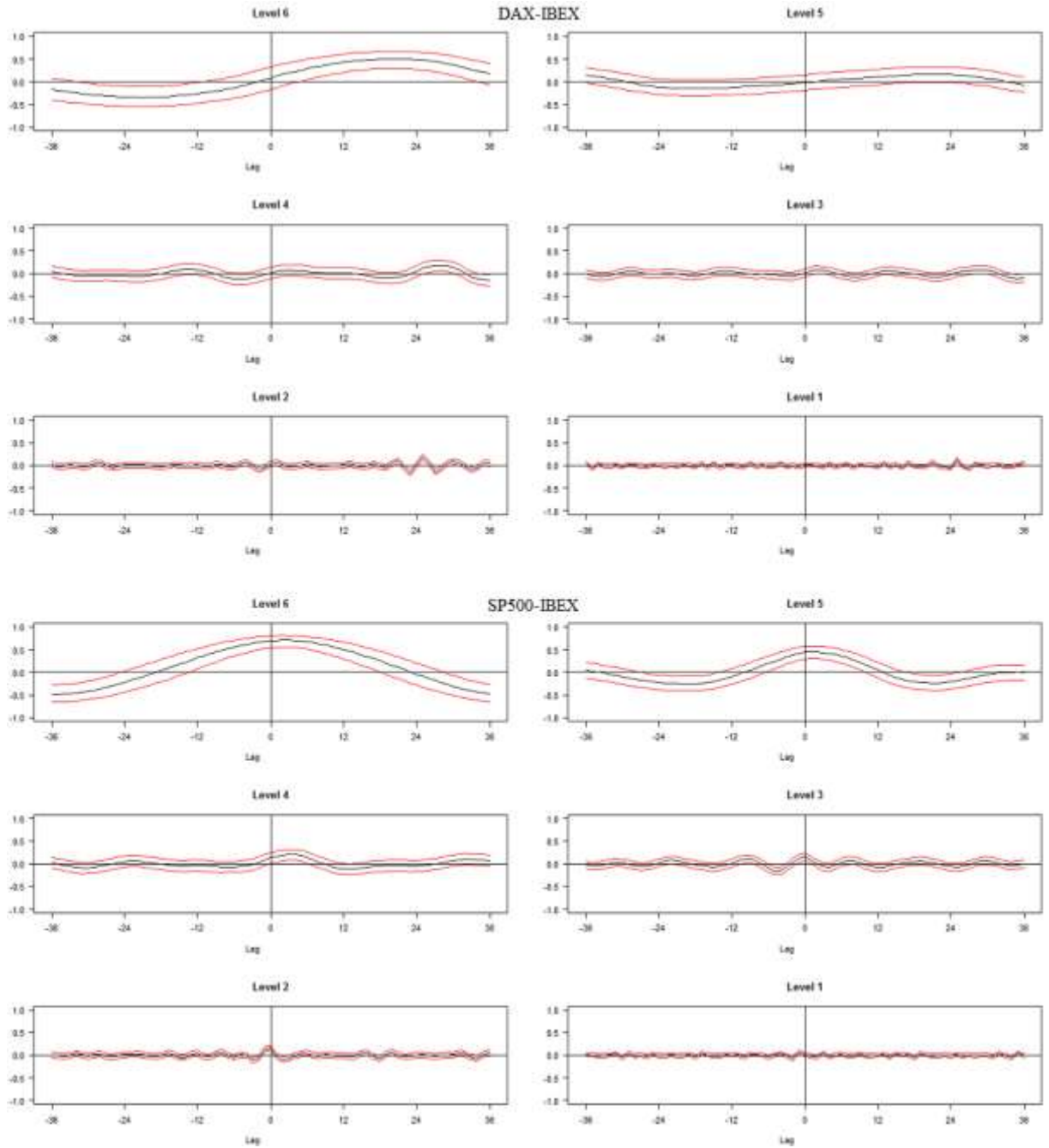

The markets of the U.S. and Spain are found to be correlated at lower time-horizons as the cross-correlation of the SP500-IBEX returns pair indicate statistically significant correlations at levels two, three and four. At the intra-weekly time horizon of 2-4 days, SP500 leads IBEX which is evident from a slight left skewness of the plot, where correlation peaks at around a lead of one day (lag of -1). This indicates that SP500 returns is correlated with the next day returns of IBEX for the intra-weekly (2-4 days) time-horizon, implying that changes in SP500 is followed by changes in IBEX 2-4 days later. Moreover, correlation among SP500 and IBEX tends to increase with the level of wavelet decomposition. Some significant contemporaneous correlation can be observed at the weekly time horizon (4-8 days). However, at the fortnightly time horizon (8-16 days) IBEX leads SP500 up to eight days lag, where cross-correlations are found to be significant during this period. Moreover, strong contemporaneous correlations between the returns of SP500 and IBEX can be observed at the monthly and two-monthly (32-64 days) time horizons. Furthermore, significant short-run comovement dynamics can be observed between stock returns of Germany and Spain where DAX is correlated to the past returns (with a monthly lag) of IBEX. IBEX leads the returns of DAX at both daily and intra-weekly time-horizons. The results from both classical wavelet correlation and wavelet cross-correlation, for all market pairs, are available upon request. However, only important and significant results for illustration purposes are reported here.

Furthermore, results from classical wavelet correlation analyses involves colossal amount of output which entails cumbersome graphical plots. For e.g. the incorporation of all possible data pairs, from the sample of markets considered in this chapter, in the bivariate wavelet correlation analysis, leads to the generation of *N(N-1)/2* graphical plots, which equals 24×(24-1)/2=276 correlation plots! Moreover, the cross-correlation plots generated would be even larger in number as the levels of wavelet decomposition (say, *J*) need to be considered too, thereby generating plots to the tune of *J × N (N-1)/2*. Therefore, a much newer technique of "*wavelet-multiple correlation and wavelet multiple cross-correlation*" (Fernandez-Macho, 2012), which can handle multivariate time-series as opposed to the bivariate classical wavelet correlation methods, is implemented for the analysis of market interdependence among some group of markets.

In essence, wavelet multiple correlation (WMCor, hereafter) allows multivariate time-series as inputs and generates significant correlation information in a single plot. This is achieved due to the fact that the output of WMCor contains a single list of wavelet correlation[6] coefficients obtained from maximum values of the square root of $R^2$. Similarly, wavelet multiple cross-correlation (WMCCor, hereafter) gives cross-correlation output in a single plot by implementing the same algorithm as above and allowing for lags. The examination of equity market interdependence and lead-lag analysis among markets carried out in the subsequent section is based on WMCor and WMCCor.

---

[6] This is obtained from linear combination of those decomposed wavelet coefficient that maximises the coefficient of determination, $R^2$ (see Fernandez-Macho, 2012; Polanco-Martinez, 2014).

*2.4.3 Results from wavelet multiple correlation methods*

The analysis of interdependence, in this section, among several groups of equity markets begins by i) pairwise computation of wavelet correlation between several pairs of equity markets and then implementing the improved graphical method of Polanco-Martinez (2014) to generate the wavelet correlation heat-map, ii) using wavelet multiple correlation methods to study the comovement among select pairs of equity market returns and iii) inferring from results the direction of returns spillover.

Markets are grouped into six different sets (Set1-Set6). The equity market contained in each set is given in Table 2.1 where the header C1 through C7 represents the seven equity market indices contained in all seven sets.

Table 2.2 Grouping of stock indices in six sets

| Grouping | Equity Indices | | | | | | |
|---|---|---|---|---|---|---|---|
| | C1 | C2 | C3 | C4 | C5 | C6 | C7 |
| Set1 | SP500 | CAC40 | DAX | NIKKEI | KOSPI | JKSE | BSE30 |
| Set2 | KOSPI | KLSE | TAIEX | SSE | STI | HSI | BSE30 |
| Set3 | FTSE | CAC40 | DAX | BEL20 | ATX | AEX | IBEX |
| Set4 | IBOV | KSE100 | BSE30 | SSE | JKSE | ---------- | ----------- |
| Set5 | BSE30 | STOXX50 | SMI | BEL20 | ATX | ---------- | ----------- |
| Set6 | NIKKEI | ASX200 | HSI | STI | TAIEX | JKSE | KLSE |

Figure 2.7 shows pairwise wavelet correlations among several combinations of equity returns of markets included in Set 1, i.e. markets from the U.S., France, Germany, Japan, South Korea, Indonesia and India. The improved graphical method of Polanco-Martinez (2014) is used to plot the pairwise wavelet correlation within a heat-map framework. Wavelet correlation is computed for eight levels of decomposition associated with the first eight wavelet details; *d1, d2, d3, d4, d5, d6, d7*, and *d8*, which correspond to the time-scale, or investment-horizon, of one-two days, two-four days, four-eight days, eight-sixteen days, sixteen-thirty two days, thirty two-sixty four days, sixty four-one hundred and twenty eight days, and one hundred and twenty eight days-two hundred and fifty six days, respectively. The vertical axis displays the wavelet level along with the legend displaying correlation strength on the right, whereas in the horizontal axis pairwise combinations are displayed.

In the figure, the indices SP500, CAC40, DAX, NIKKEI, KOSPI, JKSE, and BSE30 are labelled by *C1, C2, C3, C4, C5, C6 and C7*, respectively. The degree of wavelet correlation is given by the colour-coded heat-map where the strength of correlation rises from blue (weak) to pink (strong). It is worthwhile to note that classical wavelet correlation analysis would have entailed 7×(7-1)/2=21 correlation plots as opposed to the improved heat-map method which gives the same information in a single plot. It is evident from the plot that multiscale correlation between BSE30 (India) and all other western developed markets (in Set 1) are very weak, indicating very weak stock market integration between India and markets of the U.S., France and Germany (see labels C1-C7, C2-C7 and C3-C7 in the horizontal axis ).

Figure 2.7 Pairwise wavelet correlation among markets in Set 1

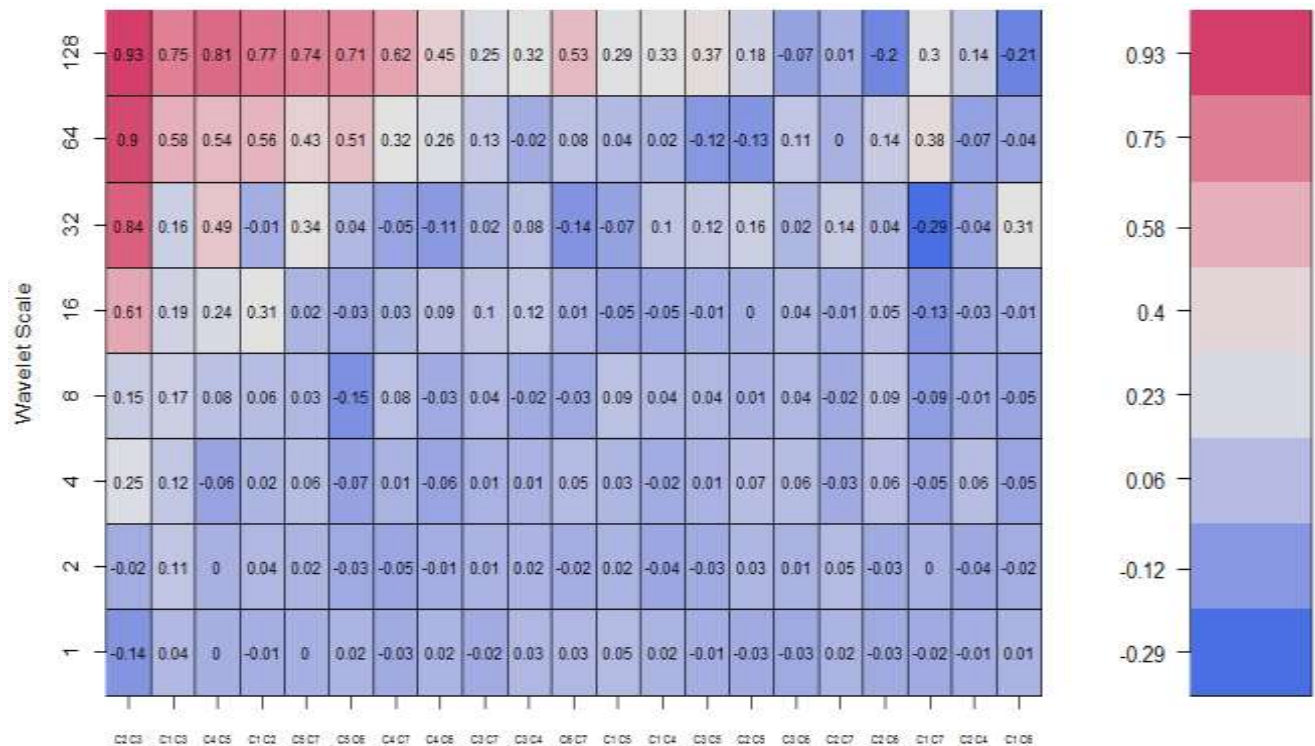

However, some significant correlation between NIKKEI-BSE30 (C4-C7) and JKSE-BSE30 (C6-C7) can be evidenced for the yearly time-horizon (128-256 days). Moreover, strong correlation between KOSPI and BSE30 (C5-C7 pair) seem to exist for both half yearly and yearly scales. In the next step, the improved version of wavelet multiple cross-correlation analysis is performed using markets from Set1. Figure 2.8 gives both improved (top panel) and classical (bottom panel) plots of wavelet multiple cross-correlation among equity returns of markets included in set 1. In the improved WMCCor plot, vertical dashed lines in bold indicate lags where cross-correlation values are strongest. The index that maximises wavelet correlation against a linear combination of other indices in the set is the one that leads all other markets in the set. Therefore, the lead-lag behaviour can be deduced, for each wavelet scale, from the variable names that are listed on the right.

Figure 2.8 Wavelet multiple cross-correlation among markets in Set 1

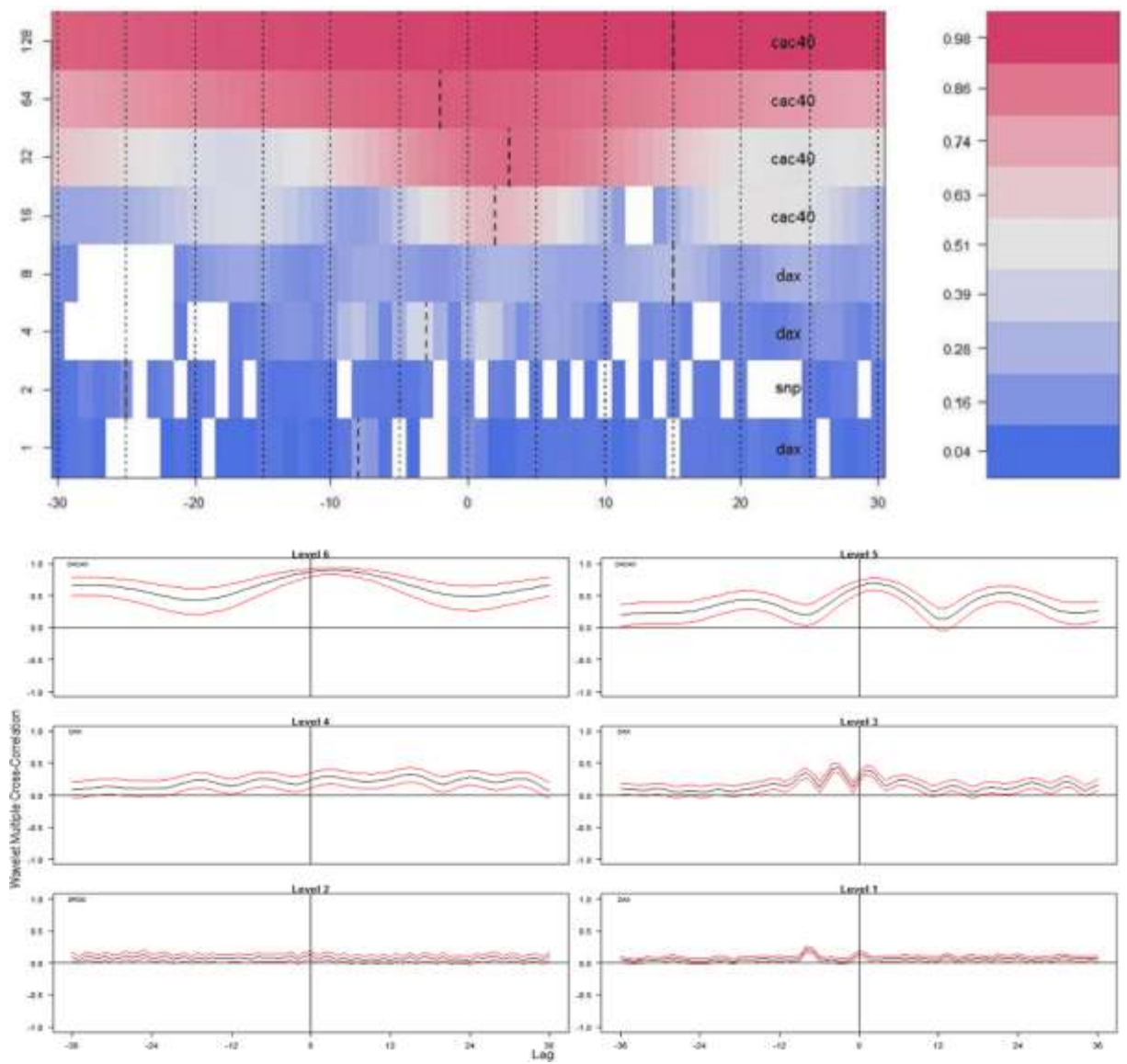

Furthermore, white areas in the colour-coded box plot indicate regions where the confidence band includes zero. As is evident from Figure 2.8, the developed markets of Germany, France and the U.S. lead all other markets in set 1. For e.g., returns of SP500 (labelled 'snp') lead all others at level 2 implying that an increase (decrease) in the returns SP500 will lead to an increase (decrease) in the returns of all other markets in the set, 2-4 days later. At level 5, where CAC40 leads all other markets in the set, correlation peaks around the lag of two days which is evident from the peak in the classical plot (bottom panel) around this lag. This information is also given in the given in the improved plot where, around the lag of two days, dashed vertical lines in bold can be seen. Moreover, scale dependent strength of correlation, where wavelet correlation increases with wavelet scale, can be observed from the plot.

Figure 2.9 shows pairwise wavelet correlations among equity returns of markets included in Set 2, i.e. markets from South Korea, Malaysia, Taiwan, China, Singapore, Hong Kong and India. This set includes major Asian markets where some are closely related in terms of regional proximity, trade and culture. In the figure, the indices KOSPI, KLSE, TAIEX, SSE, STI, HSI, and BSE30 are labelled by C1, C2, C3, C4, C5, C6 and C7, respectively.

Figure 2.9 Pairwise wavelet correlation among markets in Set 2

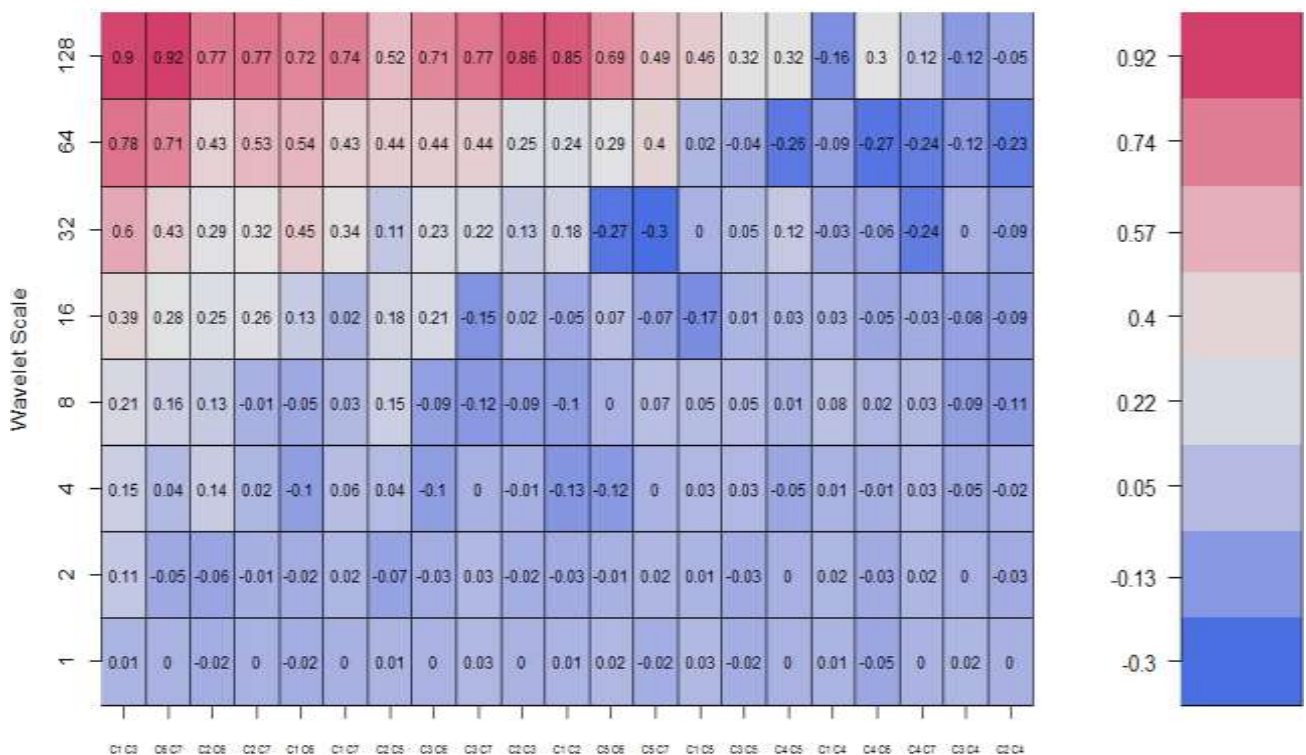

As it can be seen from the above plot, BSE30 is significantly correlated with many Asian markets. On the other hand, as observed in Figure 2.7, the Indian market is not significantly correlated with developed markets of Europe and the U.S. Statistically significant multiscale correlation, beyond 16-32 days timescale, between Malaysian and Indian market (C2-C7 pair) can be observed. The same holds for South Korea and India (C1-C7 pair). BSE30 is also significantly correlated with TAIEX during the monthly time-horizon and beyond. Furthermore, statistically significant between BSE30 and HSI (C6-C7 pair) starting from weekly time-horizon and continuing up to yearly time-horizon is evidenced from the multiscale plot. However, multiscale correlation between India and China seems to be weak indicating weak market integration. Nonetheless, the Indian equity market seems to be strongly interrelated with majority of Asian markets in the set indicating strong interdependence between Indian and some Asian markets. Figure 2.10 shows WMCCor among markets in Set2. With respect to lead-lag behaviour among markets from Set 2, the equity market of Hong Kong, Taiwan and South Korea lead all other markets in the set.

Figure 2.10 Wavelet multiple cross-correlation among markets in Set 2

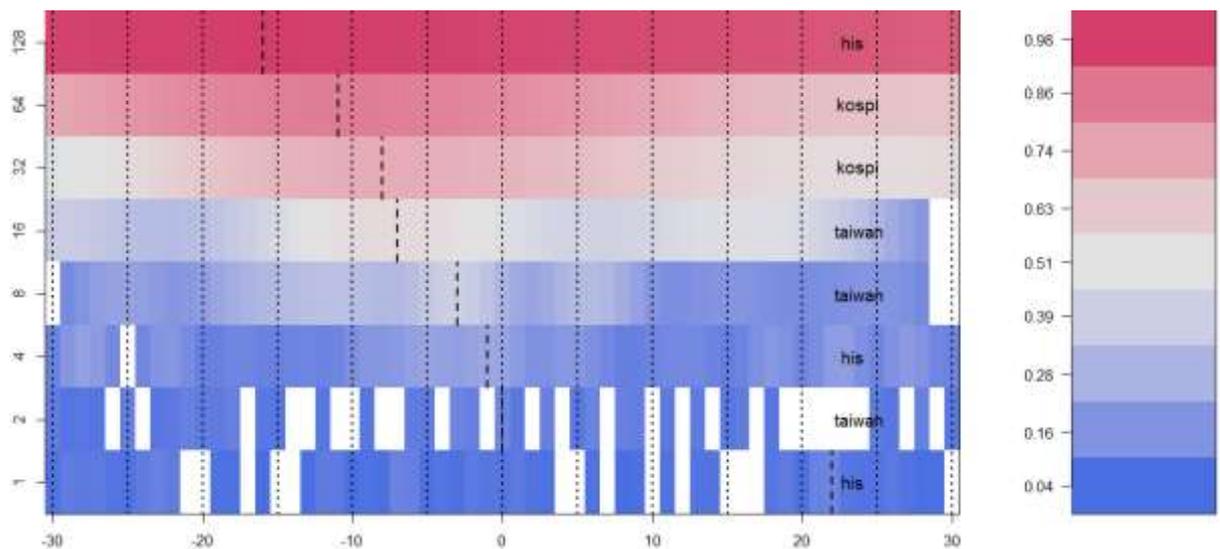

The market of Hong Kong (labelled 'his') lead all other markets at level 1, 3 and 8 corresponding to time-horizons of 1-2 days (daily), 4-8 days (weekly) and 128-256 days (yearly), respectively. On the other hand, South Korean market (KOSPI) lead all others at levels six and seven. Furthermore, the stock market of Taiwan lead all others at fortnightly (8-16 days) and monthly (16-32 days) time horizons. Set 3 comprises of the developed European equity markets from the U.K., France, Germany, Belgium, Austria, Netherland and Spain. Figure 2.11 displays the multiscale correlation among these European markets.

Figure 2.11 Pairwise wavelet correlation among markets in Set 3

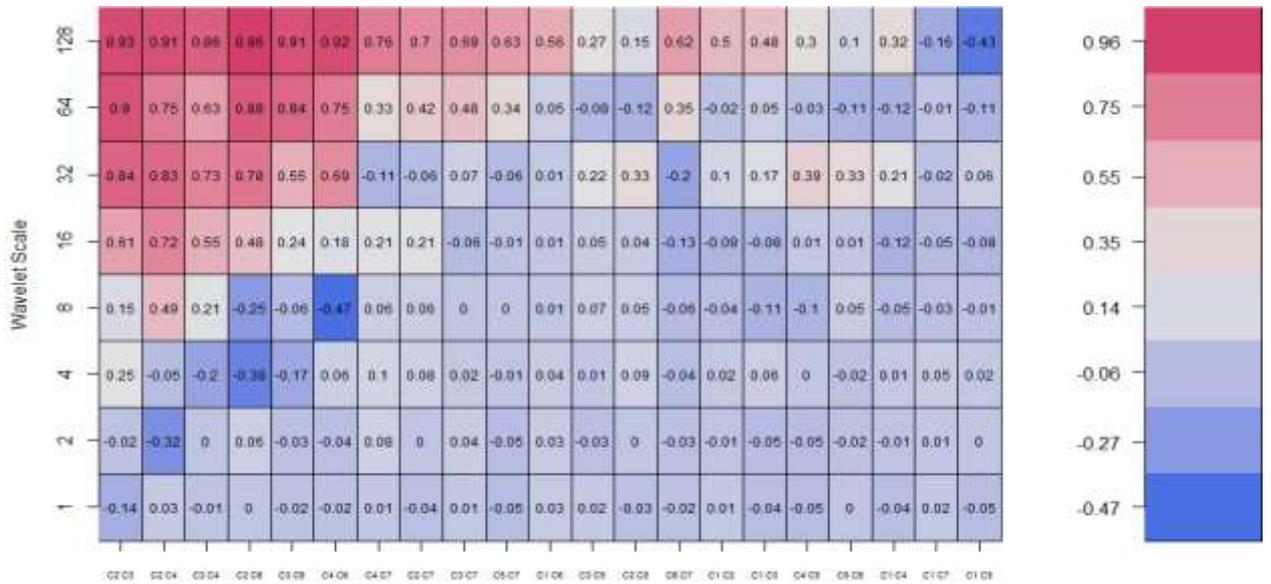

Strong market interdependence among most equity markets of Europe in Set 3 can be observed from Figure 2.11. Furthermore, from Figure 2.12 it can be observed that, markets of France lead all others at levels one, three, five, six and seven.

Figure 2.12 Wavelet multiple cross-correlation among markets in Set 3

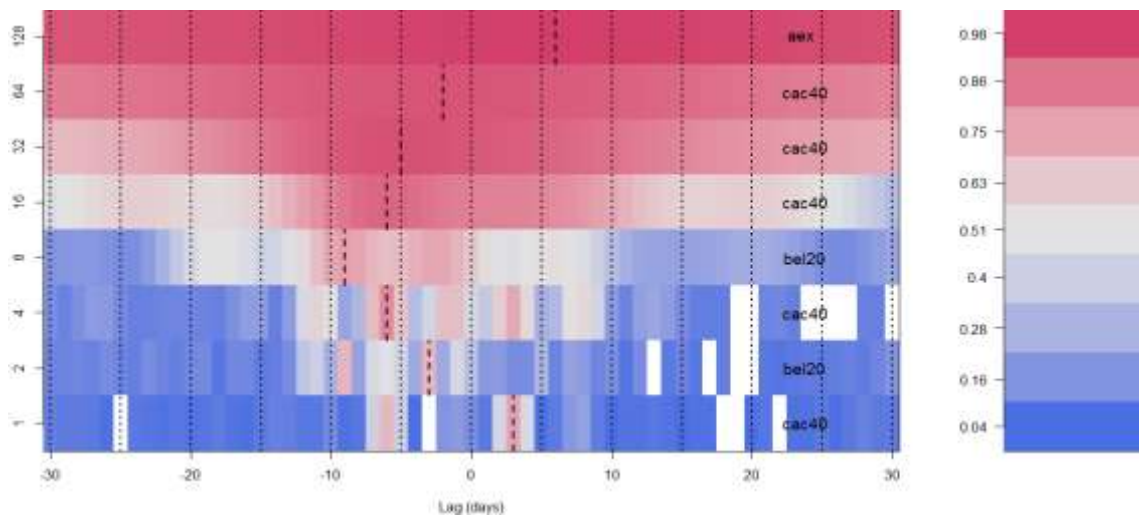

Figure 2.13 displays the pairwise multiscale correlation among some emerging markets included in Set 4. Stock markets from Brazil, Pakistan, India, China and Indonesia comprises this set.

Figure 2.13 Pairwise wavelet correlation among markets in Set 4

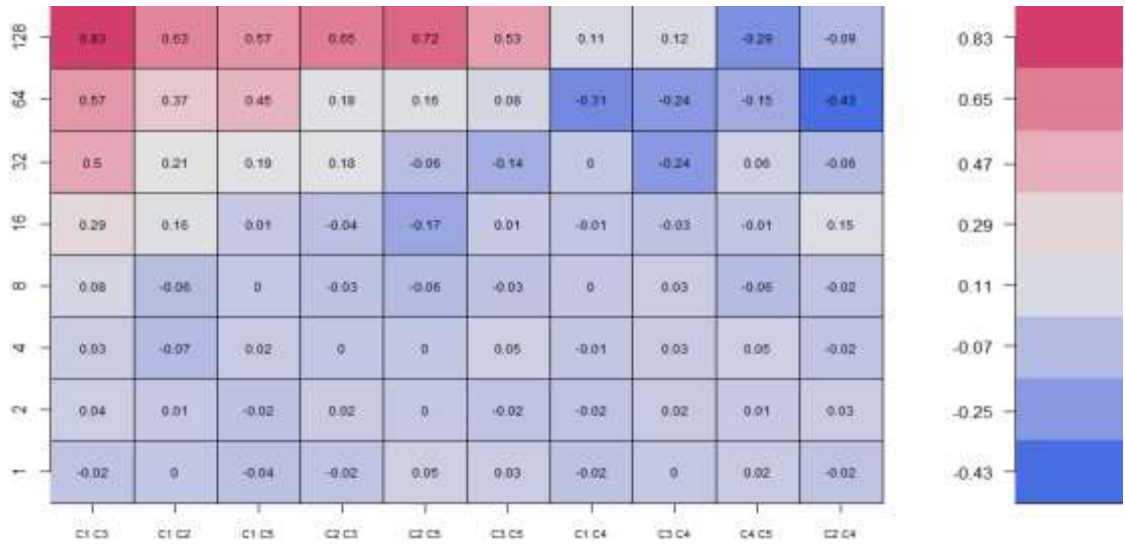

Statistically significant correlation between BSE30 and IBOV (C1-C3 pair) is evidenced from the plot where wavelet correlation increase with the increase in wavelet scale, implying good market between India and Brazil starting with the monthly scale and beyond. However, market integration of India with Pakistan and Indonesia happens only around the yearly scale (128-256 days), where correlations are very weak up to the half yearly scale. Moreover, as shown by the WMCCor plot in Figure 2.14, India, Brazil and Indonesia lead all other markets in the set.

Figure 2.14 Wavelet multiple cross-correlation among markets in Set 4

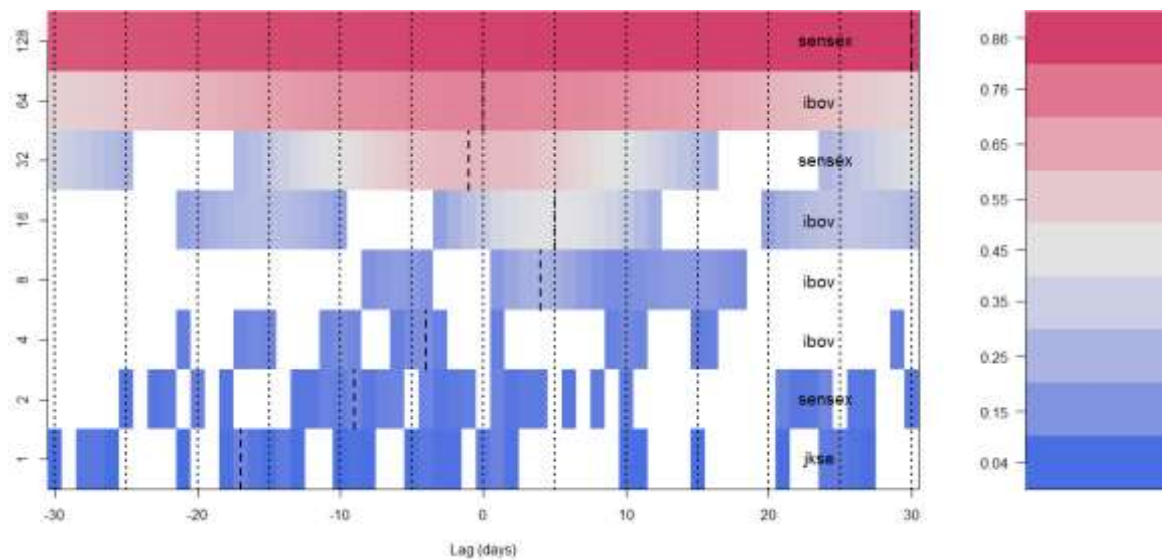

As evident from Figure 2.14, Indian stock returns (labelled 'sensex') lead returns of all other markets in the set at levels two, six and eight correspondind to the intra-weekly (2-4 days), quarterly (32-64 days) and yearly (128-256 days) time horizons. However, stock returns of Brazil (labelled 'ibov') lead all others at levels three, four , five and seven. Nonetheless, the lead-lad dynamics during lower wavelet scales suffer from bouts of insignificance as evidenced from white regions that encompass zero in the confidence interval.

Figure 2.15 shows pairwise wavelet correlations among several combinations of equity returns of markets included in Set 5, i.e. markets from India, Eurozone, Switzerland, Belgium and Austria. There exist weak integration of Indian market with many European markets as multiscale correlations, between BSE30 returns and returns of STOXX50, SMI, and BEL20, are very low. Interestingly, strong multiscale correlation exist between Indian and Austrian stock returns (C1-C5 pair) starting from fortnightly time-horizon (8-16 days) up to yearly time-horizon (128-256 days). Moreover, integration other European markets are stronger. Furthermore, the lead-lag behaviour among markets in this set is given in Figure 2.16. The stock returns of Austria, Belgium and the Eurozone index are found to be the dominant leaders at a majority of wavelet scales, with the exception of BSE30 which lead all others at the monthly time-horizon.

Figure 2.15 Pairwise wavelet correlation among markets in Set 5

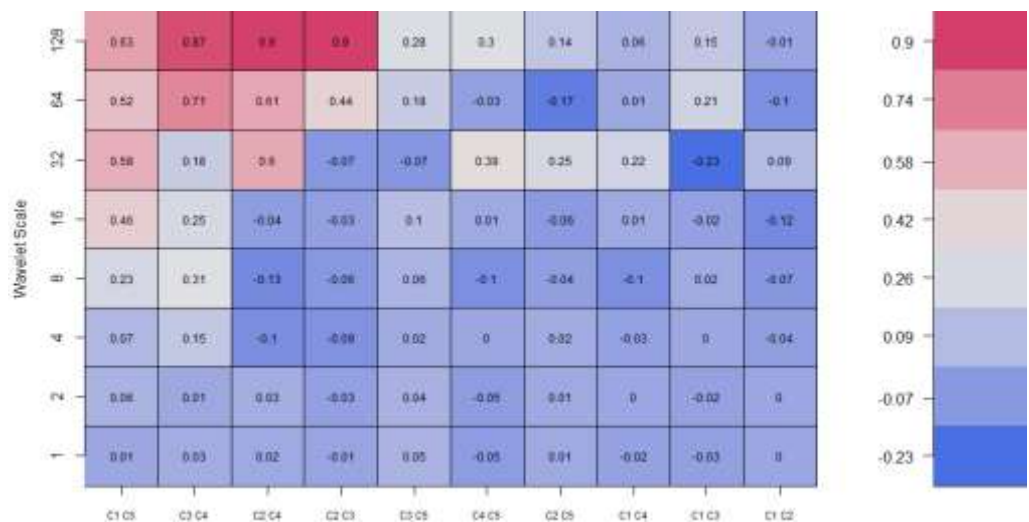

Figure 2.16 Wavelet multiple cross-correlation among markets in Set 5

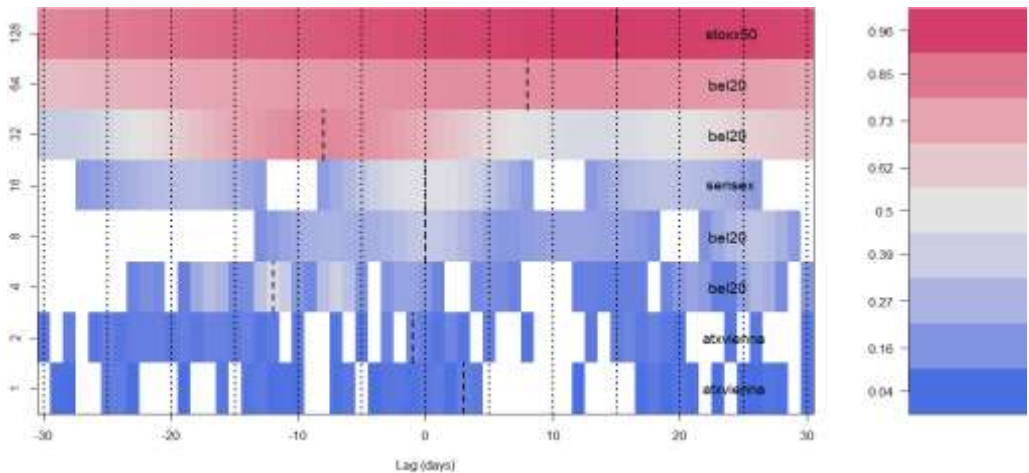

The multiscale correlation between markets from the Asia-Pacific region (Set 6) is given in Figure 2.17 where stock returns of markets from Japan, Australia, Hong Kong, Singapore, Taiwan, Indonesia and Malaysia are considered. There is an evidence of strong integration among markets in this group. The return pairs NIKKEI-TAIEX, HSI-KLSE, HSI-TAIEX, STI-KLSE and JKSE-KLSE show statistically significant wavelet correlation at majority of scales. The stock returns of Malaysia lead all others at shorter timescales whereas returns of Singapore lead at longer time-horizons. Moreover, returns of Hong Kong lead all others at the monthly time-horizon (Figure 2.18). Results from wavelet multiple cross-correlation analysis can be used to identify the direction of returns spillover.

Table 2.3 gives the direction of returns spillover for market groups in Set1-Set4. The first four set of market groups are considered as they include markets from both developed and emerging economies. Arrows indicate the direction of spillovers and are significant at the five percentage level.

There is a strong evidence of returns spillover from DAX to other markets in Set 1 on daily, weekly and fortnightly time-horizons. However, from the monthly time horizon onwards, significant returns spillover from CAC40 to other markets can be evidenced. When looking at the regional spillover dynamics, within the Asian markets contained in Set 2, HSI, TAIEX and KOSPI transmit majority of shocks. At daily, weekly and yearly time-horizons, significant spillover from HSI to other markets can be evidenced. However, spillovers from KOSPI to other markets are found to be significant at both quarterly and half-yearly time-horizons.

Figure 2.17 Pairwise wavelet correlation among markets in Set 6

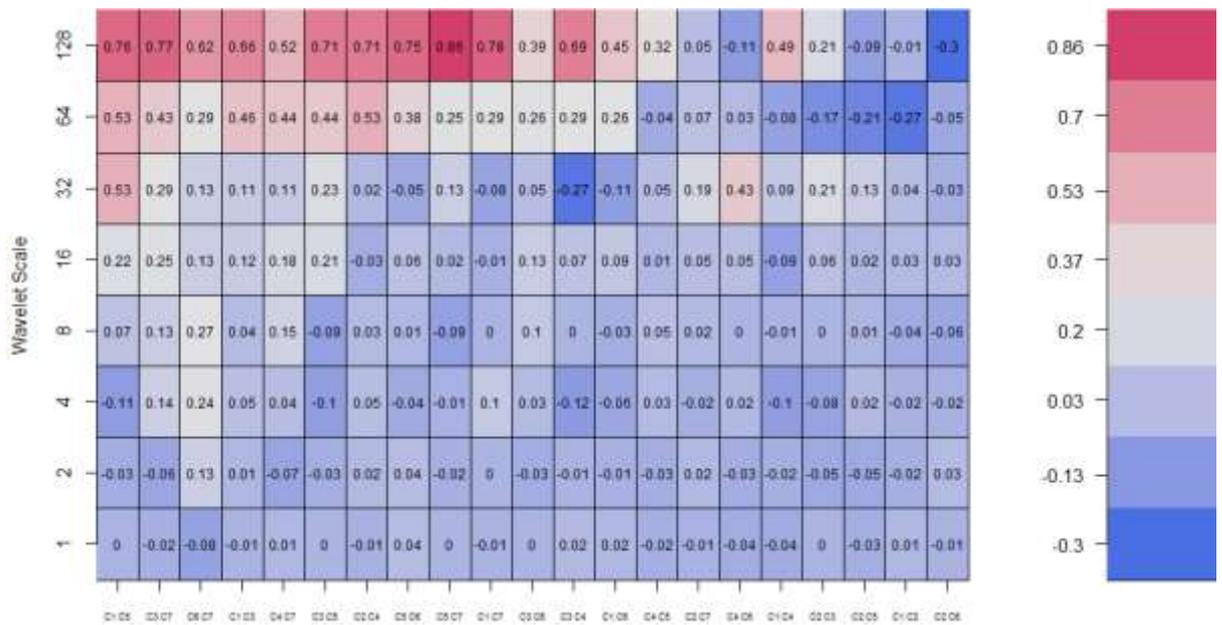

Figure 2.18 Wavelet multiple cross-correlation among markets in Set 6

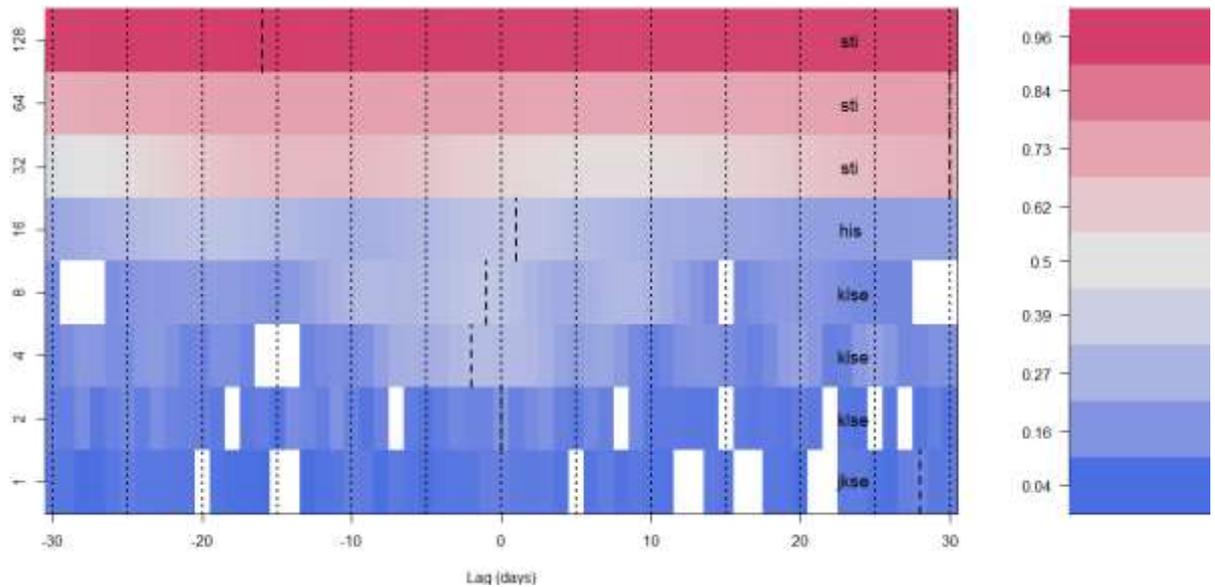

Returns spillovers from the developed market of France (CAC40) is statistically significant t levels one, three, five, six and seven. Finally, while considering emerging markets from Set 4, spillover runs from BSE30 to others at levels two, six and eight corresponding to intra-weekly, quarterly and yearly time-horizons. Nonetheless, the direction of spillover runs from Brazil to other markets in set 4 at weekly, fortnightly and monthly time-horizons.

Table 2.3 Returns spillover among equity markets

| Time Scale | Spillover Direction |
|---|---|
| 1-2 Days | DAX → Set 1  HSI → Set2  CAC → Set3  JKSE → Set4 |
| 2-4 Days | SNP → Set 1  TAIEX → Set2  BEL → Set3  BSE → Set4 |
| 4-8 Days | DAX → Set 1  HIS → Set2  CAC → Set3  IBOV → Set4 |
| 8-16 Days | DAX → Set 1  TAIEX → Set2  BEL → Set3  IBOV → Set4 |
| 16-32 Days | CAC → Set 1  TAIEX → Set2  CAC → Set3  IBOV → Set4 |
| 32-64 Days | CAC → Set 1  KOSPI → Set2  CAC → Set3  BSE → Set4 |
| 64-128 Days | CAC → Set 1  KOSPI → Set2  CAC → Set3  IBOV → Set4 |
| 128-256 Days | CAC → Set 1  HSI → Set2  AEX → Set3  BSE → Set4 |

→ Significant at 5%

In line with the objective of this chapter, i.e. to identify markets with lesser risks for Indian investors in terms of portfolio diversification, multiscale correlations of BSE30 with all developed and emerging markets in the sample are presented in Table 2.5. Entries marked in bold indicate significant correlation and therefore can be used to decide upon relevant portfolio combinations for the Indian investor operating at varying time-horizons.

Results from both Classical wavelet correlation and wavelet multiple correlation analysis document some significant wavelet correlation among BSE30 and major Asian markets. For e.g., from Table 2.5, it can be seen that BSE30 is significantly correlated at majority of wavelet scales, with HSI, KLSE, TAIEX, and KOSPI. Moreover, the Indian stock returns (BSE30) is significantly correlated with the Japanese returns at both half-yearly and yearly time-horizons. However, one can observe weak multiscale correlation between the Indian and Chinese markets. On the other hand, BSE30 is weakly correlated with the markets of the U.S. and developed markets of Europe at all time-horizons. This indicates that Indian investors, operating at varying investment holding periods, can include assets from these markets in their portfolios. One exception is the Austrian market where BSE30 is significantly correlated from level 3 onwards. Moreover, both emerging markets of India and Brazil are significantly correlated from the monthly time-horizon and beyond.

Table 2.4 Multiscale correlation of BSE30 with all markets

| Time-Scale | SNP | CAC40 | DAX | NIKKEI | KOSPI | JKSE | HSI | KLSE | TAIEX |
|---|---|---|---|---|---|---|---|---|---|
| 2-4 Days | 0 | 0.05 | 0.01 | -0.05 | 0.02 | -0.02 | -0.05 | -0.01 | 0.03 |
| 4-8 Days | -0.05 | -0.03 | 0.01 | 0.01 | 0.06 | 0.05 | 0.04 | 0.02 | 0 |
| 8-16 Days | -0.09 | -0.02 | 0.04 | 0.08 | 0.03 | -0.03 | 0.16 | -0.01 | -0.12 |
| 16-32 Days | -0.13 | -0.01 | 0.1 | 0.03 | 0.02 | 0.01 | **0.28** | **0.26** | -0.15 |
| 32-64 Days | -0.29 | 0.14 | 0.02 | -0.05 | **0.34** | -0.14 | **0.43** | **0.32** | **0.22** |
| 64-128 Days | 0.38 | 0 | 0.13 | **0.32** | **0.43** | 0.08 | **0.71** | **0.53** | **0.44** |
| 128-256 Days | 0.3 | 0.01 | **0.25** | **0.62** | **0.74** | **0.53** | **0.92** | **0.77** | **0.77** |
| | | | | | | | | | |
| Time-Scale | STI | ASX | ATX | BEL20 | SMI | STOXX | IBOV | KSE100 | SSE |
| 2-4 Days | 0.02 | 0 | 0.06 | 0 | -0.02 | 0 | 0.04 | 0.02 | 0.02 |
| 4-8 Days | 0 | 0 | 0.07 | -0.03 | 0 | -0.04 | 0.03 | 0 | 0.03 |
| 8-16 Days | 0.07 | -0.04 | **0.23** | -0.1 | 0.02 | -0.07 | 0.08 | -0.03 | 0.03 |
| 16-32 Days | -0.07 | -0.08 | **0.46** | 0.01 | -0.02 | -0.12 | **0.29** | -0.04 | -0.03 |
| 32-64 Days | -0.3 | 0.08 | **0.58** | 0.22 | -0.23 | 0.09 | **0.5** | 0.18 | -0.24 |
| 64-128 Days | 0.4 | -0.1 | **0.52** | 0.01 | 0.21 | -0.1 | **0.57** | 0.18 | -0.24 |
| 128-256 Days | 0.49 | 0.03 | **0.63** | 0.06 | 0.15 | -0.01 | **0.83** | **0.65** | 0.12 |

The information from multiscale correlation can be used by investors operating at different time-horizons to appropriately adjust their portfolio combinations. It is also important to note that a portfolio meant for shorter timescales might not yield the same risk mitigating benefits if used for other tie-horizons. Therefore, multiscale nature of correlation structure needs to be taken into account before strategizing on portfolio combinations.

## 2.5 Conclusion

This chapter investigates the multiscale linkages among global equity markets with a special focus on the Indian market. Multiscale correlation methods from the wavelet domain are used to identify interrelations between several market pairs at different time-horizons. The breakdown of correlation at different resolutions allow investors to correctly identify risks associated with assets at different time-horizons. In general, wavelet correlation among equity markets seem to increase as we move from shorter time-horizons to longer time-horizons. Correlations are significantly stronger at longer time-horizons whereas shorter time-horizons have very weak correlations. For e.g. the daily timescale (1-2 days) seems to have very weak correlations. Therefore, correlations between global equity returns are found to be dependent on investment horizons.

The separation of correlation structure at different time-horizons is very beneficial for heterogeneous investors who operate at different timescales based on their investment holding periods. Moreover, information on correlation structure at varying time-horizons will aid investors in diversifying portfolios with global asset combinations, where portfolios diversified using international assets is empirically demonstrated in the literature to reduce portfolio risks (Grubel, 1968; Agmon, 1972; Dajcman, 2012 etc.). Furthermore, the information on correlation structures at different investment holding periods will provide additional inputs for investors whose risks might not be the same for all investment decisions that they undertake. Therefore, an analysis based on these lines aids investors in internationally diversifying their portfolios while incorporating different investment holding periods, or time-horizons, into their strategy.

Multiscale correlation among developed European markets from the same region are found to be significantly strong across different time-horizons, indicating strong integration among these markets. Similarly, some Asian markets with regional proximity seem to be interdependent. This is in line with Pretorius (2002) where regional proximity, and the related trade linkage that geographical proximity engineers, plays an important role in determining market integration.

In view of the possible portfolio diversification benefits facing Indian investors, multiscale correlation structure between the Indian stock returns and returns from both developed and emerging markets are investigated. This helps in adjudicating risks engulfing heterogeneous Indian investors with varying investment holding periods. Indian investors who invest in equity markets of the U.S. and developed European markets may benefit from reduced portfolio risks as correlation between the Indian stock returns and returns of these developed western markets, at almost all time-horizons, is very low. Additionally, Indian investors might also be well off if they invest in the Chinese stock market. However, Indian investors should be cautious if they include assets from Brazilian and East Asian markets as multiscale correlation between BSE30 and markets from these regions, for a majority of investment holding periods, are very significant. Since heterogeneity of Investment horizons and corresponding information at multiple time scales allow heterogeneous Indian investors to carefully diversify their portfolio, the results obtained from this analysis might aid Indian investors in their investment decisions. Nonetheless, investors should take into consideration their investment holding periods and the associated risks when they make risk management and portfolio allocation decisions.

# Chapter 3

## A Wavelet Analysis of Contagion among Global Equity Markets

This chapter investigates the phenomenon of contagion among some select global equity markets using wavelet based time-frequency analysis. It surveys some seminal literature on contagion and examines, using both continuous and discrete wavelet methods, the effects of major financial crises on Indian markets. Strong evidence of contagion between some developed markets is revealed. Only long run comovements between Indian market and developed markets exists, revealing long run interdependence. However significant comovements in the short run, which indicates contagion, between Indian and some East Asian markets are observed, indicating diversification risks for Indian investors during periods of financial turbulence.

## 3.1 Introduction

The literature on financial contagion spans a huge body of theoretical and empirical work where there exist a substantial degree of both agreements and disagreements regarding the definition and meaning of contagion. The nature of contagion can vary with the type of financial turbulence and its spread, where the propagation of shock from a crisis hit country to other countries can occur through different channels. However, all empirical studies, trying to explain the phenomenon of contagion, agree that contagion occurs during period of financial crisis notwithstanding the differing channels through which it spreads. The transmission of shocks between two unrelated countries, with no proper linkages and different economic structures, during financial crisis can be considered as contagion. However, the same cannot be true for countries with a history of huge cross-market linkages or interdependence. This leads to difficulty in arriving at a precise definition of contagion. Forbes and Rigobon (2002) use the term shift-contagion where contagion is said to occur when shock to a country, arising out during periods of financial turmoil, leads to a substantial rise in cross-country linkages. This shift in inter-market linkages is thought to be the main carrier of shift-contagion but however the reasons for this shift are not explored.

The aforementioned definition of shift-contagion, which relies on the measurement of cross-country linkages, can be arrived at with the use of statistics like equity return correlation, the probability of shocks arising out of speculations, magnitude of volatility transmission between markets, etc. Majority of empirical literature which test for contagion rely on asset return correlation between markets. In view of the above, a substantial rise in cross-market correlation of asset returns after a financial crisis is considered as contagion, where a test for contagion, based on this strand of thought, amounts to testing the magnitude of shift in cross-country linkages after a shock. Hence, the phenomenon of contagion can be accepted or refuted by analysing the strength of this magnitude. However, the universality of this definition cannot be established as disagreements abound in literature where some economists argue that tests based on mere cross-market relationships cannot be used to arrive at a definition of contagion. Their argument centres on the dynamics of shock propagation mechanisms and how only certain channels of transmission mechanisms leads to contagion. This restrictive view of arriving at a definition of contagion is however not very popular in empirical literature. Therefore, the tests for contagion used later in this chapter is based on various tests of cross-market correlation and its strength.

An investigation of cross-market linkages, which is used to gauge the existence of contagious financial crises as described above, should enable one to distinguish between various mechanisms of shock transmission across countries. There exist a huge body of literature that explains the shock propagation mechanism across markets. The channels of shock transmission can either be through financial linkages or through fundamental linkages like trade, among others. However, testing and measurement of various transmission channels and linkages difficult. Therefore, definition of contagion based on cross-country linkages, and its measurement via estimates of asset returns correlation, should normally suffice in our understanding of contagious crises. Moreover, such tests based on cross-market linkages and asset returns correlation does not require one to distinguish between various mechanisms of shock transmission. However, a clear understanding of varying shock propagation channels is useful in analysing periods of financial turmoil in detail notwithstanding the paucity of literature that exactly arrive at a definitive conclusion with regards to these channels. Many studies on contagion, however, explicitly focus on the nature and workings of various transmission mechanism channels. The next section reviews some important literature on contagion which focuses on the international propagation of shocks.

## 3.2 Channels of Shock Propagation: A Review of Literature

Contagion, in general, can be thought of as a "spread of financial disturbances from one market to another which can be observed through downward co-movements in equity prices" (Claessens et al., 2001). However, strong co-movements between historically interdependent markets during financial crisis cannot be considered as contagion. This is due to the fact that markets with high degree of interdependence generally exhibit strong degree of both real and financial linkages. Therefore, spillovers arising out of strong market interdependence, channelled via strong real and financial linkages, cannot be described as contagion since it reflects the already existing interdependence between markets. Crisis propagated via this route is termed as "fundamentals-based contagion", as shocks are transmitted due to strong fundamental linkages between markets (Calvo and Reinhart, 1996). On the contrary, crisis can be propagated via non-fundamental channels where contagion occurs independently of fundamentals and market interdependence. This can be triggered, for example, by investors who suddenly withdraw investments from many countries notwithstanding diverse economic fundamentals. Herd behaviour, uninformed speculations, panic, loss of confidence and decisions based on imperfect information is generally attributed to contagion based on non-fundamental channels.

Many empirical literature on contagion, however, try to explain the strength of co-movements and identify the channels via which crises are transmitted.

The channels of shock transmission assumes great importance when dealing with the notion of contagion. One of the fundamental channel of transmission is attributed to common global shocks by Calvo and Reinhart (1996), where changes in commodity prices and major restructuring of advanced industrial economies can trigger crisis in emerging market economies. On the other hand, local shocks arising out of crisis in one country is also said to influence economic fundamentals in other markets. The channel of transmission in this case might involve trade linkages.

According to Claessens et al. (2001), huge currency depreciation in a crisis hit economy can affect its major trading partners through fall in asset prices and capital outflows. Currency depreciation in the crisis hit economy affects its trade relation due to decrease in imports, thereby deteriorating trade balance. Financial linkages also form an important channel of transmission especially in regions with high market interdependence. A financial crisis in one market will certainly impact markets which have high degree of interdependence with the crisis hit market. This happens via the finance route where highly interdependent markets might experience reductions in foreign direct investment and trade credit.

The transmission of shocks during times of financial turmoil can also be attributed to the herd behaviour of investors. In this type of shock propagation, information possessed by investors or market participants plays an important role in generating herd behaviour. This happens due to the highly volatile, uncertain and complex nature of financial markets where multitudes of investors operate with varying information set. As the complex nature of markets make information costly, not all investors are well informed. However, investors who are well informed specialise and operate on a particular market or geographical region. These investors, who periodically liquidate their assets to meet other demands or to reformulate their portfolios, can give wrong signals to the uninformed investors who interpret the behaviour of informed investors as an indicator of poor returns, thereby generating herding. (see Calvo, 1999; Pritsker, 2001; Kumar and Persaud, 2002). The aforementioned mode of shock transmission, where financial turbulence in one market generates shocks to other unrelated market via investors' herd instincts, sparks crises which can be ascribed to asymmetry in investors' information, false alarm, and strategies related to the reformulation of portfolios (Bayoumi et al., 2007).

Information asymmetries and differences in expectations of investors also lead to contagion. Imperfect information give rise to uninformed investors who believe that crisis in one market will similarly impact other markets. King and Wadhwani (1990), while investigating the US stock market crash of October 1987, attempt to explain the simultaneous crash in other unrelated markets via a contagion model. In this model markets are effected by contagion as a result of the information inferring behaviour of rational agents. Rational investors who wrongly infer events in other markets generate a new transmission channel where mistakes or idiosyncratic changes in one market, engendered by rational agents' incorrect interpretation of information, are transmitted to other, possibly unrelated, markets.

Calvo and Mendoza (1998) in their model of contagion conjecture that less informed investors, who find it costly to gather proper information, will mimic the behaviour of informed investors and follow their pattern of investment decision making. However, mistakes made by informed investors with regard to their portfolio decisions will influence the decision of uninformed investors due to herding. Imperfectly informed investors, due to the higher costs involved in collecting market information, find it easier and economical to follow the behaviour of informed investors. However a bad decision by informed investors, which puts them in a bad equilibrium state, will similarly move uninformed investors to a bad equilibrium state. This happens when an information cascade is generated which ultimately drives the uninformed investors to ignore their own information set and follow the behaviour of informed investors (Wermers, 1995).

Investors' behaviour based on their expectations can generate a state of multiple equilibrium, both good and bad. In this model of multiple equilibrium, crisis transmission from one emerging market to another emerging market, subsequently driving the latter to a bad equilibrium, causes contagion (Masson, 1998).

In the multiple equilibrium model of contagion, investors' expectations play a vital role in driving a market towards turmoil. According to Masson (1998), crisis in one market synchronises the expectations of investors operating in the second market, causing the equilibrium to shift from a good state to bad, thereby causing a crash in the second market. It is worthwhile to note that the shift from a good equilibrium to a bad one, transpiring after a crisis in the first market, is driven by changes in investors' expectations and not by real factors.

In theoretical studies of contagion, the above multi equilibrium model falls under the umbrella of the so-called "crisis-contingent theories" where the type of shock propagation mechanism does not normally occur during stable periods (Forbes and Rigobon, 2001). Liquidity shocks also play an important role in transmitting crisis propagation (Valdes, 1996), which also falls under crisis-contingent theories of contagion. Investors experience a reduction in their liquidity after crisis in one country. This could compel them to sell their assets in another market and reformulate their portfolios in order to, i) satisfy margin calls, ii) be able to continue their market operations, and iii) fulfil regulatory requirements.

The asymmetric information model of Calvo (1999), which also falls under the crisis-contingent category, is also a model of endogenous liquidity shocks. In both models, increased correlation among asset returns occur in the aftermath of liquidity shocks. This transmission mechanism, driven via the liquidity shock channel, does not ensue during relatively stable periods. Drazen (1998), however, evinces the existence of political factors that led to contagion during the 1992 UK exchange rate mechanism crisis, which also fall within the purview of crisis-contingent theories.

In view of the above, three different channels of transmission can be identified from the crisis-contingent theories, namely i) multiple equilibrium channel consequent on investors psychology, ii) liquidity shocks that initiate portfolio reformulation, and iii) political factors. Notwithstanding the theories and models used to arrive at the above mentioned channel, they share a common implication: the propagation mechanism during crisis is different than during stable periods.

On the other hand, theories illustrating how channels of shock propagation does not lead to shift-contagion fall under the category of non-crisis-contingent theories. In this case, the channels of transmission are the same, during both stable and non-stable periods. Highly interdependent countries with proper financial market integration experience, after a shock, high cross-correlations. This reflects the continuation of already existing channels of shock transmission where the linkages during both crisis and non-crisis periods does not significantly differ. These channels are also known as real-linkages as most of these linkages are consequent on economic fundamentals. Channels[7] like, for example, i) trade, ii) coordination of economic policies, iii) global shocks, and iii) re-evaluation of other countries' mistakes, among others, fall under the category of real-linkages. The transmission mechanism related to trade can operate when a country engages in currency devaluation which, by leading to a rise in trade competitiveness of that country and a subsequent rise in exports to its trading partner, can hurt domestic sales in the second country. Moreover, coordinated policy responses can also lead to shock transmission where a country can imitate other country's policy response to a shock. Shocks can also be transmitted when investors learn from other country's mistake and apply the re-evaluated policies to markets with similar economic policies and structures.

Researchers have, over the years, identified several channels of shock transmission which operate via propagation of bad economic news that affects cash flows in other markets. This channel of transmission leads to contagion due to the propagation of information from one market to another (see for eg. Kiyotaki and Moore, 2003; Kaminsky et al., 2003, etc.). According to Brunnermeier and Pedersen (2009), liquidity shocks also form an important channel of transmission that lead to contagion as loss incurring investors might find it difficult to obtain funds, leading to a fall in overall market liquidity. In this case liquidity crunch arises via a flight-to-quality where loss incurring investors sell their assets, which they perceive to be of higher risk, and opt for safer options. Finally, the risk-premium channel of shock transmission leads to contagion as "shock in one market leads to an increase in risk premium in another market" (see for eg. Vayanos, 2004; Acharya and Pedersen, 2005; and Longstaff, 2008). Apart from theoretical models explaining contagion and various mechanisms of shock propagation, there exists a plethora of empirical literature that tests for the existence of contagion and how shocks are transmitted between markets.

---

[7] Forbes and Rigobon (2001), Claessens et al. (2001) and Claessens et al. (2010) give a detailed theoretical survey of various channels of shock propagation mechanisms.

## 3.3 A Survey of Empirical Literature on Contagion

Empirical tests seeking to investigate the existence of contagion largely focus on co-movements in asset returns during periods of financial turbulence. A large portion of the literature employ tests based on cross-market correlation of asset returns, which amounts to comparing the correlation between asset returns during stable and turbulent periods. Under this framework a statistically significant rise in cross-market correlation reflects contagion. A significant rise in correlation during the aftermath of a shock, which normally reflects an increase in transmission mechanism, is taken to be suggestive of a phenomena indicating the contagious nature of that particular shock. Earlier empirical studies of contagion largely focus on the American stock market crash of 1987, the Mexican crisis of 1994, and the East Asian crisis of 1997.

King and Wadhwani (1990) find evidence of a significant rise in cross-market correlation between the markets of the United States, Japan and the United Kingdom after the 1987 crash. Similar results, supporting the evidence that the 1987 crash was contagious, were obtained by Lee and Kim (1993) after including more markets in their analysis. With regard to the Mexican crisis, a number of studies arrive at a conclusion supporting contagion (see for eg. Calvo and Reinhart, 1996; Frankel and Schmukler, 1996; Valdes, 1996 etc.). According to Baig and Goldfajn (1998), increased cross-correlation coefficients between many South-East Asian markets revealed the contagious nature of the East Asian financial crisis of 1997-1998.

However, increased correlation among financial markets of different economies does not necessarily imply contagion. High degree of cross-market correlation during financial crises, on the other hand, could possibly reflect the already existing strong historical interdependence and strong channels of shock transmission between certain markets during stable periods. According to Forbes and Rigobon (1998), the presence of heteroskedasticity in the movement of asset prices during volatile periods leads to higher correlation between markets. Hence a rise in cross-market correlations during financial crisis, after correcting for heteroskedasticity, manifests due to the continuation of the same strong channels of shock propagation existing during normal periods. Moreover, they demonstrate that endogenous factors, like correlation between countries' economic fundamentals, preferences and perception of risk, lead to an increase in cross-market asset price correlations.

Forbes and Rigobon (1998), after correcting for heteroskedasticity and endogeneity, do not find any evidence of contagion during American stock market crash of 1987, Mexican crisis of 1994, and the East Asian crisis of 1997. In the same vein, Arias et al. (1998) and Rigobon (1999), using similar corrective mechanisms, fail to find any evidence of contagion. Similarly Collins and Biekpe (2003) and Lee et al. (2007), after controlling for heteroskedasticity, fail to reject the null hypothesis of interdependence. However, this method has been criticised by Corsetti et al. (2005) where they prove that controlling for volatility of market-specific shock biases the result in favour of mere interdependence. Similar conclusions were derived by Pesaran and Pick (2007). Bartram and wang (2005) attribute the bias in favour of interdependence, as evidenced in Forbes and Rigobon's model, to some arbitrary restrictions and assumptions. Rodriguez (2007), on the other hand, investigates contagion using a copula based approach and finds evidence in its favour during the Mexican and Asian crises.

Eichengreen et al. (1996),employing a methodology based on conditional probabilities rather than cross-market correlations to investigate contagion, find that trade linkages play an important role in explaining contagion. Their procedure is based on estimating the probability of a crisis in one country conditional on an information set encompassing the occurrence of crisis in different market or markets. They find that speculative attacks in a foreign market leads to an increase in the probability of currency crisis back home. Similarly, Glick and Rose (1998) test for contagion during major currency crises and find trade linkages as the main channel of propagation. They however conclude that contagion is more regional than global as the intensity of trade is normally higher in regional blocks. Chan et al., (2002) find that countries with close trade ties were influenced by the Asian currency crisis. Moreover, Zhang (2008) and N'Diaye et al. (2010) demonstrate that trade linkages played a vital role in transmitting contagious shocks from the United States to Asia-pacific markets on the aftermath of 2008 financial crisis. Similar shocks, emanating from trade based linkages, were transmitted to Asian markets during the financial crisis of 2008 (Xue et al., 2012).

However, the main focus has been on idenfying and separating contagion based on pure linkages and fundamental linkages (see Kaminsky and Reinhart, 2000; and Dornbusch et al., 2002). Pure contagion, also known as excessive contagion, manifests due to the propagation of excessive shocks among countries. In this case, a crisis hit country will transmit extreme shocks, beyond any fundamental linkages and idiosyncratic instabilities, to markets in other countries (see Eichengreen et al., 1996; Forbes and Rigobon, 2002; and Bae et al., 2003). On the other hand, fundamentals based contagion occurs due to strong financial market integration during both stable and unstable periods. In this case, shocks are usually transmitted via real linkages and reflects normal interdependence between markets (Calvo and Reinhart, 1996).

Recent empirical studies on contagion use a variety of techniques to identify the contagious nature of financial crises. A plethora of studies, however, focus on investigating the contagious effect of the 2008 global financial crisis as it triggered a devastating effect on both advanced and emerging economies alike (see Claessens et al. 2010; and Longstaff, 2010).

Ait-Sahalia et al., (2010) demonstrate that shocks from pure contagion dissipates and transmits very quickly. It happens relatively faster and spreads over a short run period. Candelon et al. (2008), while investigating cross-market correlations among Asian economies during the East Asian crisis, found evidence of pure contagion as shocks were transmitted faster than usual. Contrary to pure contagion based propagation of shocks, a relatively slower and gradual transmission of shocks normally reflect financial market integration. The susceptibility of Asia-Pacific markets to contagious shocks has been confirmed by Hsin (2004) where he demonstrates that developed markets have more influence over markets from other economies. According to Bodart and Candelon (2009), contagious effects were present in both Mexican and East Asian crises. Moreover, high degree of interdependence led to the spread of crisis in Asia where spillover effects were limited to geographical region. Contagion as a regional occurrence is also supported by evidences in Kaminsky and Reinhart (1998) and Kaminsky and Schmukler (1998).

On the other hand, Chou et al. (1994) use a GARCH approach to investigate the volatility transmission between markets and find significant spillovers after the crash of 1987. However, relatively recent studies investigate the time-varying nature of correlation using different variants of multivariate GARCH models. Empirical studies on contagion and interdependence investigating the dynamic evolution of correlation, in both developed and emerging economies, are very prominent and in abundance (see for eg. Cappiello et al., 2006; Chiang, Jeon and Li, 2007; Lin, 2012; Min and Hwang, 2012; Suardi, 2012; Ahmad et al., 2013 etc.)

The empirical literature on contagion spans a huge body of work, each with diverse testing procedures, comprising of advanced modelling techniques. However, a vast majority of empirical studies on contagion rely on time domain techniques. Bodart and Candelon (2009), however, examine contagion in a frequency domain framework where high frequencies are associated with contagion. Lower frequencies, which technically reflect the long run, are associated with interdependence. They base their analysis by employing the frequency domain causality method proposed by Breitung and Candelon (2006). Similarly, Orlov (2009) uses co-spectral analysis to study exchange rate co-movements during the Asian financial crisis. However, frequency domain based spectral methods loses time information as information localisation is only a band of frequencies. Moreover, spectral methods require the data under investigation to be covariance stationary. Most often in economics and finance, this is not the case, as we encounter processes which are not stationary. Therefore, this study uses wavelet methods which can localise information from both time and frequency domains, simultaneously. A multi horizon comovement approach to identify contagion particularly in the Indian context, which can effectively identify the evolution of correlation across markets in both time and frequencies, is implemented.

The susceptibility of financial institutions to adverse effects during crises, as it was evident during the global financial crisis of 2007-08, raises doubts about market and institutional stability. Evidences of the subprime crisis spreading to even regions with lower exposure to financial instruments from US is a cause of concern (Brana and Lahet, 2010). A plethora of studies on contagion using advanced methods from time domain, each describing the nature and channels of shock transmission (see for eg. Kaminsky and Schmukler, 1999; Vayanos, 2004; Longstaff, 2008 etc.), however, fail to understand the phenomena from a multiscale dimension. Since evidences of contagion are often identified by observing the changing correlations across global markets during periods of turbulence (Candelon et al., 2008), this chapter attempts to explain contagion from a wavelet based multiscale perspective by mapping shocks to both time and frequency.

This approach is advantageous in the sense that multiscale partitioning of correlation in the time-frequency space allows one to differentiate the strength of correlation with respect to various timescales. Wavelet methods allows for a detailed analysis of interrelations as changes in the structure of comovements can be limited to a particular timescale (Rua and Nunes, 2009). This scale-dependent dynamics of comovements cannot be captured by traditional time and frequency domain methods. Therefore in the wavelet domain, short run comovements can be clearly distinguished from medium and long run comovements, allowing one to unmistakeably differentiate between contagion and interdependence. This chapter follows the approach of Ranta (2013) who, in the spirit of Forbes and Rigobon (2002), defines contagion as an increase in short timescale comovements after a financial crisis. Moreover, the implementation of wavelet methods avoids the heteroskedasticity problem of Forbes and Rigobon (2002), as volatility affects correlations from both short and long timescales.

Gallegati (2010), using a wavelet approach, finds evidence of contagion among G7 countries during the 2008 financial crisis where contagion is found to be scale-dependent where shocks were not homogenous across scales. Graham et al. (2012) studies comovements between the U.S. market and twenty two emerging markets using continuous wavelet methods. Several recent papers analyse correlation dynamics and contagion between global markets in the wavelet domain (see for eg. Benhmad, 2013; Graham et al., 2013; Tiwari et al., 2013; Aloui and Hkiri, 2014; Tiwari et al., 2016). However, none of the aforementioned studies study the contagious effects of major financial crisis on the Indian market in a wavelet based approach. Lack of studies investigating contagion from a multiscale viewpoint, with a special focus on Indian equity markets, justify the need for this study.

In this chapter, contagion between global equity markets with a special on India is examined using wavelet coherence methods of the continuous wavelet class. Analogous to the time domain measurement of correlation via the correlation coefficient, wavelet coherence gives the same information as the time domain correlation coefficient but from a time-frequency perspective, simultaneously localising information from both time and frequency domain. This chapter employs the methodology of continuous wavelet coherence developed by Torrence and Compo (1998), Torrence and Webster (1999), Grinsted et al. (2004) and Aguiar-Conraria and Soares (2014). Nevertheless, discrete wavelet methods as reviewed in the previous chapter is employed in the next step of the analysis to check the robustness of results obtained using continuous wavelet methods.

To check for robustness, the MODWT estimate of wavelet correlation is computed in a rolling window framework to obtain a time-series of wavelet correlation. This is followed by a two sample t-test to check the significant difference in correlation before and after the crisis event. The next section reviews the methodology of continuous wavelet transform and wavelet coherence.

## 3.4 Methodology

The estimator of correlation coefficient in time-frequency space, used to analyse comovements between two time domain functions, is given by wavelet coherence which is based on the "*continuous wavelet transform*". The time-frequency contour plot of wavelet coherence provides a richer description of co-movement as statistically significant areas in the plot, where the two time-series move together, can be properly identified along with a scatter of phase arrows.

*3.4.1 The continuous wavelet transform*

A wavelet is a real valued function $\psi(\cdot)$ defined on $\mathbb{R}$ such that

$$\int_{\mathbb{R}} \psi(t)\, dt = 0 \qquad (3.1)$$

$$\int_{-\infty}^{\infty} |\psi(t)|^2 \, dt = 1 \tag{3.2}$$

Wavelet analysis is performed by choosing a reference wavelet known as *mother wavelet* $\psi_{b,s}(t)$, which is defined as

$$\psi_{b,s}(t) = \frac{1}{\sqrt{s}} \psi\left(\frac{t-b}{s}\right) \tag{3.3}$$

where $s \neq 0$ and $b$ are real constants. The parameter $s$ is the scaling parameter (used to determine window widths), whereas the parameter $b$ denotes the translation parameter (used to determine the position of the window).

The "*continuous wavelet transform*" (CWT) of a time signal $x(t)$ is defined as

$$W^X(b,s) = \int_{-\infty}^{\infty} x(t) \, \overline{\psi_{b,s}(t)} \, dt \tag{3.4}$$

provided the following *admissibility condition*[8] is satisfied

$$C_\psi = \int_{-\infty}^{\infty} \frac{|\Psi(\omega)|^2}{|\omega|} d\omega < \infty \tag{3.5}$$

where $\Psi(\omega)$ is the Fourier transform[9] of the mother wavelet $\psi_{a,b}(t)$. The square of the absolute value of the CWT is known as the wavelet power and is given by $|W^X(b,s)|^2$, where the complex argument of $W^X(b,s)$ gives the local phase. Analogous to the boundary problem encountered in discrete wavelet methods, CWT too suffers from edge effects as the transform is incorrectly computed at the initial and end points of the time-series. Edge effects can be taken into consideration by introducing a *Cone of Influence* (COI). It is the area in wavelet spectrum where wavelet power at the edges generated by some discontinuity has fallen by a magnitude of $e^{-2}$ of the edge's value.

*3.4.2 Wavelet coherence*

---

[8] The admissibility allows the reconstruction of $x(t)$ from the CWT.

[9] The Fourier transform of the wavelet function $\psi(t)$ is $\Psi(\omega) = \int_{-\infty}^{\infty} \psi(t) \, e^{-i\omega t} dt$

Let $W^X(b,s)$ and $W^Y(b,s)$ be the CWT of the time signals $x(t)$ and $y(t)$, respectively. Following Torrence and Webster (1999), the wavelet coherence of the two time signals is defined as

$$R^2(b,s) = \frac{\left|S(s^{-1}W^{XY}(b,s))\right|^2}{S\left(s^{-1}\left|W^X(b,s)\right|^2\right).S\left(s^{-1}\left|W^Y(b,s)\right|^2\right)}, \qquad (3.6)$$

where $0 \leq R^2(b,s) \leq 1$. S and $s$ are the smoothing function and wavelet scale, respectively. $W^{XY}(b,s) = W^X(b,s)W^{Y*}(b,s)$ is the "*cross wavelet transform*" of $x(t)$ and $y(t)$, where $W^{Y*}$ denotes the complex conjugate of $W^Y$. The absolute value of the cross wavelet, which is known as the cross wavelet power, discloses areas with high common power (Grinsted et al., 2004). The smoothing operator S(.) is defined as $S(.) = S_{scale}(S_{time}(W(s)))$, where $S_{scale}$ and $S_{time}$ denote smoothing across both scale and time. The lead-lag behaviour of the two time series, which helps in analysing the direction of contagion, is given by the cross-wavelet phase angle which is given by

$$\varphi_{X,Y} = \tan^{-1}\left[\frac{\Im(W^{XY})}{\Re}\right], \quad \text{where } \varphi_{X,Y} \in [-\pi, \pi], \qquad (3.7)$$

where $\Im$ and $\Re$ denote respectively the imaginary and real part of cross-wavelet $W^{XY}$. The phase angle is interpreted as follows,

$$\varphi_{XY} \in \begin{bmatrix} (0, \pi/2) & \Rightarrow \text{In phase and X leads Y} \\ (-\pi/2, 0) & \Rightarrow \text{In phase and Y leads X} \\ (\pi/2, \pi) & \Rightarrow \text{Anti phase and Y leading} \\ (-\pi, -\pi/2) & \Rightarrow \text{Anti phase and X leading} \end{bmatrix} \qquad (3.8)$$

Morlet wavelet, which is a product of a complex exponential and a Gaussian function, is used for computing the wavelet coherence and is given by,

$$\psi(t) = e^{i\omega_0 t} e^{-t^2/2\sigma^2}, \qquad (3.9)$$

where $\omega_0$ denotes frequency and $\sigma$ is a measure of support (or spread) of the wavelet. The scaled (via translation) and shifted (via dilation) version of the Morlet wavelet is given by,

$$\psi(b,s)(t) = \frac{1}{s} \exp\left[i\omega_0 \left(\frac{t-b}{s}\right)\right] \exp\left[-\left(\frac{t-b}{s}\right)^2 / 2\sigma^2\right]. \tag{3.10}$$

The wavelet coherence computations used in this chapter employs the Morlet wavelet as the mother wavelet as it gives good balance between the simultaneous localisation of time and frequency. It also possesses good feature extraction properties. Monte Carlo methods are used to estimate the wavelet coherence's significance level. The suggestions given by Grinsted et al. (2004) and Aguiar-Conraria and Soares (2012) are followed.

## 3.5 Empirical Analysis of Contagion

3.5.1 *Empirical data*

The empirical data consists of twenty four major stock indices comprising both developed and emerging markets. The stock indices included are BSE 30 (India), Nasdaq (U.S.), S&P 500 (U.S.), DJIA (U.S.), FTSE 100 (Great Britain), CAC40 (France), DAX 30 (Germany), NIKKEI 225 (Japan), KOSPI (Korea), KLSE (Malaysia), JKSE (Indonesia), TAIEX (Taiwan), SSE (China), STI (Singapore), HSI (Hong Kong), BEL20 (Belgium), ATX (Austria), AEX (Netherlands), IBEX 35 (Spain), SMI (Switzerland), STOXX50 (Eurozone), ASX 200 (Australia), KSE100 (Pakistan), and IBOV (Brazil). The period of study ranges from 01-07-1997 to 20-01-2014 consisting of 4096 dyadic length observations making it suitable for various wavelet methods. Areas in the wavelet coherence plots where significant events occurred are labelled. The following table explains the events in detail.

Table 3.1.           Abbreviations used in the coherence plots depicting various stock market events

| | |
|---|---|
| EA | July 15, 1997: The East Asian crisis. |
| D+ | March 24, 2000: Peak of S&P 500 during the Dot-Com bubble. |
| WT | September 11, 2001: Terrorist attack on the world trade centre. |
| D- | October 2, 2002: Lowest point of S&P 500 due to the Dot-Com bubble burst. |
| GF | Global stock markets crash during the financial crisis of 2007-09. |
| L- | December 31, 2008: S&P 500 crash. |
| l- | S&P 500 at its historical low. |
| EU | September 23, 2010: The Eurozone crisis. |

| A1 | August 2011: CAC40 at its lowest after July 2011 rescue package to Greece. |

*3.5.2 Empirical results*

The empirical analysis begins with the investigation of comovements between several pairs of equity markets, particularly focusing on Indian markets' co-movement with other developed and emerging markets , by the means of a time-frequency domain measure of correlation coefficient, aka the wavelet coherence, as described in the previous section. A marked increase in correlation between two markets after a financial crisis, according to Forbes and Rigobon (2002), is known as contagion. As the propagation of shocks in financial markets, during contagious crises, is very fast, correlations disappear very quickly and last not more than one-two weeks (Baig and Goldfajn, 1998). However, correlations that last long term reveal the existing interdependence between markets and are not necessarily the consequence of contagious shocks. However, with the introduction of wavelet methods, this chapter analyses correlation across timescales, allowing one to differentiate between pure and fundamentals based contagion or mere interdependence. Significant increase in correlation at shorter timescales is taken to be indicative of pure contagion whereas stronger correlation structure at longer timescales suggests contagion due to fundamentals and strong market integration.

Wavelet coherence diagram helps one to distinguish between significant short and long term correlations. Information from timescales ranging from around 4-1024 days is given in the left vertical axis of coherence plot. Morlet wavelet is used as the "*mother wavelet*" in computing wavelet coherence and the significance is determined by Monte Carlo methods. The cone of influence (COI), where the coherence map is affected by boundary problem, is shown in a lighter shade. Statistically significant areas in the coherence plot, with 5% significance level, are denoted by bold black borders. The colour coded coherence map reveal strongest power at regions with red colour whereas blue regions reveal low power.  A scatter of phase arrows is also plotted to enable one determine the direction of comovement. Arrows pointing right indicate that both markets are in phase whereas anti-phase relation is depicted by arrows pointing left. BSE-NASDAQ pair used in the preliminary analysis of comovement reveal anti-phase relationship between BSE30 and NASDAQ around timescale of 64 days and an in-phase relationship between the two after timescale of 96 days. Arrows pointing down reveal that the first market index (BSE) leads the second market index (NASDAQ).

Similarly the second market (NASDAQ) leads the first market (BSE) if arrows point up. The primary focus of this analysis centres on the contagious effects of 2008 global financial crisis and the relatively recent Eurozone crisis of 2010-2012.

Figure 3.1 Wavelet coherence maps of BSE30 with NASDAQ and SP500

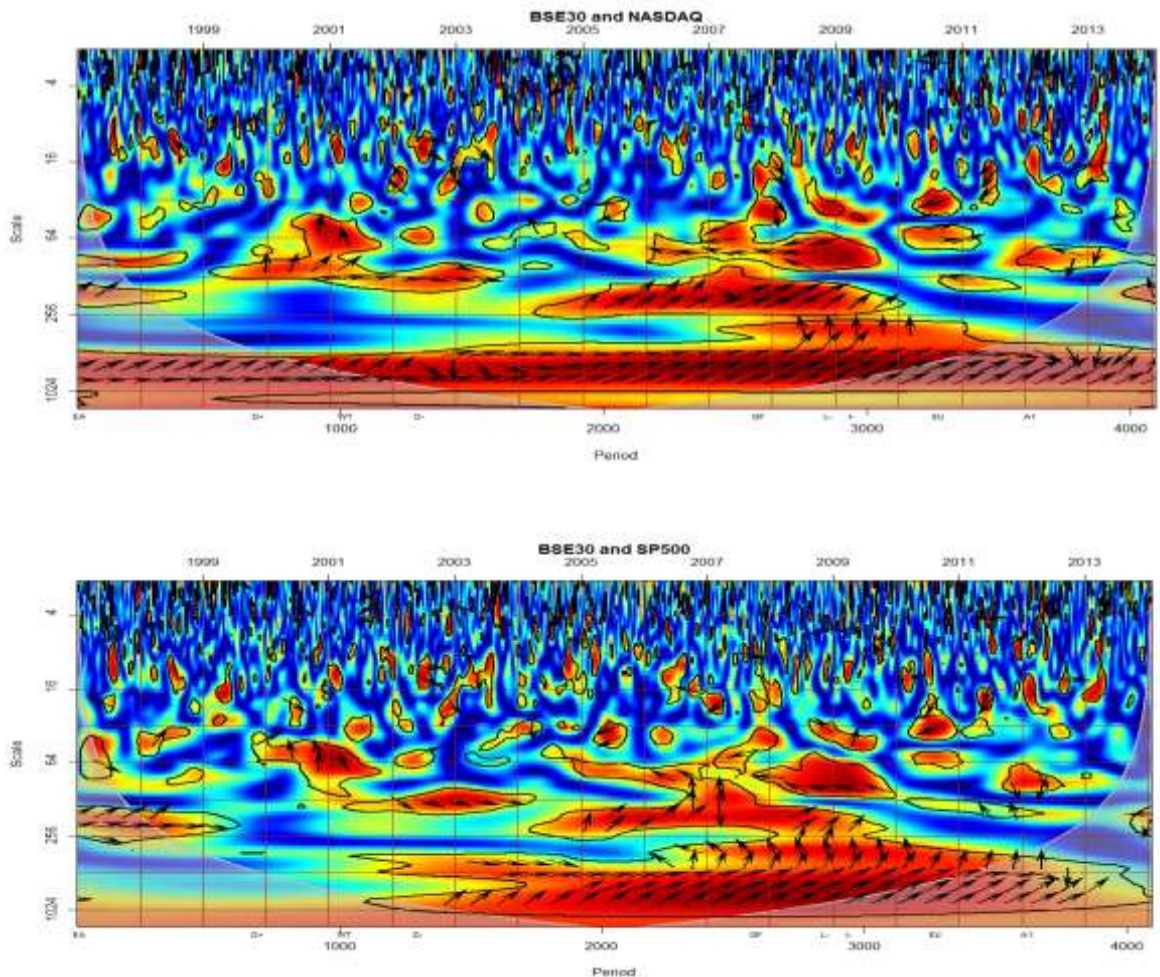

Strong correlation between BSE30 and NASDAQ (U.S.A), around the neighbourhood of global financial crisis (marked GF in the horizontal axis), is detected at higher timescale of 64 days and beyond, revealing long run integration of Indian and U.S. market. Medium-run comovement, of around 32 day oscillations, can also be observed during the neighbourhood of Eurozone crisis (marked EU). However, the absence of any significant short timescale comovement reveals the absence of pure contagion between US and Indian markets. Similar conclusions can be derived from the BSE30-S&P500 pair where correlations tend to increase in the long-run. Short run comovement of around 8-16 days oscillation can be observed around the neighbourhood of 2008 financial crisis. However correlations around this period dissipates faster and does not sustain for longer period.

The coherence map of the French (CAC40) and Indian market (BSE30) reveal long-run fundamental linkages with scant evidence of pure contagion. Long term comovements between BSE30 and CAC40, beyond 96 day oscillations, can be observed alongside significant medium-run (around 64 days) correlation during the Dow Jones crash in 2009. Similar results hold true for the BSE30 and DAX (Germany) pair, indicating long-run market integration. The direction of spillover for both BSE30-CAC40 and BSE30-DAX pairs in the long run, however, seems to run from the developed French and German markets as revealed by the scatter of phase arrows.

Figure 3.2 Wavelet coherence maps of BSE30 with CAC40 and DAX

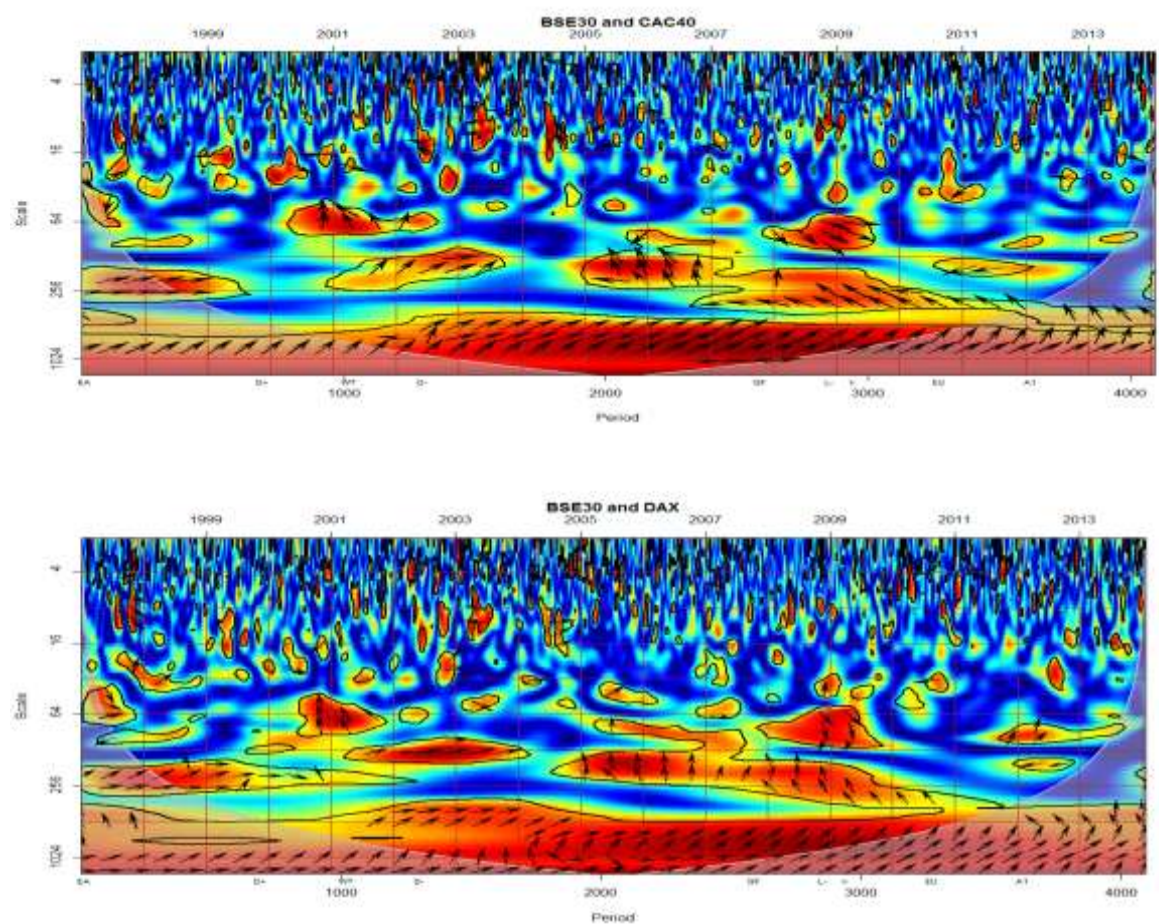

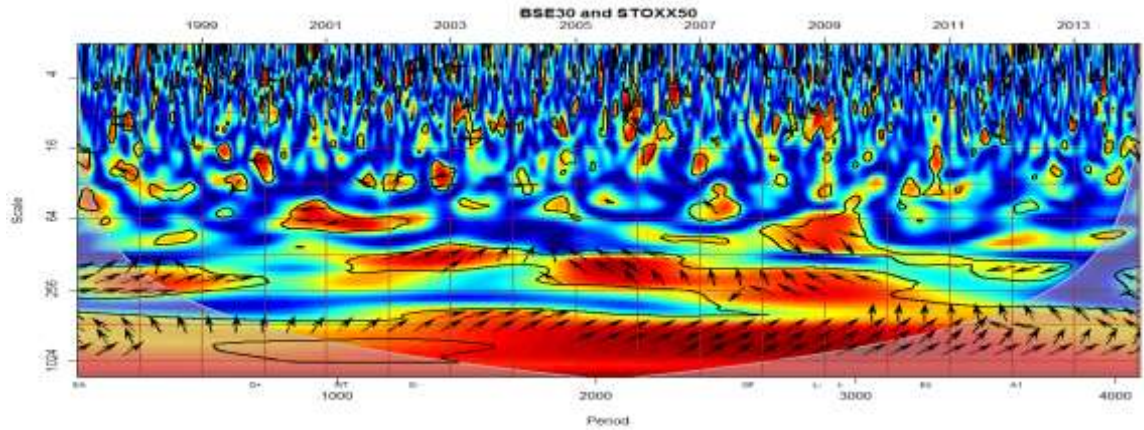

The EURO STOXX 50, which represents market leaders in the Eurozone, is positively correlated with BSE30 at longer time horizons. However, some relatively short-lived comovement with BSE30 is observed at 8-16 day frequency oscillations. Moreover, coherence map of FTSE 100 (Great Britain) and BSE 30 share common regions of significant power beyond 128 day (six month) time horizon, during both global financial crisis (GFC, hereafter) and Eurozone crisis, revealing fundamentals based contagion. The direction of shock runs from FTSE 100 to BSE 30 as FTSE 100 leads BSE 30 in the long run. No evidence of significant short-run comovement dynamics can be detected for the FTSE100-BSE30 pair. Similar results indicating no evidence of contagion can be deduced from coherence maps of BSE 30 and other developed European markets. However, pure contagion between ATX (Austria) and BSE 30 can be detected as strong coherence, around the neighbourhood of the GFC, can be observed at short run timescales.

Figure 3.3 Wavelet coherence maps of BSE30 with FTSE and ATX

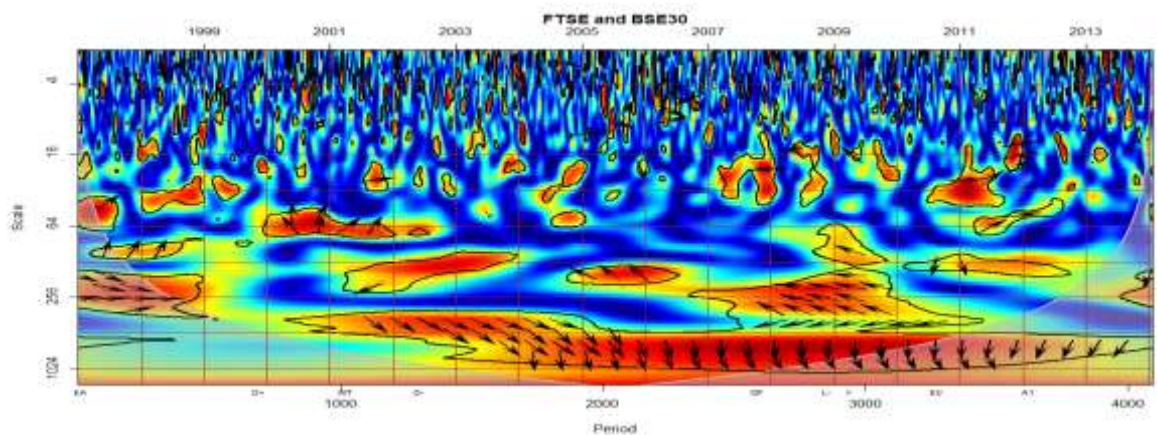

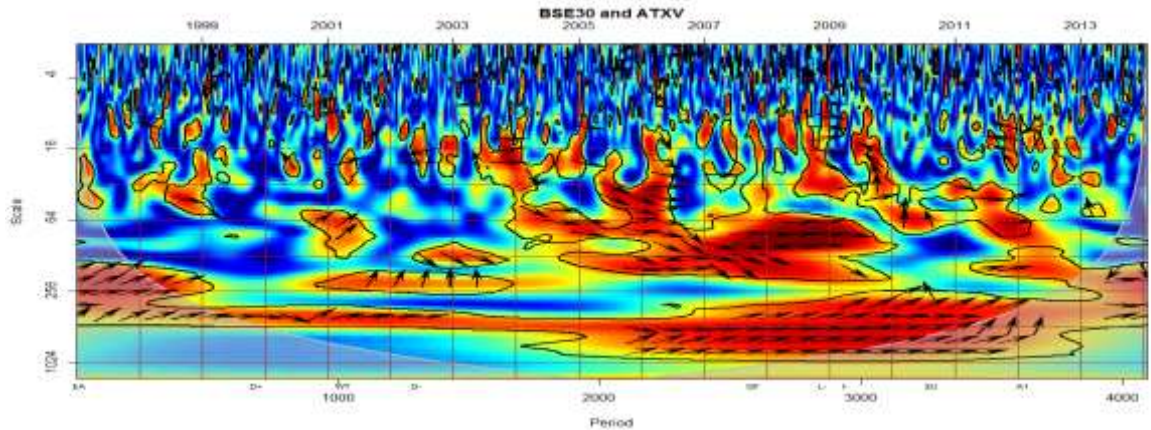

With regard to Asian markets, strong comovement between BSE 30 and NIKKEI (Japan) is observed beyond 64 days oscillation, indicating good interdependence, with NIKKEI leading BSE30. However strong comovements are recorded, in the short-run period encompassing 6-16 days oscillation, during time periods between global financial crisis and the Eurozone crisis. Areas with significant short run comovements reveal some evidence of contagion. Similar conclusions with regard to long run interdependence can be drawn from the BSE30-KOSPI (South Korea) pair. Nevertheless, comovements operating at short to medium run ranging between 8-64 days can be observed, particularly around the neighbourhood of S&P500 crash in 2009 (denoted by l- in the horizontal axis). Both Japanese and South Korean markets lead Indian market at significant regions in the coherence map. The existence of some strong short-run correlation, between both BSE30-NIKKEI and BSE30-KOSPI pairs, indicate some level of diversification risks for Indian investors.

Figure 3.4 Wavelet coherence maps of BSE30 with NIKKEI and KOSPI

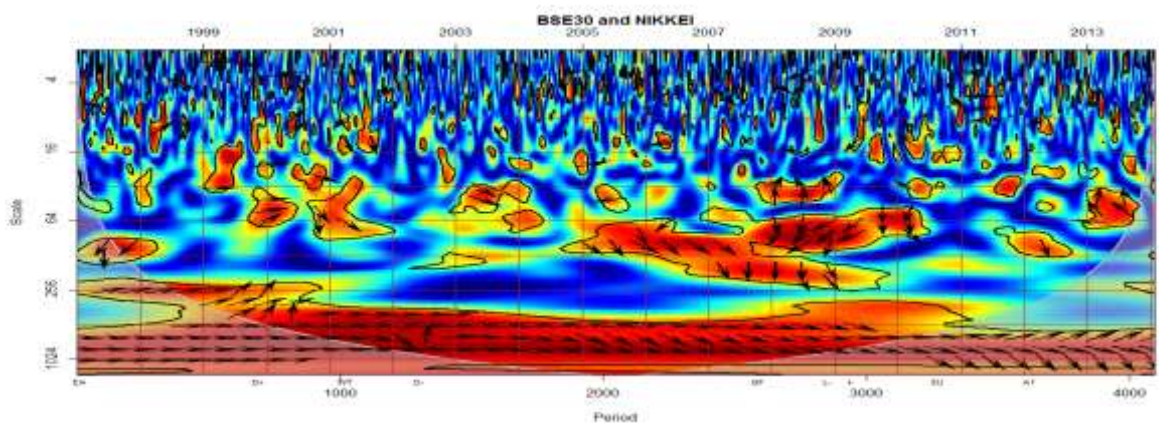

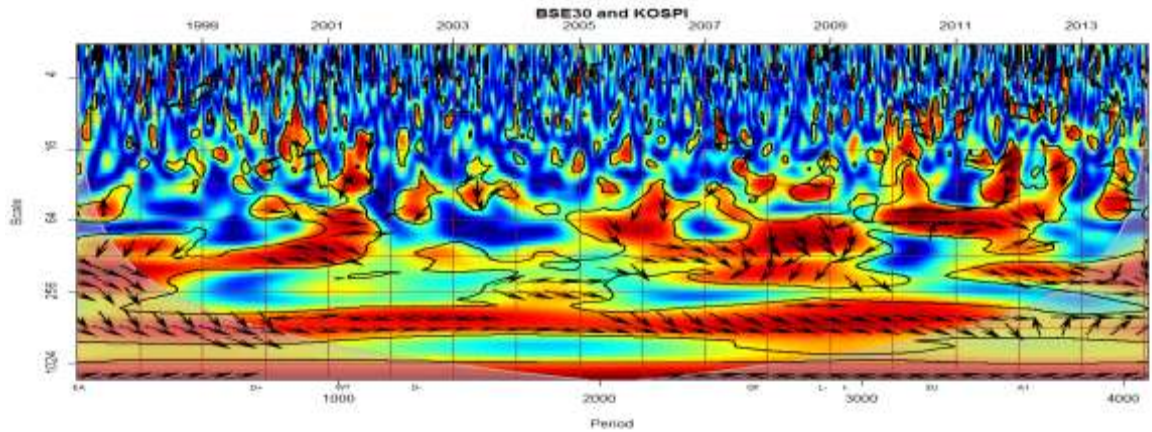

Coherence dynamics between HSI (Hong Kong) and BSE 30 reveal some evidence of pure contagion during the global financial crisis. Significant coherence at frequencies ranging between 4-16 days oscillation can be detected during this period, demonstrating the incidence of pure contagion. Moreover, strong medium and long-range interdependence can be observed as both markets move together at time horizon beginning with 32 days and beyond, thus evidencing strong long run interdependence. KLSE (Malaysia) and BSE 30 show strong regions of comovement at timescale spanning 8-32 days, evidencing some transmission of short term shocks beyond any fundamental linkages.

Figure 3.5 Wavelet coherence maps of BSE30 with HSI and KLSE

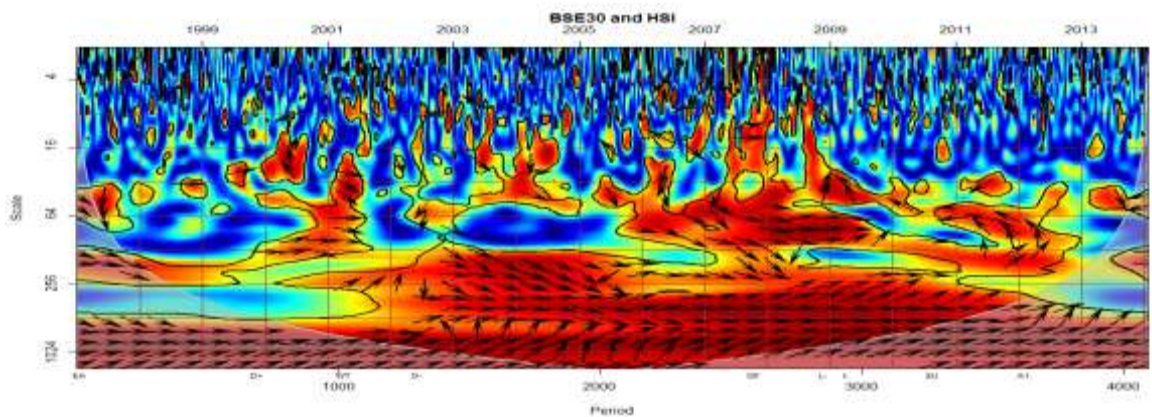

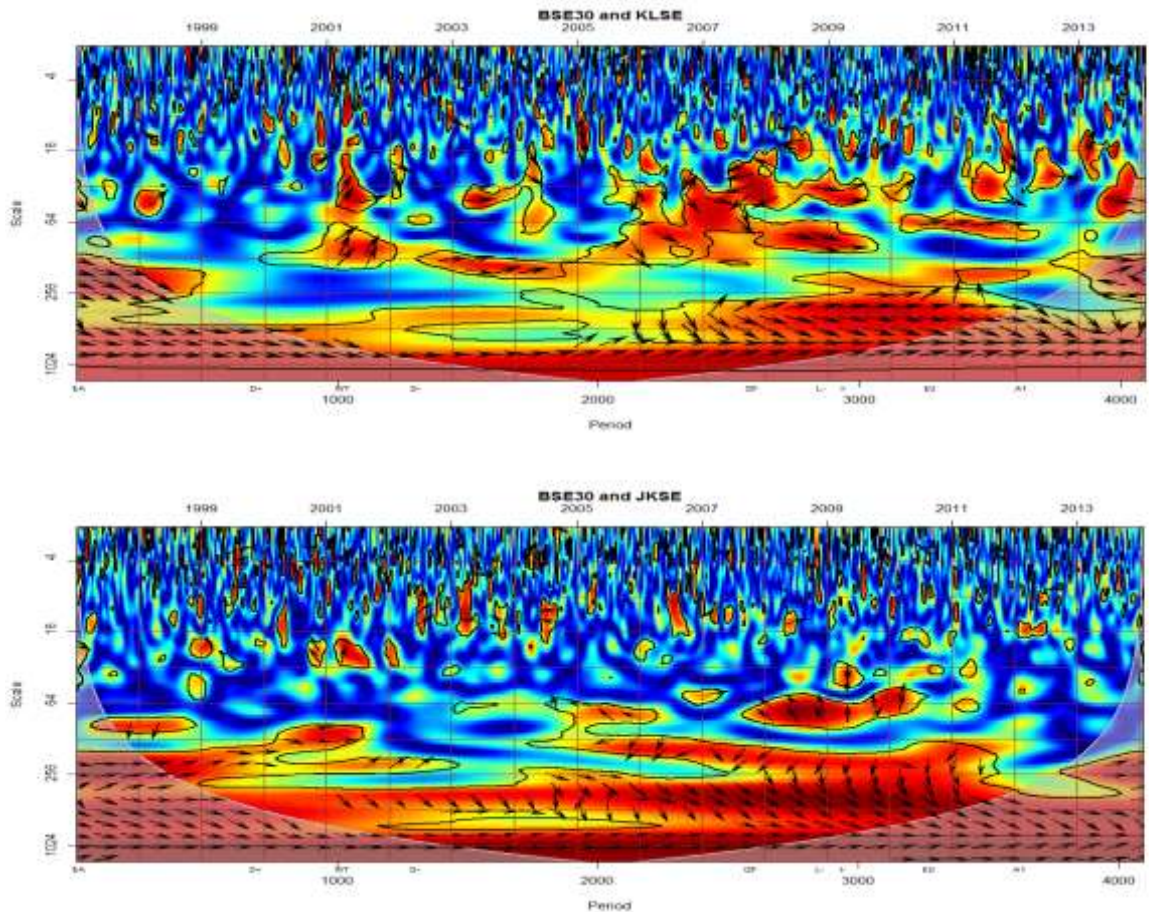

This occurrence is detected around the neighbourhood of both GFC and Eurozone crises. Similar results can be observed when looking at coherence map of BSE 30 and JKSE (Indonesia). BSE 30 also seems to be strongly correlated with TAIEX (Taiwan), during the Eurozone crisis, at shorter timescale of 8-16 days where the direction of shock runs from Taiwan to India. Moreover, strong medium-run comovement between BSE 30 and TAIEX, operating at time horizon between 64-96 days, is detected during both Eurozone crisis and GFC. Moreover, wavelet coherence map between BSE 30 and STI (Singapore) reveal some significant comovement region around the neighbourhood of Eurozone crisis at monthly time horizon. On the other hand, strong regions of common power between BSE 30 and STI, during both GFC and Eurozone crisis, can be detected at longer time horizons revealing long run market integration and fundamental linkages.

Figure 3.6 Wavelet coherence maps of BSE30 with TAIEX, STI and IBEX

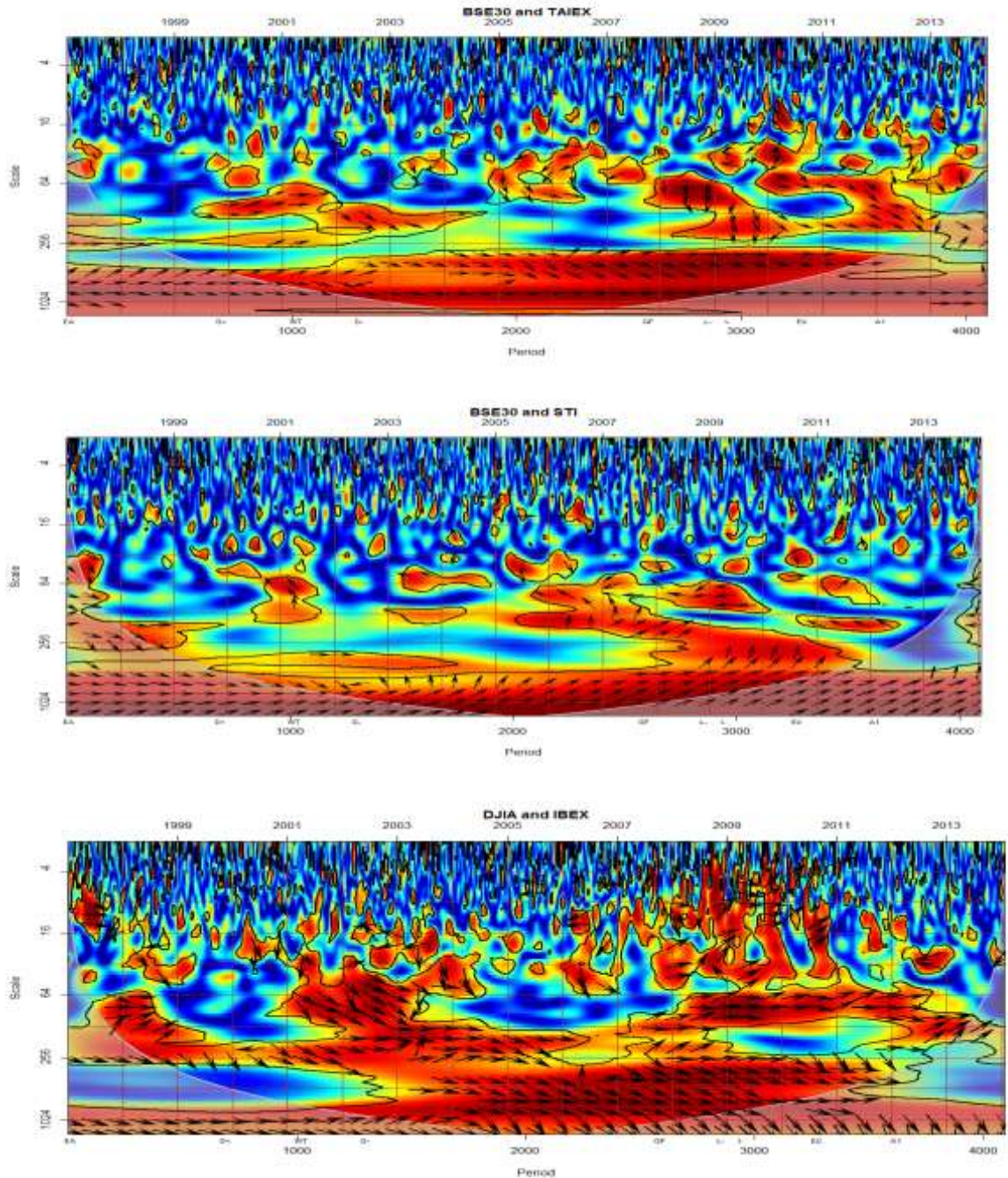

Strong interdependence, at both shorter and long run time horizons, between developed European and American markets can be seen from the coherence maps revealing good market integration between these economies. For e.g. contagion can be clearly detected between DJIA (U.S.A) and IBEX (Spain) as strong comovement can be detected at shorter time horizons spanning 4-8 days. The direction of shock runs from IBEX to DJIA at shorter time horizons whereas at time horizons beyond 256 days, shocks are transmitted from DJIA to IBEX.

The wavelet coherence maps of various market pairs gives a visual graphical tool to analyse the strength of comovement between markets. However, as in the case of discrete wavelet correlation diagrams, coherence tends to increase as we move towards longer time horizons where wider regions of significance in coherence maps can be detected. Irrespective of this fact, short-run comovements can be detected between market pairs but spread of correlation around shorter time horizons is not wide. Nevertheless, after a careful visual exploration, one can detect significant comovements at shorter time horizons too. Contagion of the pure kind is said to manifest itself during the short run, which in the wavelet case, can be thought of as the spread of significant coherence at shorter time horizons.

The analysis of coherence between various equity markets and the Indian market reveal an overall increase in coherence between the European markets and India at longer time horizons, revealing interdependence and fundamental linkages. This phenomena is more concentrated around periods of financial turbulence, especially during GFC and Eurozone crises, where developed markets lead the Indian equity market. However with the exception of the BSE30-ATX pair, short run dynamics, which reveals the absence of pure contagion between Indian and developed European markets, are not present. The same is true for the Indian and American market pairs. On the other hand, some pure contagion can be observed between Indian and select Asian markets, during both GFC and Eurozone crisis, as evidenced by short-run comovements between Indian and Asian markets. Moreover, strong comovements are detected between markets from developed economies.

The significance of the results obtained from wavelet coherence is further tested by analysing contagion between various market pairs in a wavelet based discrete correlation approach. A multiresolution analysis of stock returns of select equity markets is carried out using a MODWT decomposition. Short run and long run dynamics are captured by decomposing stock returns into six levels of resolution associated with the first six wavelet details, namely, "d1, d2, d3, d4, d5 and d6", corresponding to timescale of "1-2 days, 2-4 days, 4-8 days, 8-16 days, 16-32 days and 32-64 days", respectively. The MODWT based multiresolution analysis is carried out using reflection boundary condition with the least-asymmetric filter (LA8) of length eight. Correlation is then estimated using a rolling window method to generate a time-series of wavelet correlation at all six levels of decomposition. The time-series of wavelet correlation generated using the rolling wavelet correlation method is then used to test for contagion between Indian and major world markets. This is done by the means of a two sample t-test to compare the difference in correlation before and after crisis events. The time periods before and after crisis events have 250 days as suggested by (Ranta, 2013). Shorter time windows fail to capture long term events whereas longer window does not capture short term events. The main events selected are, i) The global financial crisis (GF), ii) The Eurozone crisis (EU), and iii) the August 2011 Eurozone crisis after rescue package to Greece. The hypothesis of the test is, therefore, to test the difference in average correlation before (say, BC) and after (say, AC) the crisis events. Therefore, the following hypothesis is formulated

$$H0: BC-AC=0$$

$$H1: BC-AC \neq 0$$

The results of t-test are given in tables 3.2-3.17. Significant increase in correlation coefficient after a crisis event, at one percent and five percent critical values, are marked in bold. The results of t-test support the findings of wavelet correlation analysis. Short-run increase in correlation, between Indian and some Asian markets, can be seen during the 2008 financial crisis. The market pairs formed from Indian (BSE 30) and some Asian markets show significant rise in short-run correlation after the 2008 crisis, suggesting pure contagion (see Table 3.2). BSE-KOSPI, BSE-NIKKEI, BSE-TAIEX, BSE-JKSE, BSE-KLSE and BSE-STI show significant rise in correlation, at 1% significance level, after crisis period indicating some contagion between BSE 30 (India) and equity markets of South Korea, Japan, Taiwan, Indonesia, Malaysia and Singapore. However, with the exception of FTSE 100 (Great Britain) and ATX (Austria), BSE30 does not seem to have significant short run correlation differences with markets from developed European economies and the United States (see Table 3.3). The robustness of the MODWT based rolling correlation method coupled with tow-sample t-test lies in the fact that coherence plots cannot significantly identify strength of correlation at shorter time scales of 1-8 days. This is mitigated by the MODWT multiresolution decomposition which allows one to capture correlation differences at finer scales. For e.g. short-run comovements between FTSE and BSE30 were not clear from the coherence map. However, strong contagion effects seems to be more prevalent among developed markets of Europe and the United States, as significant rise in short-run correlation, after the 2008 crisis, can be observed from S&P 500-DAX (Germany), S&P 500-IBEX (Spain), S&P 500-BEL20 (Belgium), S&P 500-STOXX50 (Eurozone), and FTSE 100 (Great Britain) pairs (Table 3.4). Moreover, Asian markets from China, Hong Kong, Korea, Japan and Malaysia also suffer from contagious shocks as evidenced by the significant rise in short run correlations, after the 2008 crisis, among S&P 500-SSE, S&P 500-HSI, S&P 500-KOSPI, NIKKEI, and S&P 500-KLSE pairs (Table 3.5).

In view of the European sovereign debt crisis, and with the exception of Taiwan (TAIEX) and Hong Kong (HSI), Indian market (BSE 30) seems to be immune from the contagious shocks emanating from other East Asian markets (Table 3.6). However, evidence of contagion between Indian and some European markets can be observed from the rise in short run correlations after the Eurozone crisis among BSE 30-DAX (Germany), BSE 30-SMI (Switzerland), BSE 30-STOXX50 (Eurozone index), and BSE 30-IBEX (Spain). The same is true for the Indian and Australian (ASX 200) market pair, BSE-ASX (Table 3.7). On the other hand, clear signs of contagion can be observed among some European markets as the pairs FTSE-CAC40, FTSE-ASE (Greece), DAX-STOXX50, and DAX-BEL20 reveal significant rise in short-run correlation after the Eurozone sovereign debt crisis (see Table 3.8 and Table 3.11). Similarly from the Asian market pairs of FTSE-HSI, FTSE-TAIEX, DAX-JKSE, DAX-STI, and DAX-NIKKEI, one can deduce the rise in short run correlation after the Eurozone crisis (see Table 3.9 and Table 3.10). However, with the exception of BEL20 and STOXX50, no signs of short run rise in correlation between Indian and European markets, at least for two successive timescales, can be observed after the August 2011 crash which was induced by fear of contagion spreading to Spain and Italy after the Sovereign debt crisis (Table 3.12). On the other hand, clear signs of contagious shocks can be seen transmitted between Indian and some Asian markets. The rise in short run correlation between India and markets of Japan, South Korea, Malaysia, Indonesia, Singapore and Taiwan, indicates the existence of some contagion (Table 3.13).

To summarise, the results from t-test show: (i) contagion between Indian and East Asian markets during the GFC; (ii) no sign of significant short-run comovement between Indian and developed markets from Europe and United States, with the exception of some significant interrelations between Indian market and markets from Great Britain and Austria; (iii) evidence of contagion among developed markets of Europe, the United States and Asia during the GFC; (iv) No contagion between Indian and Asian markets during Eurozone sovereign debt crisis but presence of some contagion among Indian and some European markets; (iv) no contagion between most European and Indian market during the August 2011 euro crisis after rescue package to Greece and clear signs of contagion between Indian and Asian markets during the same period.

## 3.6 Conclusion

With the pre-millennium seminal works on global shock propagation that evidenced the existence of contagion (King and Wadhwani, 1990; Calvo and Reinhart, 1996; Eichengreen et al., 1996; Baig and Goldfajn, 1998 etc.), and the subsequent rejection of contagion by Forbes and Rigobon (2002) after correcting for heteroskedasticity, there has emerged copious amount of both theoretical and empirical literature explaining the modes of shock propagation during periods of financial turmoil. Divergent conclusions on the existence or absence of contagion, however, continue to engulf the empirical literature, as do the general consensus on the exact definition of contagion in case of theoretical models explaining contagion. Moreover, majority of empirical literature on contagion use time domain models to test for contagion, clearly missing out information concerning different frequencies which can be isolated into its low and high counterparts to gauge both short-run and long-run dynamics. This chapter extends the empirical literature on contagion by analysing contagion from a time-frequency perspective, allowing one to clearly differentiate between short-run and long-run transmission of shocks. This is achieved by the use of wavelet methods, both continuous and discrete, which makes possible a thorough analysis of contagion in the time-frequency space (Rua and Nunes, 2009). In doing so, the existence of a mapping between short-run and long-run dynamics to the presence of pure and fundamentals contagion, respectively, is conjectured.

In view of the above, this chapter attempts to investigate the contagious effects of the Global financial crisis of 2007-2009 and the sovereign debt crisis of 2010-2012 on the Indian equity market from a time-frequency perspective. No evidence of pure contagion between India and major developed markets of Europe and the United States, as said to manifest itself by the sudden and significant short-run comovement (Candelon et al., 2008), is reported. However, Indian market seems to be affected by contagious shocks from some developed Asian markets during the 2008 financial crisis. Moreover, contagion between Indian and some Asian markets is also evidenced during the aftermath of the August 2011 Eurozone crisis, which supports the regional nature of contagion as evidenced in Glick and Rose (1998). On the contrary, no evidence of contagion is reported between Indian and developed western markets during this period. However, evidence of contagion is affirmed, during both financial crises, among some highly integrated developed markets of the west. The same holds true between some Asian markets and the crisis engulfed markets of United States and Europe. Moreover, wavelet coherence maps reveal the existence of strong fundamental linkages between the developed markets, as evidenced by the existence of strong long-run comovements between these markets. This phenomenon of fundamentals based contagion, which manifests in a relatively stronger manner during the studied periods of financial turmoil, can be argued to have taken place via strong trade linkages between developed markets, thereby confirming similar conclusions from other studies (see Zhang, 2008; N'Diaye et al., 2010). Similarly, long-run fundamental linkages between Indian and some Asian markets, possibly via the trade channel, is validated by inferences from coherence maps. Nevertheless, evidences of pure contagion between Indian and some Asian markets, with relatively strong trade channels, can be attributable to regional proximity of the studied markets, confirming the conclusions made in some seminal works[10] dealing with trade linkages.

As far as policy implications are concerned, this study offers several recommendations. The separation of short-run and long-run shocks alongside their relative power, in a time-frequency framework, allows investors to clearly formulate optimal investment strategies based on risks involved at various investment horizons. With respect to portfolio diversification, Indian investors are required to exercise caution, especially in the short-run, while formulating portfolios comprising of stocks from some Asian economies. However, strategic investors might benefit in the long run by including stocks from some European and Asian markets.

With respect to the mitigation of short-run contagious shocks, stabilisation policies aimed at the short-run can help infected markets to bypass speculative attacks emanating via investors' psychology and herding behaviour. Moreover, transparency in financial policies, proper disclosure of data, supervision and regulation of financial sectors etc. might help strengthen the country's financial system, thereby providing some immunity from contagious shocks.

---

[10] Glick and Rose (1998), Kaminsky and Reinhart (2000), and Diwan and Hoekman (1998) discuss the importance of intra-regional trade as a channel of shock propagation.

## 3.7 Results of t-test Comparing Difference in Wavelet Correlation

Table 3.2 T-test results comparing wavelet correlation for 2008

| Timescale | BSE-NIKKEI Before | After | Test Statistic | BSE-KOSPI Before | After | Test Statistic | BSE-JKSE Before | After | Test Statistic | BSE-KLSE Before | After | Test Statistic |
|---|---|---|---|---|---|---|---|---|---|---|---|---|
| 1-2 Days | -0.01 | -0.01 | 0.52 | -0.06 | 0.14 | -35.08** | 0.06 | 0.05 | 1.57 | -0.01 | 0.02 | -7.70** |
| 2-4 Days | -0.03 | 0.02 | -14.83** | -0.05 | 0.17 | -28.07** | -0.07 | 0.01 | -18.43** | -0.03 | 0.09 | -19.35** |
| 4-8 Days | -0.11 | -0.01 | -20.02** | -0.09 | 0.12 | -32.26** | -0.12 | 0.18 | -19.41** | 0.16 | -0.02 | 11.70** |
| 8-16 Days | -0.02 | -0.01 | -1.97* | 0.10 | -0.09 | 30.39** | -0.13 | 0.13 | -25.01** | -0.67 | -0.05 | -40.39** |
| 16-32 Days | -0.16 | -0.21 | 6.74** | -0.59 | -0.57 | -1.64** | -0.05 | -0.21 | 16.61** | -0.33 | 0.56 | -39.63** |
| 32-64 Days | -0.09 | -0.47 | 30.50** | 0.30 | 0.25 | 5.27** | -0.42 | -0.39 | -6.68** | 0.76 | 0.86 | -19.65** |
| Timescale | BSE-STI Before | After | Test Statistic | BSE-HSI Before | After | Test Statistic | BSE-TAIEX Before | After | Test Statistic | BSE-SSE Before | After | Test Statistic |
| 1-2 Days | -0.12 | -0.08 | -13.58** | 0.26 | 0.06 | 35.25** | -0.01 | -0.01 | -1.31 | -0.02 | 0.02 | -11.82** |
| 2-4 Days | -0.07 | 0.15 | -31.19** | 0.19 | -0.34 | 33.57** | -0.01 | 0.02 | -3.80** | 0.13 | 0.08 | 17.85** |
| 4-8 Days | 0.10 | -0.11 | 29.46** | 0.42 | -0.07 | 42.17** | -0.06 | 0.13 | -24.38** | 0.13 | 0.09 | 5.52** |
| 8-16 Days | 0.18 | 0.04 | 21.46** | 0.60 | 0.15 | 47.36** | 0.00 | 0.18 | -17.70** | 0.16 | -0.26 | 39.28** |
| 16-32 Days | 0.05 | -0.17 | 21.03** | 0.78 | 0.28 | 27.77** | -0.59 | -0.51 | -5.17** | -0.34 | -0.01 | -17.64** |
| 32-64 Days | -0.56 | -0.91 | 38.68** | 0.91 | 0.97 | -18.53** | -0.14 | -0.20 | 8.53** | -0.23 | -0.62 | 13.62** |
| Timescale | BSE-KSE100 Before | After | Test Statistic | BSE-IBOV Before | After | Test Statistic | BSE-DJIA Before | After | Test Statistic | BSE-SNP Before | After | Test Statistic |
| 1-2 Days | -0.04 | 0.01 | -9.29** | -0.06 | -0.18 | 11.89** | 0.00 | -0.02 | 9.08** | -0.01 | -0.03 | 13.27** |
| 2-4 Days | 0.10 | 0.04 | 6.69** | 0.14 | 0.18 | -3.42** | -0.08 | -0.05 | -7.62** | -0.04 | -0.06 | 4.62** |
| 4-8 Days | 0.05 | -0.01 | 6.38** | 0.22 | 0.21 | 0.40 | 0.09 | -0.19 | 37.58** | 0.12 | -0.18 | 40.70** |
| 8-16 Days | 0.00 | 0.44 | -28.82** | 0.50 | 0.48 | 1.88 | -0.09 | -0.05 | -4.80** | -0.07 | -0.10 | 2.81** |
| 16-32 Days | -0.15 | -0.17 | 1.59 | 0.59 | 0.63 | -5.60** | -0.25 | -0.15 | -14.59** | -0.20 | -0.21 | 0.45 |
| 32-64 Days | 0.09 | 0.68 | -25.57** | 0.78 | 0.97 | -38.38** | -0.55 | -0.94 | 25.64** | -0.57 | -0.94 | 25.45** |

Note: ** denote significance at 1% critical value and * denote significance at 5% critical value

Table 3.3 T-test results comparing wavelet correlation for 2008

| Timescale | BSE-DAX Before | After | Test Statistic | BSE-FTSE Before | After | Test Statistic | BSE-NASDAQ Before | After | Test Statistic | BSE-CAC40 Before | After | Test Statistic |
|---|---|---|---|---|---|---|---|---|---|---|---|---|
| 1-2 Days | -0.06 | -0.08 | 4.80** | 0.07 | 0.04 | 5.45** | -0.01 | -0.03 | 8.96** | 0.02 | 0.02 | 2.17* |
| 2-4 Days | 0.09 | 0.04 | 16.64** | -0.03 | -0.01 | -2.86** | -0.08 | -0.03 | -10.90** | 0.10 | 0.12 | -8.32** |
| 4-8 Days | 0.15 | 0.08 | 6.21** | -0.04 | 0.03 | -13.91** | 0.23 | -0.15 | 48.40** | -0.34 | 0.23 | -58.77** |
| 8-16 Days | -0.10 | 0.18 | -65.99** | -0.48 | -0.09 | -29.62** | -0.24 | -0.19 | -4.20** | -0.02 | -0.24 | 16.87** |
| 16-32 Days | -0.10 | 0.36 | -20.49** | -0.24 | -0.19 | -5.15** | -0.10 | -0.28 | 18.67** | -0.09 | -0.24 | 8.32** |
| 32-64 Days | -0.32 | 0.13 | -30.66** | 0.30 | -0.17 | 32.75** | -0.61 | -0.95 | 28.19** | -0.04 | 0.62 | -33.51** |

| Timescale | BSE-STOXX50 Before | After | Test Statistic | BSE-ATX Before | After | Test Statistic | BSE-IBEX Before | After | Test Statistic | BSE-BEL20 Before | After | Test Statistic |
|---|---|---|---|---|---|---|---|---|---|---|---|---|
| 1-2 Days | -0.02 | -0.04 | 2.73** | 0.13 | 0.01 | 21.47** | -0.07 | -0.13 | 14.74** | -0.07 | -0.15 | 27.20** |
| 2-4 Days | -0.04 | -0.09 | 10.23** | 0.19 | 0.28 | -13.42** | 0.07 | -0.01 | 22.15** | -0.09 | -0.03 | -12.70** |
| 4-8 Days | 0.15 | -0.36 | 50.48** | -0.23 | 0.34 | -42.44** | -0.12 | -0.24 | 22.47** | 0.05 | -0.04 | 7.20** |
| 8-16 Days | -0.13 | -0.13 | 0.61 | -0.01 | 0.56 | -41.70** | 0.12 | 0.02 | 9.85** | -0.12 | -0.31 | 23.57** |
| 16-32 Days | -0.20 | -0.32 | 16.82** | 0.28 | 0.63 | -57.05** | 0.02 | -0.34 | 43.74** | -0.22 | -0.13 | -6.60** |
| 32-64 Days | 0.02 | 0.74 | -35.87** | 0.91 | 0.98 | -12.95** | -0.29 | -0.89 | 38.94** | -0.08 | 0.59 | -40.04** |

| Timescale | BSE-AEX Before | After | Test Statistic | BSE-ASE Before | After | Test Statistic | BSE-ASX Before | After | Test Statistic | BSE-SMI Before | After | Test Statistic |
|---|---|---|---|---|---|---|---|---|---|---|---|---|
| 1-2 Days | 0.14 | 0.02 | 33.50** | 0.08 | 0.05 | 10.94** | 0.03 | -0.10 | 38.40** | -0.13 | -0.14 | 3.65** |
| 2-4 Days | 0.11 | 0.09 | 3.68** | 0.00 | 0.10 | -17.81** | 0.04 | -0.02 | 15.46** | -0.07 | -0.05 | -4.45** |
| 4-8 Days | 0.22 | -0.05 | 72.96** | 0.02 | 0.12 | -9.96** | 0.25 | -0.10 | 31.44** | 0.00 | -0.04 | 8.33** |
| 8-16 Days | 0.24 | 0.05 | 29.40** | 0.01 | -0.22 | 43.29** | -0.20 | -0.02 | -24.55** | 0.37 | 0.16 | 21.48** |
| 16-32 Days | 0.11 | -0.26 | 24.94** | 0.22 | 0.05 | 7.74** | -0.11 | 0.13 | -13.33** | 0.14 | -0.32 | 29.46** |
| 32-64 Days | 0.29 | 0.88 | -34.56** | 0.08 | -0.85 | 31.01** | -0.06 | 0.59 | -34.49** | -0.66 | -0.95 | 23.31** |

Note: ** denote significance at 1% critical value and * denote significance at 5% critical value

Table 3.4 T-test results comparing wavelet correlation for 2008

| Timescale | SNP-DAX Before | After | Test Statistic | SNP-FTSE Before | After | Test Statistic | SNP-CAC40 Before | After | Test Statistic | SNP-STOXX50 Before | After | Test Statistic |
|---|---|---|---|---|---|---|---|---|---|---|---|---|
| 1-2 Days | -0.02 | 0.22 | -64.35** | -0.06 | 0.09 | -53.11** | -0.03 | -0.06 | 7.31** | -0.10 | 0.01 | -23.30** |
| 2-4 Days | 0.06 | 0.28 | -42.01** | -0.16 | 0.00 | -27.77** | 0.04 | -0.13 | 31.02** | -0.03 | 0.05 | -15.71** |
| 4-8 Days | 0.06 | 0.00 | 13.04** | 0.14 | 0.02 | 22.53** | 0.02 | -0.15 | 41.81** | 0.10 | 0.23 | -13.90** |
| 8-16 Days | -0.09 | 0.36 | -38.47** | 0.03 | -0.08 | 11.59** | -0.14 | -0.32 | 30.78** | 0.29 | -0.20 | 53.91** |
| 16-32 Days | -0.51 | 0.04 | -51.42** | -0.41 | -0.06 | -39.34** | -0.21 | 0.52 | -58.55** | 0.22 | 0.39 | -13.92** |
| 32-64 Days | -0.25 | -0.22 | -7.04** | 0.12 | 0.14 | -2.57** | -0.55 | -0.71 | 28.18** | -0.68 | -0.85 | 32.27** |

| Timescale | SNP-ATX Before | After | Test Statistic | SNP-IBEX Before | After | Test Statistic | SNP-BEL20 Before | After | Test Statistic | SNP-ASX Before | After | Test Statistic |
|---|---|---|---|---|---|---|---|---|---|---|---|---|
| 1-2 Days | -0.10 | -0.11 | 1.61 | -0.15 | 0.20 | -48.90** | 0.13 | 0.06 | 20.11** | -0.01 | -0.05 | 10.03** |
| 2-4 Days | -0.11 | -0.09 | -5.19** | 0.18 | 0.64 | -50.97** | 0.01 | 0.08 | -17.54** | 0.15 | -0.21 | 62.94** |
| 4-8 Days | 0.08 | -0.14 | 38.30** | 0.41 | 0.74 | -60.76** | -0.05 | 0.07 | -24.08** | -0.17 | 0.26 | -74.90** |
| 8-16 Days | 0.12 | -0.03 | 21.25** | 0.69 | 0.74 | -9.08** | -0.28 | -0.24 | -2.94** | -0.07 | 0.07 | -8.49** |
| 16-32 Days | 0.34 | 0.03 | 30.87** | 0.73 | 0.65 | 10.51** | -0.36 | 0.32 | -68.84** | -0.56 | 0.18 | -54.69** |
| 32-64 Days | -0.62 | -0.96 | 23.40** | 0.51 | 0.97 | -39.96** | -0.34 | -0.70 | 40.87** | -0.37 | -0.68 | 35.28** |

| Timescale | SNP-AEX Before | After | Test Statistic | SNP-ASE Before | After | Test Statistic | SNP-SMI Before | After | Test Statistic | DAX-IBEX Before | After | Test Statistic |
|---|---|---|---|---|---|---|---|---|---|---|---|---|
| 1-2 Days | 0.10 | 0.05 | 7.58** | -0.05 | 0.11 | -52.07** | -0.02 | 0.16 | -30.98** | -0.02 | 0.06 | -29.33** |
| 2-4 Days | -0.16 | -0.02 | -20.76** | 0.14 | 0.05 | 15.93** | 0.09 | -0.20 | 39.43** | 0.07 | 0.08 | -0.90 |
| 4-8 Days | -0.04 | 0.12 | -27.80** | -0.09 | 0.30 | -40.07** | -0.34 | -0.57 | 27.58** | 0.05 | 0.17 | -30.35** |
| 8-16 Days | 0.15 | -0.06 | 12.62** | -0.06 | -0.20 | 10.96** | -0.10 | 0.17 | -67.30** | 0.02 | -0.01 | 3.42** |
| 16-32 Days | 0.22 | 0.22 | -0.19 | -0.45 | 0.01 | -29.16** | 0.79 | 0.74 | 14.31** | -0.29 | 0.09 | -23.32** |
| 32-64 Days | -0.75 | -0.95 | 32.91** | 0.38 | 0.94 | -35.27** | 0.92 | 0.95 | -8.07** | -0.45 | -0.67 | 19.77** |

Note: ** denote significance at 1% critical value and * denote significance at 5% critical value

Table 3.5 T-test results comparing wavelet correlation for 2008

|  | SNP-NIKKEI |  |  | SNP-KOSPI |  |  | SNP-STI |  |  | SNP-HSI |  |  |
|---|---|---|---|---|---|---|---|---|---|---|---|---|
| Timescale | Before | After | Test Statistic | Before | After | Test Statistic | Before | After | Test Statistic | Before | After | Test Statistic |
| 1-2 Days | -0.02 | 0.22 | -64.35** | -0.06 | 0.09 | -53.11** | -0.03 | -0.06 | 7.31** | -0.10 | 0.01 | -23.30** |
| 2-4 Days | 0.06 | 0.28 | -42.01** | -0.16 | 0.00 | -27.77** | 0.04 | -0.13 | 31.02** | -0.03 | 0.05 | -15.71** |
| 4-8 Days | 0.06 | 0.00 | 13.04** | 0.14 | 0.02 | 22.53** | 0.02 | -0.15 | 41.81** | 0.10 | 0.23 | -13.90** |
| 8-16 Days | -0.09 | 0.36 | -38.47** | 0.03 | -0.08 | 11.59** | -0.14 | -0.32 | 30.78** | 0.29 | -0.20 | 53.91** |
| 16-32 Days | -0.51 | 0.04 | -51.42** | -0.41 | -0.06 | -39.34** | -0.21 | 0.52 | -58.55** | 0.22 | 0.39 | -13.92** |
| 32-64 Days | -0.25 | -0.22 | -7.04** | 0.12 | 0.14 | -2.57** | -0.55 | -0.71 | 28.18** | -0.68 | -0.85 | 32.27** |
|  | SNP-JKSE |  |  | SNP-KLSE |  |  | SNP-SSE |  |  | SNP-IBOV |  |  |
| Timescale | Before | After | Test Statistic | Before | After | Test Statistic | Before | After | Test Statistic | Before | After | Test Statistic |
| 1-2 Days | -0.10 | -0.11 | 1.61 | -0.15 | 0.20 | -48.90** | 0.13 | 0.06 | 20.11** | -0.01 | -0.05 | 10.03** |
| 2-4 Days | -0.11 | -0.09 | -5.19** | 0.18 | 0.64 | -50.97** | 0.01 | 0.08 | -17.54** | 0.15 | -0.21 | 62.94** |
| 4-8 Days | 0.08 | -0.14 | 38.30** | 0.41 | 0.74 | -60.76** | -0.05 | 0.07 | -24.08** | -0.17 | 0.26 | -74.90** |
| 8-16 Days | 0.12 | -0.03 | 21.25** | 0.69 | 0.74 | -9.08** | -0.28 | -0.24 | -2.94** | -0.07 | 0.07 | -8.49** |
| 16-32 Days | 0.34 | 0.03 | 30.87** | 0.73 | 0.65 | 10.51** | -0.36 | 0.32 | -68.84** | -0.56 | 0.18 | -54.69** |
| 32-64 Days | -0.62 | -0.96 | 23.40** | 0.51 | 0.97 | -39.96** | -0.34 | -0.70 | 40.87** | -0.37 | -0.68 | 35.28** |
|  | SNP-TAIEX |  |  | SNP-KSE100 |  |  | SNP-BSE |  |  | FTSE-BSE |  |  |
| Timescale | Before | After | Test Statistic | Before | After | Test Statistic | Before | After | Test Statistic | Before | After | Test Statistic |
| 1-2 Days | -0.11 | 0.01 | -42.12** | -0.07 | 0.03 | -23.65** | -0.01 | -0.03 | 13.27** | -0.07 | -0.07 | -0.82 |
| 2-4 Days | 0.15 | -0.06 | 48.99** | 0.00 | -0.05 | 21.24** | -0.04 | -0.06 | 4.62** | -0.10 | 0.00 | -25.69** |
| 4-8 Days | 0.18 | -0.04 | 38.19** | -0.05 | -0.05 | -0.97 | 0.12 | -0.18 | 40.70** | -0.02 | -0.13 | 22.68** |
| 8-16 Days | 0.02 | 0.17 | -18.77** | -0.07 | -0.04 | -2.05* | -0.07 | -0.10 | 2.81** | 0.19 | 0.26 | -6.68** |
| 16-32 Days | 0.30 | 0.21 | 13.08** | -0.12 | 0.23 | -25.62** | -0.20 | -0.21 | 0.45 | -0.02 | 0.46 | -25.40** |
| 32-64 Days | -0.39 | 0.15 | -42.34** | -0.07 | -0.67 | 21.11** | -0.56 | -0.94 | 25.45** | -0.18 | 0.03 | -14.89** |

Note: ** denote significance at 1% critical value and * denote significance at 5% critical value

Table 3.6 T-test results comparing wavelet correlation for EU

|  | BSE-NIKKEI |  |  | BSE-KOSPI |  |  | BSE-JKSE |  |  |
|---|---|---|---|---|---|---|---|---|---|
| Timescale | Before | After | Test Statistic | Before | After | Test Statistic | Before | After | Test Statistic |
| 1-2 Days | 0.00 | 0.10 | -23.65** | 0.02 | -0.05 | 10.29** | 0.01 | -0.01 | 4.52** |
| 2-4 Days | 0.03 | -0.18 | 22.43** | 0.10 | 0.10 | -0.78 | 0.06 | 0.01 | 10.46** |
| 4-8 Days | -0.09 | -0.18 | 18.12** | 0.02 | 0.00 | 2.14* | 0.02 | -0.08 | 9.63** |
| 8-16 Days | 0.05 | 0.11 | -8.40** | 0.04 | -0.06 | 4.98** | 0.35 | 0.38 | -3.22** |
| 16-32 Days | -0.13 | -0.11 | -1.90 | 0.04 | 0.54 | -18.49** | -0.25 | -0.04 | -21.27** |
| 32-64 Days | -0.35 | 0.13 | -23.78** | 0.22 | 0.84 | -34.41** | -0.18 | -0.02 | -12.92** |
|  | BSE-STI |  |  | BSE-HSI |  |  | BSE-TAIEX |  |  |
| Timescale | Before | After | Test Statistic | Before | After | Test Statistic | Before | After | Test Statistic |
| 1-2 Days | -0.06 | -0.06 | -1.15 | 0.03 | 0.02 | 2.29* | 0.11 | 0.15 | -6.09** |
| 2-4 Days | 0.01 | 0.02 | -0.98 | -0.24 | -0.03 | -24.06** | -0.04 | 0.01 | -4.11** |
| 4-8 Days | 0.13 | -0.05 | 12.03** | 0.04 | 0.36 | -31.12** | 0.15 | 0.18 | -4.76** |
| 8-16 Days | 0.28 | 0.08 | 12.37** | -0.03 | 0.29 | -43.04** | -0.21 | -0.38 | 10.58** |
| 16-32 Days | -0.02 | 0.33 | -31.78** | -0.32 | -0.64 | 20.32** | -0.44 | 0.02 | -23.22** |
| 32-64 Days | -0.63 | 0.54 | -42.26** | 0.62 | -0.40 | 37.38** | 0.18 | 0.63 | -15.29** |
|  | BSE-KSE100 |  |  | BSE-SSE |  |  | BSE-KLSE |  |  |
| Timescale | Before | After | Test Statistic | Before | After | Test Statistic | Before | After | Test Statistic |
| 1-2 Days | -0.07 | -0.05 | -5.11** | 0.00 | 0.04 | -7.89** | 0.01 | 0.11 | -18.92** |
| 2-4 Days | -0.09 | 0.16 | -27.03** | 0.11 | 0.04 | 20.43** | 0.03 | 0.04 | -1.52 |
| 4-8 Days | 0.19 | 0.08 | 13.01** | -0.03 | 0.10 | -20.32** | 0.08 | -0.19 | 11.47** |
| 8-16 Days | 0.34 | -0.10 | 38.66** | -0.06 | -0.10 | 4.89** | 0.29 | 0.20 | 11.39** |
| 16-32 Days | -0.22 | 0.01 | -24.91** | 0.09 | 0.17 | -2.81** | 0.73 | 0.44 | 32.90** |
| 32-64 Days | 0.12 | 0.06 | 1.71 | -0.40 | -0.30 | -6.78** | 0.62 | 0.45 | 12.03** |

Note: ** denote significance at 1% critical value and * denote significance at 5% critical value

Table 3.7 T-test results comparing wavelet correlation for EU

| | BSE-DAX | | | BSE-FTSE | | | BSE-SMI | | | BSE-CAC40 | | |
|---|---|---|---|---|---|---|---|---|---|---|---|---|
| Timescale | Before | After | Test Statistic | Before | After | Test Statistic | Before | After | Test Statistic | Before | After | Test Statistic |
| 1-2 Days | 0.00 | 0.10 | -15.87** | -0.07 | -0.07 | -0.82 | -0.02 | -0.03 | 2.12* | 0.07 | 0.00 | 20.73** |
| 2-4 Days | -0.07 | -0.05 | -1.97* | -0.10 | 0.00 | -25.69** | -0.17 | 0.03 | -23.26** | 0.07 | 0.07 | 0.41 |
| 4-8 Days | 0.15 | -0.15 | 42.68** | -0.02 | -0.13 | 22.68** | 0.06 | 0.11 | -5.81** | 0.09 | -0.15 | 23.79** |
| 8-16 Days | 0.18 | -0.15 | 30.43** | 0.19 | 0.26 | -6.68** | 0.12 | -0.03 | 19.42** | -0.23 | 0.10 | -29.91** |
| 16-32 Days | 0.46 | -0.44 | 56.87** | -0.02 | 0.46 | -25.40** | -0.32 | 0.40 | -58.81** | 0.02 | -0.28 | 21.73** |
| 32-64 Days | 0.12 | -0.14 | 43.70** | -0.18 | 0.03 | -14.89** | -0.62 | 0.21 | -24.17** | 0.37 | -0.47 | 38.27** |
| | BSE-STOXX50 | | | BSE-ATX | | | BSE-IBEX | | | BSE-BEL20 | | |
| Timescale | Before | After | Test Statistic | Before | After | Test Statistic | Before | After | Test Statistic | Before | After | Test Statistic |
| 1-2 Days | -0.04 | 0.02 | -14.16** | -0.08 | -0.19 | 18.49** | -0.14 | 0.02 | -30.96** | -0.06 | -0.06 | -1.71 |
| 2-4 Days | -0.19 | -0.05 | -31.72** | 0.02 | -0.10 | 12.46** | -0.02 | 0.09 | -19.09** | -0.03 | -0.05 | 2.85** |
| 4-8 Days | -0.24 | 0.16 | -34.38** | 0.06 | -0.19 | 16.41** | 0.02 | 0.25 | -13.62** | -0.16 | 0.06 | -35.06** |
| 8-16 Days | 0.04 | 0.29 | -20.28** | 0.11 | -0.02 | 6.67** | 0.12 | 0.03 | 10.11** | -0.35 | -0.25 | -17.95** |
| 16-32 Days | -0.15 | -0.04 | -6.80** | 0.41 | 0.46 | -2.74** | -0.25 | -0.17 | -7.46** | 0.02 | -0.52 | 29.81** |
| 32-64 Days | 0.47 | -0.38 | 30.61** | 0.74 | 0.06 | 34.04** | -0.57 | 0.42 | -34.65** | 0.45 | -0.40 | 44.02** |
| | BSE-AEX | | | BSE-ASE | | | BSE-ASX | | | BSE-SNP | | |
| Timescale | Before | After | Test Statistic | Before | After | Test Statistic | Before | After | Test Statistic | Before | After | Test Statistic |
| 1-2 Days | 0.03 | 0.03 | 0.19 | 0.04 | -0.05 | 17.97** | -0.04 | -0.02 | -2.31* | -0.03 | -0.05 | 6.97** |
| 2-4 Days | 0.15 | 0.07 | 17.25** | 0.09 | 0.00 | 19.37** | -0.05 | 0.13 | -34.80** | 0.02 | 0.01 | 1.21 |
| 4-8 Days | 0.08 | 0.16 | -10.30** | 0.02 | -0.06 | 9.95** | -0.11 | 0.08 | -17.19** | 0.03 | 0.23 | -11.06** |
| 8-16 Days | 0.18 | 0.15 | 3.97** | -0.27 | -0.10 | -28.45** | -0.05 | -0.26 | 23.64** | 0.04 | -0.08 | 14.24** |
| 16-32 Days | -0.22 | 0.19 | -28.47** | 0.13 | 0.07 | 2.56** | 0.29 | -0.66 | 66.62** | -0.10 | -0.22 | 17.58** |
| 32-64 Days | 0.60 | -0.46 | 37.00** | -0.54 | -0.03 | -22.72** | 0.22 | -0.50 | 33.15** | -0.66 | 0.48 | -44.26** |

Note: ** denote significance at 1% critical value and * denote significance at 5% critical value

Table 3.8 T-test Results comparing wavelet correlation for EU

| | FTSE-SNP | | | FTSE-DAX | | | FTSE-CAC40 | | |
|---|---|---|---|---|---|---|---|---|---|
| Timescale | Before | After | Test Statistic | Before | After | Test Statistic | Before | After | Test Statistic |
| 1-2 Days | 0.13 | 0.09 | 7.50** | -0.10 | 0.02 | -23.09** | 0.02 | 0.17 | -41.44** |
| 2-4 Days | -0.01 | 0.16 | -26.57** | 0.02 | -0.08 | 18.86** | -0.10 | 0.15 | -36.46** |
| 4-8 Days | 0.02 | -0.02 | 12.22** | 0.15 | -0.08 | 25.92** | 0.06 | 0.11 | -10.95** |
| 8-16 Days | -0.03 | -0.02 | -0.43 | -0.30 | 0.04 | -23.38** | 0.17 | 0.13 | 5.39** |
| 16-32 Days | 0.08 | -0.25 | 22.84** | 0.02 | -0.37 | 16.36** | 0.12 | -0.13 | 35.51** |
| 32-64 Days | -0.09 | -0.29 | 10.98** | 0.86 | 0.45 | 49.84** | 0.59 | 0.58 | 1.44 |
| | FTSE-STOXX50 | | | FTSE-ATX | | | FTSE-IBEX | | |
| Timescale | Before | After | Test Statistic | Before | After | Test Statistic | Before | After | Test Statistic |
| 1-2 Days | 0.18 | -0.09 | 65.55** | -0.09 | 0.04 | -20.91** | 0.06 | -0.10 | 25.29** |
| 2-4 Days | 0.14 | -0.02 | 35.57** | -0.06 | -0.07 | 1.51 | 0.24 | 0.20 | 4.28** |
| 4-8 Days | -0.10 | -0.21 | 17.55** | -0.04 | -0.03 | -1.44 | 0.03 | -0.09 | 22.47** |
| 8-16 Days | 0.31 | -0.19 | 36.02** | -0.08 | 0.30 | -28.61** | -0.12 | -0.26 | 14.00** |
| 16-32 Days | 0.14 | 0.12 | 1.79 | -0.04 | 0.08 | -11.10** | 0.25 | -0.16 | 20.26** |
| 32-64 Days | 0.41 | 0.63 | -31.00** | 0.10 | 0.04 | 5.75** | -0.16 | -0.35 | 11.64** |
| | FTSE-NASDAQ | | | FTSE-ASE | | | BSE-ASX | | |
| Timescale | Before | After | Test Statistic | Before | After | Test Statistic | Before | After | Test Statistic |
| 1-2 Days | 0.13 | 0.09 | 8.48** | -0.01 | 0.04 | -10.94** | 0.02 | 0.03 | -0.72 |
| 2-4 Days | -0.03 | 0.15 | -23.37** | -0.08 | 0.02 | -22.04** | -0.02 | 0.01 | -5.80** |
| 4-8 Days | 0.09 | 0.00 | 20.15** | -0.01 | -0.04 | 2.85** | -0.12 | -0.20 | 5.93** |
| 8-16 Days | -0.09 | -0.06 | -2.47** | 0.04 | -0.23 | 17.46** | -0.27 | 0.12 | -32.45** |
| 16-32 Days | 0.17 | -0.27 | 29.39** | -0.28 | -0.20 | -5.75** | 0.20 | 0.64 | -30.63** |
| 32-64 Days | -0.09 | -0.30 | 11.45** | -0.10 | -0.40 | 24.99** | 0.17 | -0.22 | 27.41** |

Note: ** denote significance at 1% critical value and * denote significance at 5% critical value

Table 3.9 T-test Results comparing wavelet correlation for EU

| Timescale | FTSE-NIKKEI Before | After | Test Statistic | FTSE-KOSPI Before | After | Test Statistic | FTSE-JKSE Before | After | Test Statistic |
|---|---|---|---|---|---|---|---|---|---|
| 1-2 Days | 0.12 | 0.02 | 26.51** | **-0.08** | **-0.06** | **-4.95**** | 0.03 | 0.03 | 0.67 |
| 2-4 Days | -0.01 | -0.06 | 13.62** | 0.14 | 0.06 | 9.86** | 0.10 | -0.08 | 24.73** |
| 4-8 Days | 0.27 | 0.28 | -0.63 | 0.04 | 0.00 | 7.15** | -0.07 | -0.16 | 7.55** |
| 8-16 Days | 0.00 | -0.11 | 17.93** | -0.09 | -0.31 | 15.73** | 0.21 | 0.02 | 25.88** |
| 16-32 Days | 0.12 | -0.31 | 42.30** | **-0.13** | **0.11** | **-31.43**** | **-0.35** | **0.21** | **-43.62**** |
| 32-64 Days | 0.19 | 0.08 | 2.74** | 0.75 | 0.37 | 22.51** | **-0.19** | **0.29** | **-20.46**** |
| | FTSE-KLSE Before | After | Test Statistic | FTSE-STI Before | After | Test Statistic | FTSE-HSI Before | After | Test Statistic |
| 1-2 Days | -0.03 | -0.03 | -0.20 | **-0.03** | **-0.02** | **-3.64**** | **-0.13** | **0.03** | **-23.42**** |
| 2-4 Days | 0.03 | -0.06 | 19.54** | 0.03 | -0.14 | 36.08** | **0.00** | **0.17** | **-39.89**** |
| 4-8 Days | -0.13 | -0.12 | -0.64 | **0.16** | **0.19** | **-2.75**** | 0.15 | 0.06 | 15.49** |
| 8-16 Days | **-0.20** | **0.19** | **-29.53**** | 0.10 | -0.37 | 38.75** | 0.07 | 0.03 | 2.96** |
| 16-32 Days | **-0.16** | **0.09** | **-34.84**** | 0.14 | 0.32 | -22.25** | 0.29 | 0.04 | 24.60** |
| 32-64 Days | 0.14 | 0.02 | 23.55** | 0.08 | -0.51 | 29.75** | **0.04** | **0.36** | **-14.12**** |
| | FTSE-SSE Before | After | Test Statistic | FTSE-TAIEX Before | After | Test Statistic | FTSE-BSE Before | After | Test Statistic |
| 1-2 Days | 0.17 | 0.04 | 18.80** | **-0.03** | **0.03** | **-14.06**** | -0.07 | -0.07 | -0.82 |
| 2-4 Days | 0.03 | 0.02 | 2.47** | **-0.08** | **-0.02** | **-12.16**** | **-0.10** | **0.00** | **-25.69**** |
| 4-8 Days | **-0.23** | **-0.10** | **-30.57**** | -0.03 | 0.00 | -2.08* | -0.02 | -0.13 | 22.68** |
| 8-16 Days | 0.20 | -0.20 | 42.31** | 0.16 | -0.09 | 15.90** | **0.19** | **0.26** | **-6.68**** |
| 16-32 Days | **0.01** | **0.31** | **-23.35**** | 0.24 | 0.03 | 12.18** | **-0.02** | **0.46** | **-25.40**** |
| 32-64 Days | -0.08 | -0.32 | 13.32** | 0.42 | 0.28 | 4.03** | **-0.18** | **0.03** | **-14.89**** |

Note: ** denote significance at 1% critical value and * denote significance at 5% critical value

Table 3.10 T-test Results comparing wavelet correlation for EU

| Timescale | DAX-NIKKEI Before | After | Test Statistic | DAX-KOSPI Before | After | Test Statistic | DAX-JKSE Before | After | Test Statistic | DAX-KLSE Before | After | Test Statistic |
|---|---|---|---|---|---|---|---|---|---|---|---|---|
| 1-2 Days | -0.02 | -0.06 | 5.81** | -0.03 | -0.03 | 0.36** | 0.10 | 0.16 | -6.22** | 0.10 | -0.06 | 59.81** |
| 2-4 Days | **-0.07** | **-0.05** | **-5.21**** | -0.12 | -0.02 | -14.44** | -0.09 | 0.01 | -23.96** | 0.10 | 0.04 | 9.57** |
| 4-8 Days | **0.00** | **0.18** | **-20.58**** | 0.04 | 0.06 | -5.78** | 0.01 | 0.23 | -13.30** | 0.13 | 0.14 | -0.50 |
| 8-16 Days | 0.07 | -0.04 | 11.03** | **-0.04** | **0.05** | **-6.64**** | 0.18 | -0.01 | 16.57** | **-0.17** | **-0.09** | **-6.92**** |
| 16-32 Days | 0.43 | 0.13 | 16.91** | -0.12 | -0.21 | 3.29** | **-0.14** | **0.06** | **-9.79**** | 0.28 | -0.17 | 36.60** |
| 32-64 Days | 0.19 | -0.17 | 13.37** | 0.78 | 0.11 | 56.88** | **-0.37** | **0.10** | **-15.29**** | 0.41 | 0.08 | 19.27** |
| | DAX-STI | | | DAX-HSI | | | DAX-TAIEX | | | DAX-SSE | | |
| 1-2 Days | 0.08 | 0.03 | 8.72** | 0.04 | -0.12 | 32.77** | **0.05** | **0.10** | **-8.86**** | **-0.01** | **0.13** | **-27.93**** |
| 2-4 Days | **-0.01** | **0.07** | **-9.60**** | 0.14 | -0.04 | 22.19** | 0.09 | -0.03 | 16.26** | 0.04 | 0.00 | 6.54** |
| 4-8 Days | -0.12 | -0.04 | -12.75** | -0.09 | 0.00 | -6.79** | 0.01 | -0.01 | 2.58** | 0.00 | -0.12 | 16.11** |
| 8-16 Days | -0.05 | -0.11 | 5.27** | -0.09 | -0.29 | 12.11** | -0.01 | 0.02 | -3.49** | 0.16 | 0.15 | 1.59 |
| 16-32 Days | -0.32 | -0.15 | -19.12** | 0.08 | 0.32 | -15.42** | **-0.50** | **-0.14** | **-14.31**** | 0.05 | -0.29 | 19.19** |
| 32-64 Days | -0.14 | -0.52 | 38.23** | 0.24 | 0.72 | -28.78** | 0.50 | 0.13 | 20.13** | -0.24 | -0.22 | -0.85 |
| | DAX-KSE100 | | | DAX-IBOV | | | DAX-BSE | | | DAX-NIKKEI | | |
| 1-2 Days | 0.03 | -0.01 | 6.42** | **-0.11** | **0.02** | **-21.76**** | **0.00** | **0.10** | **-15.87**** | -0.02 | -0.06 | 5.81** |
| 2-4 Days | -0.05 | -0.08 | 4.09** | 0.03 | 0.00 | 4.93** | **-0.07** | **-0.05** | **-1.97**** | **-0.07** | **-0.05** | **-5.21**** |
| 4-8 Days | 0.04 | -0.07 | 16.56** | 0.07 | 0.16 | -15.65** | 0.15 | -0.15 | 42.68** | **0.00** | **0.18** | **-20.58**** |
| 8-16 Days | **0.28** | **0.42** | **-12.97**** | 0.16 | 0.32 | -11.97** | 0.18 | -0.15 | 30.43** | 0.07 | -0.04 | 11.03** |
| 16-32 Days | **0.14** | **0.50** | **-33.01**** | -0.04 | 0.03 | -5.19** | 0.46 | -0.44 | 56.87** | 0.43 | 0.13 | 16.91** |
| 32-64 Days | **-0.26** | **0.07** | **-13.90**** | 0.16 | 0.56 | -33.80** | 0.12 | -0.14 | 43.70** | 0.19 | -0.17 | 13.37** |

Table 3.11 T-test Results comparing wavelet correlation for EU

|  | DAX-SNP |  |  | DAX-NASDAQ |  |  | DAX-FTSE |  |  | DAX-IBEX |  |  |
|---|---|---|---|---|---|---|---|---|---|---|---|---|
| Timescale | Before | After | Test Statistic | Before | After | Test Statistic | Before | After | Test Statistic | Before | After | Test Statistic |
| 1-2 Days | 0.06 | 0.01 | 6.31** | 0.07 | 0.00 | 8.03** | -0.10 | 0.02 | -23.09** | -0.02 | 0.06 | -29.33** |
| 2-4 Days | 0.21 | -0.15 | 45.80** | 0.23 | -0.15 | 51.33** | 0.02 | -0.08 | 18.86** | 0.07 | 0.08 | -0.90 |
| 4-8 Days | 0.08 | -0.06 | 20.79** | 0.11 | -0.04 | 27.15** | 0.15 | -0.08 | 25.92** | 0.05 | 0.17 | -30.35** |
| 8-16 Days | 0.07 | -0.19 | 12.75** | 0.10 | -0.08 | 8.40** | -0.30 | 0.04 | -23.38** | 0.02 | -0.01 | 3.42** |
| 16-32 Days | -0.13 | 0.25 | -23.41** | -0.17 | 0.27 | -29.90** | 0.02 | -0.37 | 16.36** | -0.29 | 0.09 | -23.32** |
| 32-64 Days | -0.38 | -0.56 | 17.40** | -0.36 | -0.60 | 23.91** | 0.86 | 0.45 | 49.84** | -0.45 | -0.67 | 19.77** |
|  | DAX-CAC40 |  |  | DAX-STOXX50 |  |  | DAX-AEX |  |  | DAX-BEL20 |  |  |
| 1-2 Days | -0.01 | 0.02 | -9.17** | -0.04 | 0.00 | -18.12** | 0.13 | -0.05 | 44.69** | -0.06 | 0.05 | -28.32** |
| 2-4 Days | -0.05 | -0.10 | 9.05** | -0.02 | 0.23 | -57.61** | -0.06 | -0.09 | 5.01** | 0.06 | 0.25 | -23.06** |
| 4-8 Days | -0.04 | -0.21 | 16.18** | 0.12 | 0.38 | -30.92** | 0.07 | 0.11 | -4.45** | -0.58 | -0.68 | 26.38** |
| 8-16 Days | -0.56 | -0.52 | -6.59** | -0.75 | -0.63 | -27.98** | -0.47 | -0.32 | -15.33** | 0.05 | 0.21 | -11.55** |
| 16-32 Days | 0.20 | 0.21 | -1.22 | -0.32 | -0.25 | -9.88** | -0.50 | -0.54 | 4.82** | 0.47 | 0.65 | -27.77** |
| 32-64 Days | 0.77 | 0.74 | 16.78** | 0.63 | 0.53 | 35.53** | 0.20 | 0.23 | -3.08** | 0.79 | 0.86 | -23.62** |
|  | DAX-ASX |  |  | DAX-ASE |  |  | DAX-ATX |  |  | DAX-SMI |  |  |
| 1-2 Days | -0.01 | 0.11 | -18.97** | 0.12 | 0.02 | 30.58** | 0.04 | -0.11 | 23.95** | 0.09 | 0.08 | 2.00* |
| 2-4 Days | -0.46 | -0.07 | -30.00** | 0.14 | 0.01 | 23.76** | -0.17 | -0.03 | -42.20** | 0.05 | -0.05 | 15.67** |
| 4-8 Days | -0.22 | 0.45 | -71.60** | 0.10 | 0.12 | -1.97* | -0.03 | 0.16 | -16.31** | -0.18 | -0.09 | -16.31** |
| 8-16 Days | 0.40 | 0.58 | -25.72** | 0.03 | 0.18 | -6.97** | 0.10 | 0.03 | 3.92** | 0.26 | 0.38 | -13.81** |
| 16-32 Days | 0.72 | 0.80 | -18.60** | 0.08 | 0.22 | -11.12** | 0.11 | -0.05 | 14.84** | -0.59 | -0.66 | 10.96** |
| 32-64 Days | 0.89 | 0.82 | 9.25** | -0.30 | 0.11 | -41.97** | 0.45 | 0.50 | -2.66** | -0.09 | -0.29 | 17.26** |

Table 3.12 T-test Results comparing wavelet correlation for A1

|  | BSE-DAX |  |  | BSE-FTSE |  |  | BSE-SMI |  |  | BSE-CAC40 |  |  |
|---|---|---|---|---|---|---|---|---|---|---|---|---|
| Timescale | Before | After | Test Statistic | Before | After | Test Statistic | Before | After | Test Statistic | Before | After | Test Statistic |
| 1-2 Days | 0.05 | -0.04 | 21.05** | -0.06 | -0.07 | 2.89** | -0.05 | -0.13 | 20.14** | -0.01 | 0.00 | -4.25** |
| 2-4 Days | 0.04 | 0.04 | 0.63** | 0.03 | 0.02 | 4.40** | 0.14 | 0.11 | 6.29** | 0.01 | -0.06 | 18.01** |
| 4-8 Days | -0.15 | -0.05 | -20.96** | -0.10 | 0.13 | -44.39** | 0.05 | 0.16 | -12.88** | -0.09 | 0.09 | -24.19** |
| 8-16 Days | -0.28 | -0.17 | -9.64** | 0.40 | 0.27 | 13.59** | -0.13 | -0.19 | 4.74** | 0.16 | 0.02 | 14.35** |
| 16-32 Days | -0.36 | -0.12 | -19.76** | 0.58 | 0.54 | 7.19** | 0.37 | 0.40 | -3.53** | -0.13 | 0.00 | -10.25** |
| 32-64 Days | -0.10 | 0.09 | -17.62** | 0.31 | 0.22 | 4.79** | -0.08 | 0.21 | -12.22** | -0.25 | -0.17 | -4.96** |
|  | BSE-STOXX50 |  |  | BSE-ATX |  |  | BSE-IBEX |  |  | BSE-BEL20 |  |  |
| 1-2 Days | 0.07 | 0.08 | -5.54** | -0.14 | -0.18 | 9.65** | 0.00 | 0.01 | -2.50** | -0.09 | -0.09 | -2.15** |
| 2-4 Days | -0.08 | -0.02 | -14.64** | -0.11 | -0.06 | -15.77** | 0.06 | -0.14 | 24.56** | 0.00 | 0.04 | -12.88** |
| 4-8 Days | 0.09 | -0.14 | 21.22** | -0.20 | -0.49 | 40.30** | 0.14 | 0.11 | 3.20** | 0.08 | 0.15 | -9.52** |
| 8-16 Days | 0.24 | 0.10 | 16.15** | 0.10 | -0.05 | 13.49** | 0.03 | 0.38 | -41.57** | -0.16 | -0.18 | 2.75** |
| 16-32 Days | 0.15 | -0.04 | 12.98** | 0.66 | 0.48 | 10.45** | -0.19 | 0.18 | -48.55** | -0.36 | -0.03 | -22.25** |
| 32-64 Days | -0.11 | -0.38 | 13.46** | 0.24 | 0.90 | -28.19** | 0.33 | 0.24 | 10.10** | -0.19 | 0.37 | -32.31** |
|  | BSE-AEX |  |  | BSE-ASE |  |  | BSE-ASX |  |  | BSE-SNP |  |  |
| 1-2 Days | 0.04 | -0.03 | 20.50** | 0.01 | 0.09 | -17.41** | -0.07 | 0.00 | -19.39** | 0.00 | -0.05 | 8.61** |
| 2-4 Days | 0.09 | 0.12 | -8.16** | -0.02 | -0.04 | 2.92** | 0.13 | 0.14 | -3.76** | 0.04 | 0.23 | -34.04** |
| 4-8 Days | 0.11 | -0.12 | 29.71** | -0.10 | 0.00 | -21.32** | -0.01 | 0.12 | -21.32** | 0.10 | -0.02 | 17.52** |
| 8-16 Days | 0.10 | 0.08 | 3.02** | -0.08 | 0.15 | -41.11** | -0.25 | 0.27 | -35.86** | -0.16 | -0.43 | 50.58** |
| 16-32 Days | 0.38 | 0.22 | 15.84** | 0.35 | 0.30 | 3.82** | -0.68 | -0.48 | -25.97** | -0.26 | 0.01 | -19.47** |
| 32-64 Days | -0.29 | -0.47 | 10.82** | -0.12 | 0.57 | -66.73** | -0.57 | 0.12 | -51.99** | 0.50 | 0.16 | 40.43** |

Table 3.13 T-test Results comparing wavelet correlation for A1

|  | BSE-NIKKEI | | | BSE-KOSPI | | | BSE-JKSE | | | BSE-KLSE | | |
|---|---|---|---|---|---|---|---|---|---|---|---|---|
| Timescale | Before | After | Test Statistic | Before | After | Test Statistic | Before | After | Test Statistic | Before | After | Test Statistic |
| 1-2 Days | 0.13 | -0.14 | 77.43** | **-0.04** | -0.02 | -6.59** | -0.06 | -0.05 | -0.62 | 0.17 | 0.10 | 14.47** |
| 2-4 Days | **-0.28** | **-0.13** | -25.33** | 0.02 | 0.00 | 2.33* | -0.01 | -0.03 | 4.96** | -0.02 | -0.15 | 12.27** |
| 4-8 Days | **-0.21** | **-0.14** | -29.57** | 0.09 | 0.09 | 0.12 | **0.01** | **0.07** | -10.91** | 0.00 | -0.02 | 2.56** |
| 8-16 Days | 0.09 | 0.08 | 1.65 | -0.35 | -0.38 | 1.96* | **0.33** | 0.05 | 29.11** | **0.25** | **0.43** | -32.81** |
| 16-32 Days | 0.09 | -0.09 | 11.12** | 0.34 | 0.11 | 15.87** | **0.09** | **0.36** | -29.67** | 0.34 | 0.31 | 3.88** |
| 32-64 Days | -0.04 | 0.08 | -5.76** | 0.83 | 0.77 | 8.80** | -0.01 | -0.40 | 40.55** | **0.46** | **0.67** | -19.29** |
|  | BSE-STI | | | BSE-HSI | | | BSE-TAIEX | | | BSE-SSE | | |
| 1-2 Days | **-0.09** | **-0.01** | -12.45** | 0.00 | -0.05 | 11.20** | 0.06 | -0.07 | 27.58** | -0.05 | -0.11 | 7.86** |
| 2-4 Days | **0.11** | **0.12** | -2.74** | -0.03 | -0.11 | 8.70** | **-0.08** | **-0.01** | -12.27** | -0.03 | -0.18 | 36.22** |
| 4-8 Days | 0.09 | -0.13 | 19.63** | 0.36 | 0.23 | 23.97** | 0.13 | -0.08 | 25.99** | 0.09 | -0.03 | 21.15** |
| 8-16 Days | **-0.03** | **0.23** | -45.03** | 0.23 | 0.01 | 19.83** | -0.31 | -0.51 | 15.39** | -0.18 | -0.17 | -0.11 |
| 16-32 Days | 0.44 | 0.17 | 19.23** | **-0.59** | **-0.54** | -9.24** | -0.17 | -0.12 | -3.31** | **0.31** | **0.37** | -3.11** |
| 32-64 Days | **0.55** | **0.62** | -5.46** | -0.62 | -0.57 | -4.59** | **0.54** | **0.58** | -4.70** | -0.13 | 0.52 | -58.28** |
|  | BSE-KSE100 | | | BSE-IBOV | | | BSE-DJIA | | | BSE-NASDAQ | | |
| 1-2 Days | **-0.01** | **0.02** | -6.43** | 0.01 | -0.08 | 24.17** | -0.02 | -0.04 | 4.69** | 0.04 | -0.05 | 14.77** |
| 2-4 Days | 0.09 | 0.02 | 9.35** | **0.06** | **0.12** | -12.20** | **0.06** | **0.19** | -36.06** | **0.06** | **0.28** | -30.53** |
| 4-8 Days | 0.04 | -0.16 | 32.82** | -0.31 | -0.50 | 30.70** | 0.06 | -0.02 | 10.41** | 0.17 | 0.02 | 23.58** |
| 8-16 Days | **-0.23** | **0.00** | -14.23** | 0.18 | -0.02 | 14.36** | -0.20 | -0.42 | 35.25** | -0.11 | -0.39 | 41.74** |
| 16-32 Days | **0.08** | **0.41** | -32.76** | 0.43 | 0.47 | -3.43** | -0.18 | -0.05 | -11.22** | -0.27 | 0.05 | -18.86** |
| 32-64 Days | 0.25 | -0.30 | 25.66** | **0.58** | **0.64** | -4.11** | 0.53 | 0.31 | 27.82** | 0.52 | 0.02 | 48.59** |

Table 3.14 T-test Results comparing wavelet correlation for A1

|  | DAX-SNP | | | DAX-NASDAQ | | | DAX-FTSE | | | DAX-IBEX | | |
|---|---|---|---|---|---|---|---|---|---|---|---|---|
| Timescale | Before | After | Test Statistic | Before | After | Test Statistic | Before | After | Test Statistic | Before | After | Test Statistic |
| 1-2 Days | 0.00 | -0.01 | 3.80** | -0.01 | -0.03 | 4.84** | 0.04 | 0.05 | -1.51 | **0.01** | **0.06** | -9.85** |
| 2-4 Days | **-0.09** | **-0.04** | -5.63** | -0.11 | -0.03 | -10.48** | -0.05 | 0.13 | -32.29** | 0.07 | 0.17 | -18.68** |
| 4-8 Days | **0.01** | **0.11** | -8.77** | 0.03 | 0.09 | -5.72** | -0.14 | 0.06 | -31.20** | 0.09 | 0.03 | 5.30** |
| 8-16 Days | **0.02** | **0.20** | -12.48** | 0.12 | 0.24 | -9.61** | -0.06 | -0.05 | -0.30 | -0.10 | -0.12 | 1.35 |
| 16-32 Days | 0.24 | 0.14 | 7.00** | 0.24 | 0.16 | 5.75** | -0.36 | -0.34 | -1.49 | 0.18 | 0.04 | 9.86** |
| 32-64 Days | -0.55 | -0.80 | 23.75** | -0.63 | -0.85 | 27.42** | 0.29 | 0.18 | 6.39** | -0.59 | -0.78 | 18.23** |
|  | DAX-CAC40 | | | DAX-STOXX50 | | | DAX-AEX | | | DAX-BEL20 | | |
| 1-2 Days | **0.00** | **0.05** | -10.95** | 0.00 | -0.17 | 35.88** | **-0.04** | **0.04** | -20.29** | 0.04 | 0.01 | 9.03** |
| 2-4 Days | -0.10 | -0.15 | 11.23** | 0.20 | 0.06 | 20.59** | **-0.09** | **-0.05** | -7.52** | 0.28 | 0.35 | -17.16** |
| 4-8 Days | **-0.11** | **0.00** | -13.43** | 0.29 | 0.26 | 5.24** | 0.05 | 0.03 | 2.53** | -0.68 | -0.72 | 14.53** |
| 8-16 Days | -0.56 | -0.68 | 32.75** | **-0.61** | **-0.60** | -3.16** | -0.24 | -0.17 | -14.11** | 0.14 | 0.01 | 24.12** |
| 16-32 Days | **0.26** | **0.50** | -26.40** | -0.15 | 0.10 | -19.00** | -0.50 | -0.32 | -22.20** | **0.70** | **0.73** | -7.30** |
| 32-64 Days | **0.73** | **0.87** | -39.30** | 0.51 | 0.64 | -28.32** | 0.32 | 0.54 | -26.29** | **0.78** | **0.86** | -9.32** |
|  | DAX-ASX | | | DAX-ASE | | | DAX-ATX | | | DAX-SMI | | |
| 1-2 Days | 0.04 | -0.11 | 21.59** | 0.03 | 0.04 | -1.49 | **-0.07** | **-0.01** | -9.92** | 0.03 | 0.02 | 4.51** |
| 2-4 Days | 0.07 | 0.04 | 4.32** | 0.08 | 0.08 | 0.70 | 0.03 | -0.15 | 31.60** | **-0.02** | **0.07** | -16.35** |
| 4-8 Days | 0.29 | -0.01 | 14.89** | 0.15 | -0.03 | 27.02** | 0.14 | 0.00 | 30.07** | **-0.17** | **-0.02** | -16.94** |
| 8-16 Days | 0.36 | -0.44 | 37.17** | 0.09 | 0.00 | 17.79** | **-0.22** | **0.13** | -14.94** | 0.38 | 0.28 | 24.80** |
| 16-32 Days | 0.67 | 0.07 | 38.14** | **0.12** | **0.32** | -14.33** | -0.18 | -0.22 | 4.85** | **-0.63** | **-0.57** | -11.77** |
| 32-64 Days | **0.66** | **0.87** | -15.87** | 0.18 | 0.17 | 1.23 | 0.18 | 0.11 | 2.89** | -0.31 | -0.67 | 23.84** |

Table 3.15 T-test Results comparing wavelet correlation for A1

|  | DAX-NIKKEI | | | DAX-KOSPI | | | DAX-JKSE | | | DAX-KLSE | | |
|---|---|---|---|---|---|---|---|---|---|---|---|---|
| Timescale | Before | After | Test Statistic | Before | After | Test Statistic | Before | After | Test Statistic | Before | After | Test Statistic |
| 1-2 Days | -0.10 | -0.20 | 26.77** | -0.03 | -0.09 | 15.14** | 0.06 | -0.08 | 21.14** | -0.06 | 0.10 | -33.92** |
| 2-4 Days | -0.03 | -0.03 | -1.87 | 0.06 | 0.00 | 9.98** | -0.01 | -0.07 | 16.67** | 0.11 | 0.15 | -9.04** |
| 4-8 Days | 0.16 | -0.01 | 33.48** | 0.08 | 0.08 | 0.37 | 0.06 | -0.03 | 6.45** | 0.00 | 0.09 | -8.52** |
| 8-16 Days | 0.06 | -0.08 | 18.92** | 0.14 | 0.04 | 7.98** | -0.08 | -0.17 | 13.29** | -0.09 | -0.25 | 11.24** |
| 16-32 Days | -0.10 | 0.03 | -8.82** | 0.14 | 0.05 | 3.44** | -0.21 | -0.44 | 7.86** | -0.25 | -0.25 | -0.01 |
| 32-64 Days | -0.04 | -0.63 | 30.85** | 0.11 | 0.24 | -9.76** | -0.27 | -0.63 | 18.19** | -0.01 | 0.43 | -26.44** |
|  | DAX-STI | | | DAX-HSI | | | DAX-TAIEX | | | DAX-SSE | | |
| Timescale | Before | After | Test Statistic | Before | After | Test Statistic | Before | After | Test Statistic | Before | After | Test Statistic |
| 1-2 Days | 0.00 | 0.05 | -13.96** | -0.10 | 0.09 | -32.59** | 0.05 | 0.01 | 12.39** | 0.10 | -0.10 | 60.59** |
| 2-4 Days | -0.02 | 0.20 | -27.02** | -0.01 | 0.11 | -14.70** | -0.07 | -0.03 | -7.32** | 0.04 | 0.02 | 2.52** |
| 4-8 Days | 0.02 | 0.07 | -10.62** | -0.01 | -0.19 | 20.02** | -0.10 | 0.07 | -20.89** | 0.01 | 0.00 | 0.96 |
| 8-16 Days | -0.22 | -0.05 | -18.95** | -0.30 | 0.02 | -23.97** | 0.00 | 0.27 | -29.45** | 0.15 | 0.33 | -20.19** |
| 16-32 Days | -0.16 | -0.22 | 4.76** | 0.20 | 0.29 | -6.38** | 0.12 | -0.19 | 13.64** | -0.46 | -0.05 | -26.54** |
| 32-64 Days | -0.49 | -0.31 | -22.45** | 0.51 | 0.61 | -6.23** | 0.09 | -0.36 | 22.30** | 0.11 | 0.26 | -5.47** |
|  | DAX-KSE100 | | | DAX-IBOV | | | DAX-BSE | | | CAC40-BSE | | |
| Timescale | Before | After | Test Statistic | Before | After | Test Statistic | Before | After | Test Statistic | Before | After | Test Statistic |
| 1-2 Days | -0.04 | -0.04 | 0.27 | 0.03 | -0.11 | 29.64** | 0.05 | -0.04 | 21.05** | -0.01 | 0.00 | -4.25** |
| 2-4 Days | -0.06 | 0.02 | -22.18** | 0.02 | -0.23 | 42.22** | 0.04 | 0.04 | 0.63 | 0.01 | -0.06 | 18.01** |
| 4-8 Days | -0.06 | -0.05 | -2.48** | 0.13 | -0.03 | 26.51** | -0.15 | -0.05 | -20.96** | -0.09 | 0.09 | -24.19** |
| 8-16 Days | 0.37 | 0.28 | 11.43** | 0.19 | 0.25 | -3.50** | -0.28 | -0.17 | -9.64** | 0.16 | 0.02 | 14.35** |
| 16-32 Days | 0.48 | -0.03 | 32.76** | 0.14 | 0.27 | -19.49** | -0.36 | -0.12 | -19.76** | -0.13 | 0.00 | -10.25** |
| 32-64 Days | -0.13 | -0.50 | 19.16** | 0.49 | 0.59 | -10.95** | -0.10 | 0.09 | -17.62** | -0.25 | -0.17 | -4.96** |

Note: ** denote significance at 1% critical value and * denote significance at 5% critical value

Table 3.16 T-test Results comparing wavelet correlation for A1

| Timescale | Before | After | Test Statistic | Before | After | Test Statistic | Before | After | Test Statistic | Before | After | Test Statistic |
|---|---|---|---|---|---|---|---|---|---|---|---|---|
| 1-2 Days | -0.02 | 0.02 | -19.09** | -0.02 | -0.05 | 4.64** | 0.12 | -0.06 | 36.70** | -0.07 | 0.05 | -30.72** |
| 2-4 Days | -0.02 | 0.12 | -53.66** | -0.04 | -0.06 | 5.19** | -0.01 | 0.07 | -19.88** | -0.21 | -0.12 | -24.70** |
| 4-8 Days | -0.04 | -0.01 | -7.67** | 0.17 | -0.04 | 33.73** | 0.00 | 0.11 | -28.42** | 0.00 | -0.07 | 16.95** |
| 8-16 Days | -0.14 | -0.05 | -13.37** | -0.05 | -0.12 | 12.18** | 0.01 | 0.26 | -23.34** | -0.13 | 0.10 | -13.92** |
| 16-32 Days | -0.19 | 0.02 | -27.12** | 0.02 | 0.04 | -0.81 | -0.48 | 0.13 | -39.72** | -0.13 | 0.10 | -23.71** |
| 32-64 Days | -0.17 | -0.74 | 47.92** | 0.09 | 0.18 | -4.83** | -0.33 | -0.43 | 3.69** | 0.21 | 0.33 | -13.80** |
|  | CAC40-STI | | | CAC40-HSI | | | CAC40-SSE | | | CAC40-SNP | | |
|  | Before | After | Test Statistic | Before | After | Test Statistic | Before | After | Test Statistic | Before | After | Test Statistic |
| 1-2 Days | 0.03 | -0.13 | 34.21** | 0.03 | -0.12 | 33.07** | 0.08 | 0.12 | -15.95** | -0.05 | -0.03 | -3.82** |
| 2-4 Days | -0.08 | -0.02 | -16.18** | 0.08 | -0.22 | 36.57** | -0.10 | -0.06 | -9.02** | -0.03 | 0.14 | -35.66** |
| 4-8 Days | -0.06 | -0.28 | 25.38** | -0.29 | -0.09 | -25.97** | 0.17 | -0.13 | 37.83** | 0.12 | -0.23 | 32.46** |
| 8-16 Days | -0.02 | -0.04 | 2.08* | 0.02 | -0.31 | 18.85** | -0.06 | -0.14 | 9.44** | 0.21 | -0.15 | 32.23** |
| 16-32 Days | 0.15 | 0.04 | 11.88** | 0.37 | 0.21 | 14.87** | -0.05 | -0.05 | -0.92 | 0.34 | 0.19 | 16.29** |
| 32-64 Days | -0.74 | -0.63 | -14.41** | 0.35 | 0.65 | -15.17** | -0.10 | -0.05 | -1.78 | -0.68 | -0.77 | 10.02** |
|  | CAC40-DAX | | | CAC40-IBEX | | | CAC40-FTSE | | | CAC40-NASDAQ | | |
|  | Before | After | Test Statistic | Before | After | Test Statistic | Before | After | Test Statistic | Before | After | Test Statistic |
| 1-2 Days | 0.00 | 0.05 | -10.95** | 0.06 | -0.08 | 16.13** | 0.19 | 0.04 | 45.68** | -0.05 | -0.04 | -2.24* |
| 2-4 Days | -0.10 | -0.15 | 11.23** | 0.10 | -0.08 | 25.29** | 0.23 | -0.05 | 76.50** | -0.03 | 0.09 | -21.51** |
| 4-8 Days | -0.11 | 0.00 | -13.43** | 0.17 | 0.11 | 8.29** | 0.03 | -0.10 | 18.96** | 0.11 | -0.25 | 33.61** |
| 8-16 Days | -0.56 | -0.68 | 32.75** | -0.04 | 0.15 | -18.94** | 0.10 | -0.02 | 7.26** | 0.17 | -0.17 | 31.27** |
| 16-32 Days | 0.26 | 0.50 | -26.40** | 0.20 | 0.42 | -27.76** | -0.21 | -0.44 | 28.03** | 0.35 | 0.14 | 16.44** |
| 32-64 Days | 0.73 | 0.87 | -39.30** | -0.81 | -0.83 | 2.59** | 0.37 | 0.33 | 2.00* | -0.64 | -0.76 | 13.15** |

Note: ** denote significance at 1% critical value and * denote significance at 5% critical value

Table 3.17 T-test Results comparing wavelet correlation for A1

| Timescale | FTSE-NIKKEI Before | After | Test Statistic | FTSE-KOSPI Before | After | Test Statistic | FTSE-JKSE Before | After | Test Statistic | FTSE-KLSE Before | After | Test Statistic |
|---|---|---|---|---|---|---|---|---|---|---|---|---|
| 1-2 Days | -0.03 | -0.05 | 3.70** | 0.00 | 0.06 | -11.43** | 0.03 | 0.01 | 4.81** | -0.09 | 0.02 | -22.18** |
| 2-4 Days | -0.03 | 0.07 | -17.94** | -0.04 | 0.02 | -13.23** | -0.14 | 0.05 | -35.58** | -0.06 | -0.10 | 6.30** |
| 4-8 Days | 0.30 | 0.18 | 8.24** | -0.09 | -0.01 | -7.56** | 0.03 | 0.14 | -8.77** | -0.16 | 0.06 | -35.26** |
| 8-16 Days | -0.11 | -0.20 | 8.22** | -0.45 | -0.08 | -26.49** | 0.05 | -0.01 | 11.06** | 0.37 | 0.44 | -5.11** |
| 16-32 Days | -0.14 | -0.40 | 11.07** | -0.02 | 0.26 | -12.96** | 0.22 | 0.22 | 0.77 | -0.05 | -0.01 | -2.50** |
| 32-64 Days | 0.45 | -0.07 | 37.44** | 0.58 | 0.50 | 5.69** | 0.12 | 0.13 | -0.93 | 0.06 | 0.26 | -28.98** |

| Timescale | FTSE-STI Before | After | Test Statistic | FTSE-BSE Before | After | Test Statistic | FTSE-SMI Before | After | Test Statistic | FTSE-SNP Before | After | Test Statistic |
|---|---|---|---|---|---|---|---|---|---|---|---|---|
| 1-2 Days | -0.04 | 0.07 | -26.81** | -0.06 | -0.07 | 2.89** | 0.04 | -0.07 | 39.53** | 0.04 | -0.08 | 28.88** |
| 2-4 Days | -0.12 | -0.03 | -32.49** | 0.03 | 0.02 | 4.40** | 0.04 | 0.06 | -5.43** | 0.06 | -0.05 | 12.29** |
| 4-8 Days | 0.07 | -0.02 | 11.29** | -0.10 | 0.13 | -44.39** | -0.01 | 0.11 | -8.33** | -0.07 | -0.05 | -2.74** |
| 8-16 Days | -0.40 | -0.10 | -40.62** | 0.40 | 0.27 | 13.59** | 0.07 | -0.06 | 8.22** | -0.07 | -0.11 | 4.69** |
| 16-32 Days | 0.23 | -0.16 | 38.05** | 0.58 | 0.54 | 7.19** | 0.51 | 0.23 | 21.39** | -0.42 | -0.06 | -20.59** |
| 32-64 Days | -0.26 | -0.15 | -5.86** | 0.31 | 0.22 | 4.79** | -0.06 | 0.17 | -19.19** | 0.06 | 0.10 | -1.76 |

| Timescale | FTSE-DAX Before | After | Test Statistic | FTSE-IBEX Before | After | Test Statistic | FTSE-CAC40 Before | After | Test Statistic | FTSE-NASDAQ Before | After | Test Statistic |
|---|---|---|---|---|---|---|---|---|---|---|---|---|
| 1-2 Days | 0.04 | 0.05 | -1.51 | -0.04 | -0.02 | -4.12** | 0.19 | 0.04 | 45.68** | 0.02 | -0.07 | 22.80** |
| 2-4 Days | -0.05 | 0.13 | -32.29** | 0.08 | 0.00 | 11.36** | 0.23 | -0.05 | 76.50** | 0.04 | -0.08 | 13.42** |
| 4-8 Days | -0.14 | 0.06 | -31.20** | -0.04 | 0.10 | -24.37** | 0.03 | -0.10 | 18.96** | -0.09 | -0.08 | -0.92 |
| 8-16 Days | -0.06 | -0.05 | -0.30 | -0.17 | -0.07 | -6.14** | 0.10 | -0.02 | 7.26** | -0.09 | -0.17 | 8.53** |
| 16-32 Days | -0.36 | -0.34 | -1.49 | -0.47 | -0.25 | -15.04** | -0.21 | -0.44 | 28.03** | -0.44 | -0.11 | -18.18** |
| 32-64 Days | 0.29 | 0.18 | 6.39** | -0.02 | -0.05 | 1.14 | 0.37 | 0.33 | 2.00* | 0.01 | -0.01 | 1.14 |

Note: ** denote significance at 1% critical value and * denote significance at 5% critical value

# Chapter 4

## Long Memory among Global Equity Markets

### 4.1 Introduction

The estimation and the analysis of long memory parameters have mainly focused on the analysis of long-range dependence in stock return volatility using traditional time and spectral domain estimators of long memory. The definitive ubiquity and existence of long memory in the volatility, estimated or generated using various methods, of stock returns is an established stylized fact. The presence of long memory requires major revisions in the standard estimation procedures without which the estimated results can be seriously biased. In this chapter on long memory among global equity markets, several wavelet based estimators are applied to test for the presence of long memory in the global equity returns and returns volatility. The presence of long memory in the volatility of the stock returns as well as some returns themselves is demonstrated from the empirical evidences. Furthermore, phases of efficiency and inefficiency of markets, as adjudicated by presence of both long memory and no-memory, is evidenced when the analysis is performed using rolling windows. The existence or absence of long memory in stock returns can be used to determine the stage of market development in terms of efficiency and inefficiency. According to the weak-form version of the *Efficient Market Hypothesis* (EMH), equity prices contain all available information about the equity price, acquired from past trading. This suggests that prediction of prices, when the EMH hold, is not possible. On the other hand, the presence of long-memory in equity returns and volatility implies that distant observations in the equity returns and volatility series are related to each other. This implication leads to the rejection of efficient markets as the presence of long range dependence is incompatible with the basic tenets of efficient market hypothesis (EMH).

The analysis of long memory is further extended to estimate long-run correlation matrix of global equity returns using wavelet based multivariate long memory estimator. The estimates generated from multivariate long memory model allows one to detect mechanisms that give rise to long memory. Long memory among several groups of equity markets either be the result of some same underlying process generating the data or it can be a product of multiple mechanisms (Wendt et al., 2009). The long-run correlation matrix, also known as the fractal connectivity matrix, generated from the multivariate long-memory model helps in determining the convergence of wavelet correlations of long-range dependent processes. The convergence to an asymptotic value over a range of low-frequency wavelet scales helps one in determining regimes of fractal connectivity (Achard et al., 2008; Achard and Gannaz, 2016). In doing so, associations and similarities between the processes that generate equity market returns of various markets can be highlighted. Furthermore, a hierarchical clustering algorithm is implemented on the elements of the generated fractal connectivity matrix to group markets having similar long-run correlation behavior. Significant rise in long-run correlations is evidenced during the subprime crisis period. However, long-run correlations among all equity markets are very low[11]. Nonetheless, comparisons can be drawn with regard to the long-memory behavior of global equity markets during both normal and crisis-hit periods. In this chapter, the issue of multifractality of equity returns is also highlighted via the implementation of a rolling window long memory procedure. The resulting estimates of long memory parameters, with varying degrees of fractal structures, are not always stable and fluctuate between regimes of efficiency and inefficiency. This implies that markets are not always efficient in the weak sense and arbitrage opportunities exists. The pattern of evolution of long memory parameter, as verified from the time-series of Hurst exponents, is in agreement with the adaptive markets hypothesis which allow interplay of "*[...] complex market dynamics*, *with cycles as well as trends, and panics, manias, bubbles, crashes, and other phenomena that are routinely witnessed in natural market ecologies.*" (Lo, 2004, p. 24). In the next section some relevant works related to long memory behaviour of global equity markets are reviewed.

## 4.2 Literature Review

Since the groundbreaking work of Hurst (1951), where he investigated the flow of river Nile and found evidence of long range dependence, there has been significant interest, spanning researchers across disciplines, in the phenomena of long memory. Mandelbrot and Van Ness (1968), using the idea of Hurst exponent, employed the idea of long-memory processes in conjunction with fractional Brownian motion and related stochastic processes. However, in the field of time series analysis, Granger and Joyeux (1980) and Hosking (1981) were among the first to integrate long memory processes with time series methods. Since then, a plethora of time-series based models of long memory has been developed to analyze long-range dependence in stochastic processes. However, a majority of research articles that focuses on the estimation of long memory parameters and detection of the same relies on the traditional rescaled range (R/S) approaches of Mandelbrot (1965) and its modified version developed by Lo (1991). The spectral domain approach proposed by Geweke and Porter-Hudak (1983) to estimate the long memory parameter has been used by many researchers too. This section reviews some important works on long memory concerning the analysis of global equity markets. Numerous studies have been carried out to test the presence of long run dependence in stock markets. Works related to the estimation and analysis of long memory parameters have mainly focused on the analysis of long-range dependence in stock return volatility. The concurrent use of squared returns and absolute returns as a measure of volatility is very evident from the literature that focuses on the analysis of stock returns volatility (see for eg. Ding et al., 1993; Granger and Ding, 1995; Lobato and Velasco, 2000). Studies which analyzes the long memory parameters and confirms the existence of long memory in stock returns volatility are abundant. However, since the prime focus of this chapter will be in investigating long memory in global equity returns, importance will be given to studies analyzing long memory in equity returns instead of returns volatility.

---

[11] This is not to be confused with the regular wavelet correlation where correlation tend to be strong in the long-run. Correlations based on fractal connectivity are used to determine the similarity in mechanisms that generate the underlying long memory behaviour among markets.

There is no clear consensus, among studies that attempt to detect long memory in financial data, on the existence of the phenomenon of long-range dependence. A plethora of studies that support the presence of long memory in financial time series is documented in the literature with commensurate number of articles rejecting the presence of long memory. The presence of long memory in squared daily returns of S&P 500 index is evident in the works of Ding et al. (1993) where significant autocorrelation for lags up to ten years were present. Similarly, Lobato and Savin (1998) also demonstrated the presence of long memory in the squared returns of the S&P 500 dataset spanning three decades.

Ray and Tsay (2000) unearthed the presence of strong *long-range dependence* in the volatilities of selected companies of the S&P 500 index. Granger and Ding (1995) also detected the presence of long memory in the absolute value of stock returns. Furthermore, Lobato and Velasco (2000) using a frequency domain tapering procedure in a multi stage semi-parametric method unearthed the presence of long memory in stock returns and volatility of returns. The presence of long memory in the returns of Brazilian equity market is documented in Assaf and Cavalcante (2005).

Barkoulas et al. (2000), while investigating the long memory properties of the Athens stock exchange, find evidence of long-range persistence in the returns of the Athens stock market. Moreover, the forecast performance of a long memory incorporated model significantly outdid forecasts generated from a regular random walk model.

Similarly, Panas (2001), using a spectral measure of fractality along with the Levy index, found nonlinearities in Greek equity returns and unearthed the existence of long memory, thereby rejecting the weak-form efficiency of the Greek equity market. Henry (2002), using a mixture of semi-parametric and spectral estimators, found evidence of long memory in the returns of South Korean stock market. Moreover, some evidence of weak long memory was unearthed in the markets of Germany and Taiwan.

In their analysis of the EMH, Jagric et al. (2005) employed a wavelet method to test for long memory in the returns of some select European markets. The empirical investigations documented the presence of long-range dependence in four emerging eastern European markets, thereby rejecting evidence in favour of the efficient market hypothesis. Similar analysis using wavelet based methods to detect long memory in the returns of the Dow Jones Industrial average (DJI) were employed by Elder and Serletis (2007) where no evidence of long memory was detected, thereby supporting results from a vast number of studies that reject the presence of long memory in the developed markets of the U.S. However, the presence of long memory in the equity returns of some developed markets of Europe, the U.S., and Japan is documented in Ozdemir (2007). Furthermore, Ozun and Cifter (2007), also using a wavelet based estimator of long memory, found some evidence of long-range dependence in the returns of the Istanbul Stock Index, thereby rejecting the weak form efficiency of Istanbul share prices. Similarly, evidence of long memory in the equity markets of G7 countries is documented in Bilal and Nadhem (2009). On the other hand, Mariani et al (2010), using detrended fluctuation analysis and truncated Levy flight method, found evidence of long memory in several eastern European markets. However, among the countries that are part of the Organisation for Economic Co-operation and Development (OECD), long-memory, as investigated by Tolvi (2003), was only evidenced in the smaller equity markets of Denmark and Finland.

Jefferis and Thupayagale (2008), using a long-memory variant of the GARCH model, investigated long memory behaviour of some select African equity markets and found evidence supporting the presence of long-memory in the developing markets of Botswana and Zimbabwe. The presence of long memory in the developing markets of Central and Eastern European countries (CEE) is documented in the studies of Jagric et al. (2006) and Kasman et al. (2009), where the presence of long memory in equity returns is specifically limited to the developing markets of Hungary, Czech, Slovenia and Croatia.

Kristoufek and Vosvrda (2012) constructed a measure of efficiency by measuring the distance between an efficient case and a vector containing long memory and other measures of fractality. Long memory is evidenced in many developing and emerging markets whereas all developed markets show signs of efficiency, with the Japanese NIKKIEI leading all other developed markets in terms of efficiency.

Cont (2005) attempted to identify economic intuition and mechanisms behind the existence of fractality and long memory in returns and returns volatility. The possible economic factors underlying the existence of long memory in volatility are, i) heterogeneous investment horizons of market agents, ii) evolutionary trading models that employ genetic algorithms, iii) market fluctuations arising out of investors' sudden switch between several trading strategies, and iv) the inactivity of investors, operating at certain time periods and market regimes, based on trading strategies or behavioral aspects.

The presence of heterogeneous investment horizons can be one of the most important factors that generate long memory behavior in equity markets. Investment horizons, which can be successfully disaggregated into several microunits using wavelet methods to delineate price behaviors from varying time-horizons, contain varying returns and volatility structures. The aggregation of all microunits, ranging from very short run to long run, is said to produce long memory properties in the aggregate series (see Granger, 1980; Davidson and Sibbertsen, 2005). However, contemporaneous aggregation of microunits having both short-memory and long-memory can lead to spurious long-memory in the aggregate, thereby biasing the results in favour of long-range dependence. Granger and Ding (1996) attempt to theoretically explain this bias arising out of aggregation but however fail to empirically demonstrate that long memory in returns volatility of stock indices is due to aggregating volatility of individual stocks containing short-memory. Furthermore, Andersen and Bollerslev (1997) theoretically demonstrated volatility to be an assortment of various heterogeneous information structures in the short-run and concluded that the underlying volatility processes contain long memory. Nonetheless, in some major studies, estimates of long-memory are found to be uncontaminated by aggregation effects thereby supporting evidence in favour of fractality in equity returns (see Han, 2005; Souza, 2007; Kang et al., 2010), thereby rejecting any indication of spurious long memory.

Studies examining long memory in financial time series are relatively few. Jensen (1999) estimated the "long memory parameter of a fractionally integrated process" using a wavelet based OLS method. By projecting volatility in the time-frequency domain, Jensen and Whitcher (2000) demonstrated the effectiveness of the wavelet based estimator in capturing nonstationary long-memory behavior. Vuorenmaa (2005) investigated the time-varying long memory of Nokia Oyj returns using the wavelet OLS method and found significantly strong long memory during the dot-com bubble period. Ozun and Ciftr (2007), demonstrating the superiority of wavelet OLS method as compared to the spectral long memory estimator of Geweke and Porter-Hudak (1983), found significant long memory in the returns of Istanbul stock exchange. Similarly, DiSario et al. (2008), on investigating the volatility structure of S&P 500 returns using the wavelet OLS method, found evidence of long-memory in the S&P 500 returns volatility.

In the same vein as the aforementioned studies, Tan et al. (2012) while examining the fractal structure of emerging economies using wavelet OLS method demonstrated significant long memory in the returns of larger firms as compared to smaller firms. Likewise, Tan et al. (2014), using the wavelet estimator of Jensen (1999) and detrended fluctuation analysis, examined long memory behavior of equity returns and volatility of ten markets from both developing and developed economies. On the other hand, Power and Turvey (2010) investigated long memory structure of fourteen commodity futures using the Hurst estimator of Veitch and Abry (1999) and demonstrated long-range dependence in all commodities. The presence of long-memory in the equity markets of both developing and developed economies was demonstrated. Boubaker and Peguin-Feissolle (2013) proposed semiparametric wavelet base long memory estimators and demonstrated its superiority, with respect to several non-wavelet estimators, using simulation experiments. More recently, Pascoal and Monteiro (2014), investigating the predictability of the Portuguese stock returns using wavelet estimators of long memory, fractal dimension and the Holder exponent, found no evidence of long memory in the PSI20 returns, thereby confirming the efficiency of the Portuguese equity market.

This chapter investigates long memory among global equity markets using estimators from the wavelet domain. Studies investigating long memory in global financial markets based on wavelet based long memory methods are relatively fewer as compared to traditional time and spectral domain estimators of long memory. Furthermore, empirical studies based on wavelet domain estimators of long-range dependence are practically nonexistent in the case of Indian equity markets.

This chapter, however, implements the wavelet based approaches of Abry and Veitch (1998) and Abry et al. (2003) to examine the Hurst exponents, and its time-varying structure, of global equity markets. Moreover, an analysis of multivariate long memory of global equity markets using the recent method of Achard and Gannaz (2016) is carried out, which possibly is the first application of wavelet domain multivariate long memory technique in finance and economics. The aforementioned multivariate method allows one to analyze the long-run correlation among several markets exhibiting fractal structures.

## 4.3 Methodology

In this chapter, wavelet based measures of long memory parameters are applied to analyze long memory behavior of global equity returns. There are several classes of wavelet based long memory estimators that can measure long-term correlations present in a time series. The wavelet based Hurst estimator of Abry and Veitch (1998) is used in a rolling window algorithm to analyze the time varying structure of the Hurst parameter and the evolution of Hurst parameter and long range dependence over time.

The long range dependent phenomenon is associated with a slow power law decay of the autocorrelation function of a stationary process $x$. The covariance function $\gamma_x(k)$ of the long memory process $x$ takes the following form,

$$\gamma_x(k) \square \quad ^{-H}, \quad k \to +\infty \tag{4.1}$$

where $c_\gamma$ is a positive constant and $H \in (0, 0.5)$. The Hurst parameter $H$ is used to measure the presence of long memory. The spectrum $\Gamma_x(\nu)$ of the long memory process $x$ is given by,

$$\Gamma_x(\nu) \square \quad ^{H}, \quad \nu \to 0 \tag{4.2}$$

where $\nu$ is the frequency, $c_f = \pi^{-1} c_\gamma \Lambda(2H-1)\sin(\pi - \pi H)$, and the Gamma function is given by $\Lambda$. This mathematical structure of long memory processes is the reason for its inclusion in a class of stochastic processes which have the $1/|\nu|^\alpha$ form. The property of long memory also finds some close association with the phenomenon of scale invariance, self-similarity and fractals. Hence, statistically self-similar processes like fractional Brownian motion (FBM) is closely related to long memory phenomenon.

Let $\gamma_0$ be an arbitrary reference frequency selected by the choice of $\psi_0$, the mother wavelet. The amount of energy in the signal during scaled time $2^j k$ and scaled frequency $2^{-j}\nu_0$ is measured by the squared absolute value of the detail wavelet coefficient $|d_x(j,k)|^2$. A wavelet based spectral estimator of Abry et al. (1993) is constructed by taking a time average of $|d_x(j,k)|^2$ at a given scale, and is given by,

$$\hat{\Gamma}_x(2^{-j}\nu_0) = \frac{1}{n_j}\sum_k |d_x(j,k)|^2 \tag{4.3}$$

where $n_j$ is the "number of wavelet coefficients" at level $j$, and $n_j = 2^{-j} n$, where $n$ is the data length. Therefore, $\hat{\Gamma}_x(\nu)$ captures the amount of energy that lies within a given bandwidth and around some frequency $\nu$. Hence, $\hat{\Gamma}_x(\nu)$ can be regarded as an estimator for the spectrum $\Gamma_x(\nu)$ of $x$.

The wavelet based estimator of the Hurst exponent $\hat{H}$ is designed by performing a simple linear regression of $\log_2(\hat{\Gamma}_x(2^{-j}\nu_0))$ on $j$, i.e.,

$$\log_2(\hat{\Gamma}_x(2^{-j}\nu_0)) = \log_2\left(\frac{1}{n_j}\sum_k |d_x(j,k)|^2\right) = (2\hat{H}-1)j + \hat{c} \tag{4.3}$$

where $\hat{c}$ estimates $\log_2(c_f \int |\nu|^{(1-2H)} |\Psi_0(\nu)|^2 d\nu)$, where $\Psi_0$ is the Fourier transform of the mother wavelet $\psi_0$. A weighted least square estimator is constructed by performing a WLS fit between the wavelet scales $j_1$ and $j_2$ which gives the estimator of the "*Hurst exponent*", H.

$$\hat{H}(j_1, j_2) \equiv \frac{1}{2} \left[ \frac{\sum_{j=j_1}^{j_2} S_j j n_j - \sum_{j=j_1}^{j_2} S_j j \sum_{j=j_1}^{j_2} S_j \eta_j}{\sum_{j=j_1}^{j_2} S_j \sum_{j=j_1}^{j_2} S_j j^2 - \left(\sum_{j=j_1}^{j_2} S_j j\right)^2} + 1 \right] \quad (4.4)$$

where $\eta_j = \log_2 \left( \frac{1}{n_j} \sum_k |d_x(j,k)|^2 \right)$ and the weight $S_j = (n \ln^2 2)/2^{j+1}$ is the inverse of the theoretical asymptotic variance of $\eta_j$. The estimators of multivariate long memory and the related "*fractal connectivity matrix*", based on the above univariate estimator is given in Achard et al. (2008) and Achard and Gannaz (2016).

## 4.4 Empirical Results

The presence of long memory in the volatility of select equity returns, as given by the absolute value of equity returns, is investigated by applying the wavelet based estimator of the Hurst exponent developed by Abry and Veitch (1998) and Abry et al. (2003). The estimates of the Hurst parameter *H,* obtained from a wavelet based estimator, mostly lie within the range of (0.5, 1) for all stock indices taken into consideration, signifying the presence of "long-range dependence" in volatility of all of the studied volatility indices (see Table 4.1). The presence of long-memory in all of the studied stock indices implies that distant observations in each of the volatility series are related to each other. This result is in confirmation with the *stylized fact* of volatility i.e. volatility of financial returns contain long memory.

However, the analysis of long memory in this chapter largely focuses on the detection of long memory in equity returns instead of equity returns volatility. The results obtained from the long memory analysis of global equity returns is given in Table 4.2. The estimates of Hurst parameters of the selected equity returns reveals the existence of very weak long memory in the returns of KLSE (Malaysia), TAIEX (Taiwan), Pakistan (KSE100), China (SSE), Indonesia (JKSE) and Austria (ATX).

Table 4.1 Hurst estimates of Equity returns volatility

| Indices | Hurst | Std.Err | t-value | Pr(>\|t\|) |
|---|---|---|---|---|
| SENSEX | 0.782 | 0.0333 | 23.4798 | 0 |
| FTSE | 0.8713 | 0.0482 | 18.0794 | 0 |
| SNP | 0.9542 | 0.0783 | 12.1947 | 0.0001 |
| CAC40 | 0.8932 | 0.0729 | 12.2522 | 0.0001 |
| DAX | 0.9134 | 0.0711 | 12.8511 | 0.0001 |
| DJIA | 0.9406 | 0.0746 | 12.6135 | 0.0001 |
| NASDAQ | 0.914 | 0.0717 | 12.7389 | 0.0001 |
| NIKKEI | 0.7963 | 0.0454 | 17.5312 | 0 |
| KOSPI | 0.7457 | 0.0385 | 19.3547 | 0 |
| JKSE | 0.715 | 0.0368 | 19.4176 | 0 |
| KLSE | 0.6441 | 0.0889 | 7.2436 | 0.0008 |
| TAIWAN | 0.7508 | 0.0479 | 15.6628 | 0 |
| SSE | 0.7379 | 0.0258 | 28.5976 | 0 |
| STI | 0.8147 | 0.0215 | 37.9186 | 0 |
| STOXX50 | 0.903 | 0.0705 | 12.7997 | 0.0001 |

Furthermore, equity returns of the U.S. and majority of developed European markets contain no long-memory. This can be due to the efficient nature of these markets indicating the level of equity market development. However, in the next step, the improved Hurst estimator of Abry et al. (2003) is used to analyze long memory in select global equity returns. This enables one to graphically analyze long memory from the log-log plot of the wavelet regression which contains additional information about the fractal nature of equity returns.

*4.4.1 Wavelet LRD analysis using logscale diagram*

The generated *logscale* diagram is a plot of wavelet variance at each scale against the wavelet scale. Formally, the plot of the logarithm of $v_j = \frac{1}{n_j}\sum_{k=1}^{n_j}|d_X(j,k)|^2$ against the wavelet scale *j* gives the *logscale* diagram. Here $n_j$ is the "number of wavelet coefficients" at scale *j* and $d_X(j,k)$ is the wavelet details of the process $X(t)$. The visualization of the *logscale* diagram can help one detect regions of long-range dependence via the help of an *alignment region* in the graph. The range of wavelet scales where log ( $v_j$ ) falls on a straight line is known as the *alignment region* (Abry et al. 2003) and perfect alignment, mostly at higher scales, normally constitutes long memory. In the logscale plot, perfect alignment requires the red straight line to cross (or touch) the vertical lines depicting the confidence band in an upward sloping manner.

Table 4.2 Hurst exponent of equity returns using Wavelet method of Abry and Veitch (1998)

| Indices | Hurst | Std.Err | t-value | Pr(>|t|) | Indices | Hurst | Std.Err | t-value | Pr(>|t|) |
|---|---|---|---|---|---|---|---|---|---|
| SENSEX | 0.516686 | 0.00899852 | 57.41893 | 0 | STI | 0.502436 | 0.029433 | 17.07027861 | 0 |
| FTSE100 | 0.460122 | 0.02227633 | 20.65521 | 0 | HSI | 0.485181 | 0.023468 | 20.67440063 | 0 |
| NASDAQ | 0.502911 | 0.03093422 | 16.25742 | 0 | BEL20 | 0.500705 | 0.029556 | 16.94074032 | 0 |
| DJIA | 0.410379 | 0.02117717 | 19.37838 | 0 | ATX | 0.538935 | 0.033613 | 16.03345271 | 0 |
| CAC40 | 0.465823 | 0.03026815 | 15.38987 | 0 | AEX | 0.521217 | 0.028657 | 18.18782209 | 0 |
| DAX | 0.502207 | 0.03777552 | 13.2945 | 0 | IBEX | 0.426761 | 0.042622 | 10.01263228 | 0 |
| NIKKEI | 0.476575 | 0.03361141 | 14.17898 | 0 | SMI | 0.45866 | 0.024831 | 18.47155662 | 0 |
| KOSPI | 0.477189 | 0.04337997 | 11.00021 | 0 | STOXX50 | 0.485231 | 0.033351 | 14.54904112 | 0 |
| JKSE | 0.534115 | 0.02687113 | 19.87692 | 0 | KSE100 | 0.561315 | 0.029925 | 18.75763747 | 0 |
| KLSE | 0.53596 | 0.0327541 | 16.36316 | 0 | IBOV | 0.461728 | 0.029753 | 15.51856677 | 0 |
| TAIWAN | 0.556594 | 0.03011372 | 18.48305 | 0 | ISEQ | 0.449336 | 0.014646 | 30.68049376 | 0 |
| SSE | 0.539689 | 0.03039683 | 17.75476 | 0 | ASX200 | 0.497898 | 0.028383 | 17.54221723 | 0 |

If the alignment region includes the largest scales in the logscale plot, then the returns exhibit long-range dependence. Furthermore, the value of the self-similar parameter[12] $\alpha$ should lie in the interval (0, 1). Correspondingly, the value of the Hurst exponent $H$ should lie in the interval (0.5, 1) for the data to exhibit long memory. Figure 4.1 gives the logscale diagram[13] of the equity returns of select developed markets. It can be observed from the figure that straight line slopes downward and the corresponding Hurst exponents for all six developed markets of Europe and the U.S. lie within the interval (0, 0.5) indicating short-memory.

Figure 4.1 Logscale diagram of equity returns from developed markets

---

[12] $\alpha$ is also known as the scaling exponent of self-similarity. The Hurst parameter $H$ and $\alpha$ are related by the expression: $H=(1+\alpha)/2$

[13] After repeated simulations, the optimal lower cut-off scale is taken to be 2 and the highest scale is taken to be 8.

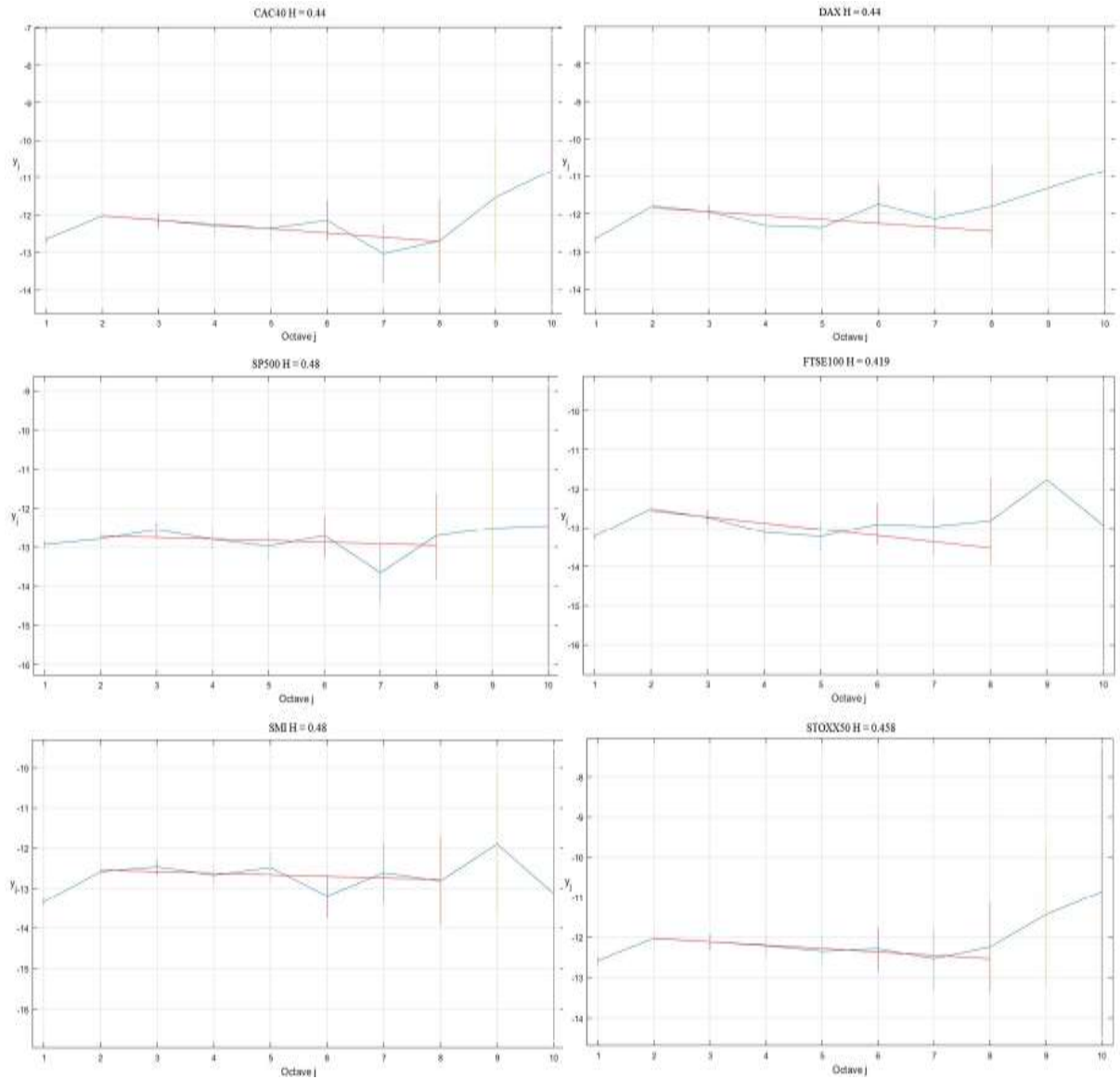

The absence of long memory in the returns of developed markets is in confirmation with results from a vast majority of literature that rejects long memory in developed financial markets. Figure 4.2 gives the logscale diagram of the equity returns of some select emerging markets. It can be noticed that the Hurst exponents of emerging markets' equity returns lie within the interval (0.5, 1).

Figure 4.2 Logscale diagram of equity returns from emerging markets

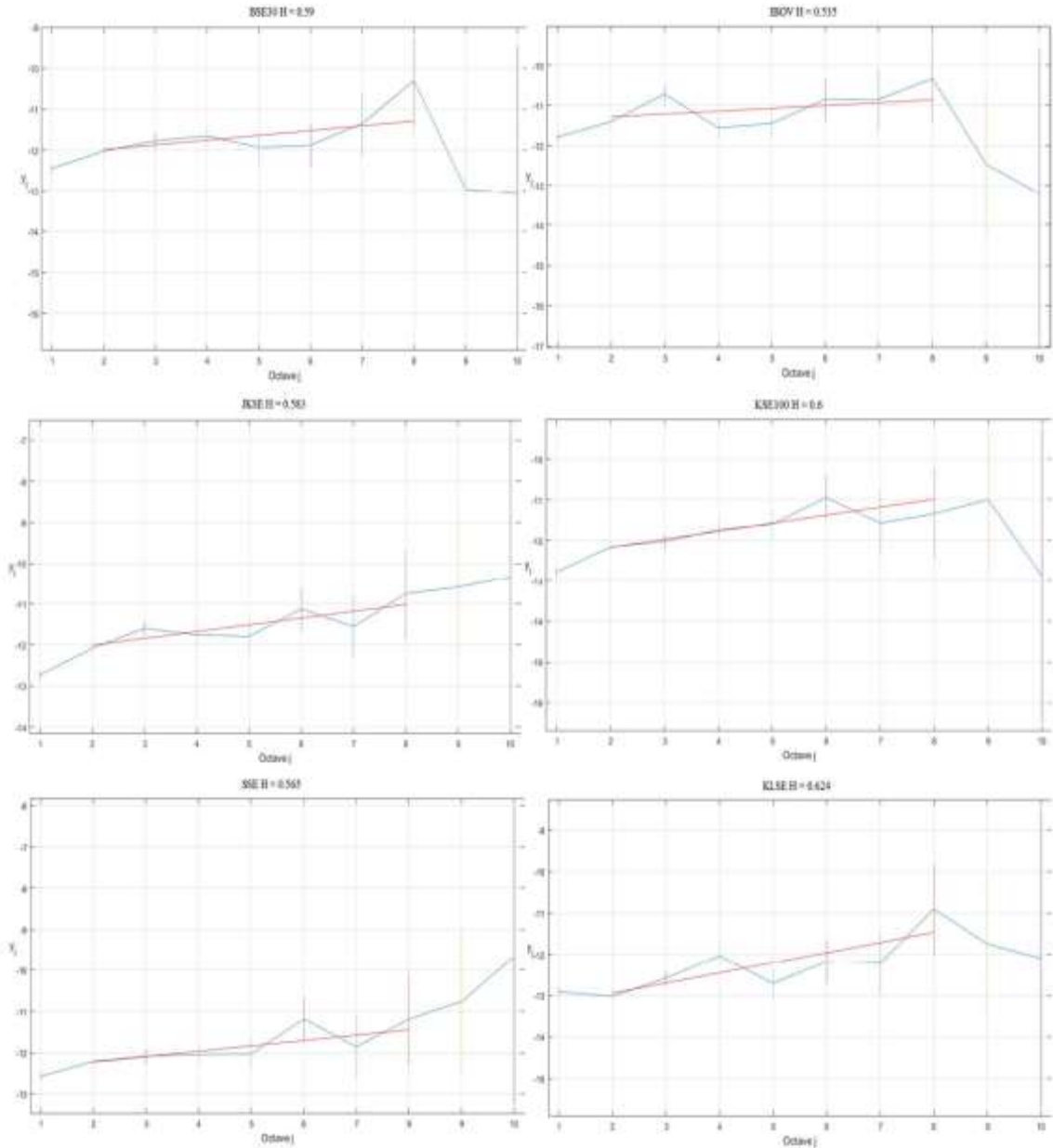

It can be observed from the above figure that the upward sloping alignment of the straight red line includes all higher scales, i.e. scales five up to eight, indicating the presence of "long-range dependence". However, among the six emerging markets, equity returns of India (BSE 30), Pakistan (KSE 30) and Malaysia (KLSE) exhibit relatively stronger long-memory. The corresponding values of several measures of fractality, including self-similarity parameter, Hurst exponent, Holder exponent and fractal dimension, are reported in Table 4.3. However, analyses based on this chapter particularly focuses on the scaling parameter of LRD i.e. the Hurst exponent.

Table 4.3 Scaling parameters of select equity returns

| BSE30 | Scaling parameters are: | alpha (LRD) | H (LRD rewrite) | H=h (ss,Holder) | D (frac dim, if alpha in (1,3)) | cf |
|---|---|---|---|---|---|---|
| | Estimates: | 0.117 | 0.558 | -0.442 | 2.442 | 0.0002 |
| | CI's: | [0.042, 0.192] | [0.521, 0.596] | [-0.479, -0.404] | [2.404, 2.479] | [0.00017, 0.00025] |
| NIKKEI | Scaling parameters are: | alpha (LRD) | H (LRD rewrite) | H=h (ss,Holder) | D (frac dim, if alpha in (1,3)) | cf |
| | Estimates: | -0.012 | 0.494 | -0.506 | 2.506 | 0.0002 |
| | CI's: | [-0.087, 0.063] | [0.457, 0.532] | [-0.543, -0.468] | [2.468, 2.543] | [0.00018, 0.00026] |
| JKSE | Scaling parameters are: | alpha (LRD) | H (LRD rewrite) | H=h (ss,Holder) | D (frac dim, if alpha in (1,3)) | cf |
| | Estimates: | 0.166 | 0.583 | -0.417 | 2.417 | 0.0002 |
| | CI's: | [0.091, 0.240] | [0.545, 0.620] | [-0.455, -0.380] | [2.380, 2.455] | [0.00016, 0.00022] |
| KLSE | Scaling parameters are: | alpha (LRD) | H (LRD rewrite) | H=h (ss,Holder) | D (frac dim, if alpha in (1,3)) | cf |
| | Estimates: | 0.247 | 0.624 | -0.376 | 2.376 | 0.0001 |
| | CI's: | [0.172, 0.322] | [0.586, 0.661] | [-0.414, -0.339] | [2.339, 2.414] | [0.00007, 0.00010] |
| HSI | Scaling parameters are: | alpha (LRD) | H (LRD rewrite) | H=h (ss,Holder) | D (frac dim, if alpha in (1,3)) | cf |
| | Estimates: | 0.107 | 0.554 | -0.446 | 2.446 | 0.0002 |
| | CI's: | [0.032, 0.182] | [0.516, 0.591] | [-0.484, -0.409] | [2.409, 2.484] | [0.00017, 0.00024] |
| SSE | Scaling parameters are: | alpha (LRD) | H (LRD rewrite) | H=h (ss,Holder) | D (frac dim, if alpha in (1,3)) | cf |
| | Estimates: | 0.13 | 0.565 | -0.435 | 2.435 | 0.0002 |
| | CI's: | [0.055, 0.205] | [0.528, 0.603] | [-0.472, -0.397] | [2.397, 2.472] | [0.00014, 0.00020] |
| SP500 | Scaling parameters are: | alpha (LRD) | H (LRD rewrite) | H=h (ss,Holder) | D (frac dim, if alpha in (1,3)) | cf |
| | Estimates: | -0.04 | 0.48 | -0.52 | 2.52 | 0.0002 |
| | CI's: | [-0.115, 0.035] | [0.443, 0.517] | [-0.557, -0.48] | [2.483, 2.557] | [0.00013, 0.00019] |
| ATX | Scaling parameters are: | alpha (LRD) | H (LRD rewrite) | H=h (ss,Holder) | D (frac dim, if alpha in (1,3)) | cf |
| | Estimates: | 0.008 | 0.504 | -0.496 | 2.496 | 0.0002 |
| | CI's: | [-0.067, 0.083] | [0.466, 0.541] | [-0.534, -0.45] | [2.459, 2.534] | [0.00018, 0.00026] |
| FTSE100 | Scaling parameters are: | alpha (LRD) | H (LRD rewrite) | H=h (ss,Holder) | D (frac dim, if alpha in (1,3)) | cf |
| | Estimates: | -0.162 | 0.419 | -0.581 | 2.581 | 0.0002 |
| | CI's: | [-0.237, -0.087] | [0.381, 0.456] | [-0.619, -0.544] | [2.544, 2.619] | [0.00017, 0.00025] |
| DAX | Scaling parameters are: | alpha (LRD) | H (LRD rewrite) | H=h (ss,Holder) | D (frac dim, if alpha in (1,3)) | cf |
| | Estimates: | -0.102 | 0.449 | -0.551 | 2.551 | 0.0003 |
| | CI's: | [-0.177, -0.027] | [0.411, 0.486] | [-0.589, -0.514] | [2.514, 2.589] | [0.00026, 0.00038] |
| CAC40 | Scaling parameters are: | alpha (LRD) | H (LRD rewrite) | H=h (ss,Holder) | D (frac dim, if alpha in (1,3)) | cf |
| | Estimates: | -0.113 | 0.443 | -0.557 | 2.557 | 0.0003 |
| | CI's: | [-0.188, -0.038] | [0.406, 0.481] | [-0.594, -0.519] | [2.519, 2.594] | [0.00023, 0.00034] |
| IBEX | Scaling parameters are: | alpha (LRD) | H (LRD rewrite) | H=h (ss,Holder) | D (frac dim, if alpha in (1,3)) | cf |
| | Estimates: | 0.013 | 0.506 | -0.494 | 2.494 | 0.0002 |
| | CI's: | [-0.062, 0.088] | [0.469, 0.544] | [-0.531, -0.456] | [2.456, 2.531] | [0.00018, 0.00026] |

### 4.4.2 Rolling windows Hurst analysis of time varying market efficiency

Nevertheless, long memory in equity returns can vary with due to the change in the efficiency of equity markets over time. The advancement of equity markets, coupled with varying phases of market development, policy decisions and financial turbulence, can significantly alter long memory structure of financial markets. Therefore, the estimates of long memory parameters are not always stable for all markets. In view of the changing structures and efficiency of equity markets, the next step in the analysis of long memory behavior of equity returns constitutes an analysis of time-varying long memory behavior of equity returns. Consequently, the Hurst exponents of select equity returns are estimated in a rolling window framework. The length of the window contains 260 observations which approximately is equal to one year. The window is moved forward by an increment of twenty four day i.e. a one month increment. Finally, the estimation of wavelet based Hurst exponent in rolling window framework generates a time-series of Hurst exponent.

The plots of the time-series of Hurst exponents of select equity returns can be visually inspected to determine the phases of efficiency and inefficiency, as measured by the drift of the Hurst exponent from the threshold value of 0.5, given by the horizontal line around the Hurst value of 0.5 in Figure 4.3-4.4 which plots the generated Hurst series against time given in the horizontal axis in years. The vertical axis in Figure 4.3 shows the Hurst values.

The plot of Hurst series given in Figures 4.3-4.4 reveal the time-varying nature of Hurst exponents. The developed equity markets of Europe show relatively less degree of persistence in returns with Hurst exponent below 0.5 for most of the time period. However, equity returns of France (CAC 40) and Germany (DAX) exhibited long-range dependence during the first three quarters of 2004, thereby allowing some possibility for returns predictability during that period. Nevertheless, equity returns of France and Germany has been relatively unpredictable throughout the studied time period. The same holds true for the Eurozone (STOXX 50) equity returns. On the other hand, the emerging markets equity returns seems to exhibit varying phases of return predictability with high values of Hurst exponent during some time intervals. For example, some indication of persistence in the returns of the Indian equity market (BSE 30) can be observed during the one year period of January 1999-January 2000 which is then followed by a sharp drop in Hurst exponent around February 2000, which can be attributed to market fluctuations arising out of the dot-com bubble. However, long-memory rises again after March 2000 extending up to January 2001 indicating some evidence of returns predictability during this period. Some evidence of returns predictability, as indicated by the presence of long-memory with Hurst value above 0.5, is also observed during Feb 2006-November 2006 and the last half of 2012. Moreover, the Asian markets of Hong Kong (HSI), China (SSE), Indonesia (KLSE) and Taiwan (TAIEX) exhibit evidence of returns predictability. Persistence in equity returns can be evidenced for, i) HSI during mid 2011-mid 2012, ii) SSE during mid 2005- early 2006, mid 2008-January 2009 and late 2010-late 2011, iii) KLSE during mid 1999-February 2000 and 2001-2002 and, iv) TAIEX during January 1999-mid 2000.

Figure 4.3 Time varying Hurst Estimates of select equity returns

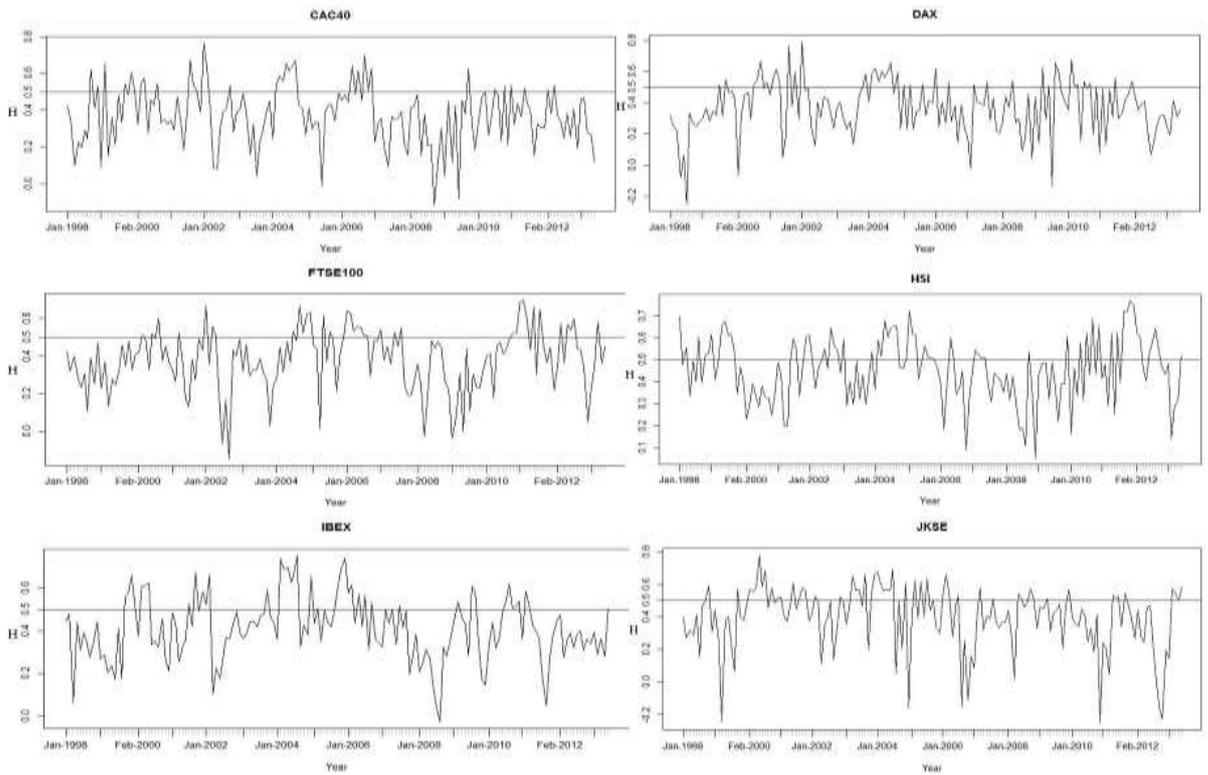

Figure 4.4 Time varying Hurst Estimates of select equity returns

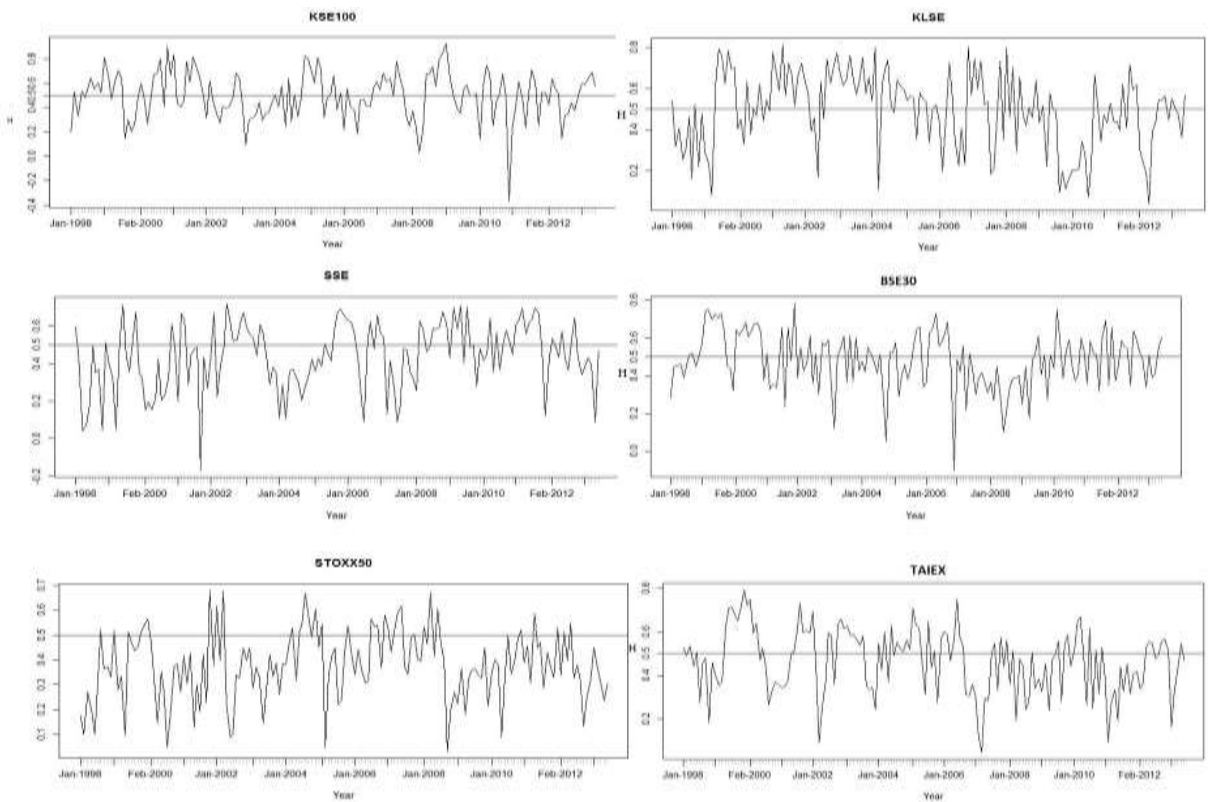

Interestingly, with the exception of equity returns of Pakistan (KSE 100) and China (SSE), returns markets from both developed and emerging economies exhibit anti-persistence (short-memory) during the financial crisis period of 2008, thereby eliminating any scope for returns predictability during this period. Moreover, barring periods of abrupt changes in Hurst parameter beyond and within the threshold range of 0.5, the phases of market efficiency are more pronounced for the developed equity markets where Hurst exponents of these markets' equity returns tend to lie below the threshold range of 0.5. However, efficiency of both developed and emerging equity markets is not stable throughout the studied time-period, allowing investors some arbitrage opportunities. Nonetheless, investors operating in emerging equity markets have more scope for arbitrage as these markets exhibit relatively more phases of long-range dependence.

*4.4.3 Long-range correlation among global equity returns*

In this section, an attempt is made to investigate the long-range correlation among global equity returns using the newly developed multivariate long-memory estimators of Achard and Gannaz (2016) which offer a more efficient way to estimate long-memory and evaluate correlation structure. The resulting long-run correlation matrix, estimated using the aforementioned multivariate method, aids in scrutinizing the correlation structure among equity returns operating at long-range frequencies. The long run correlation matrix, also known as the fractal connectivity matrix, furthermore assists in analyzing similarity of fractal structures among equity markets. The elements of long-range correlation matrix, of equity returns exhibiting LRD, is clustered using the hierarchical clustering algorithm to analyze the structure of equity returns correlations during both stable and turbulent financial phases, thereby assisting in identifying *fractally similar* market groups.

Figure 4.5 Fractal connectivity matrix of select equity returns

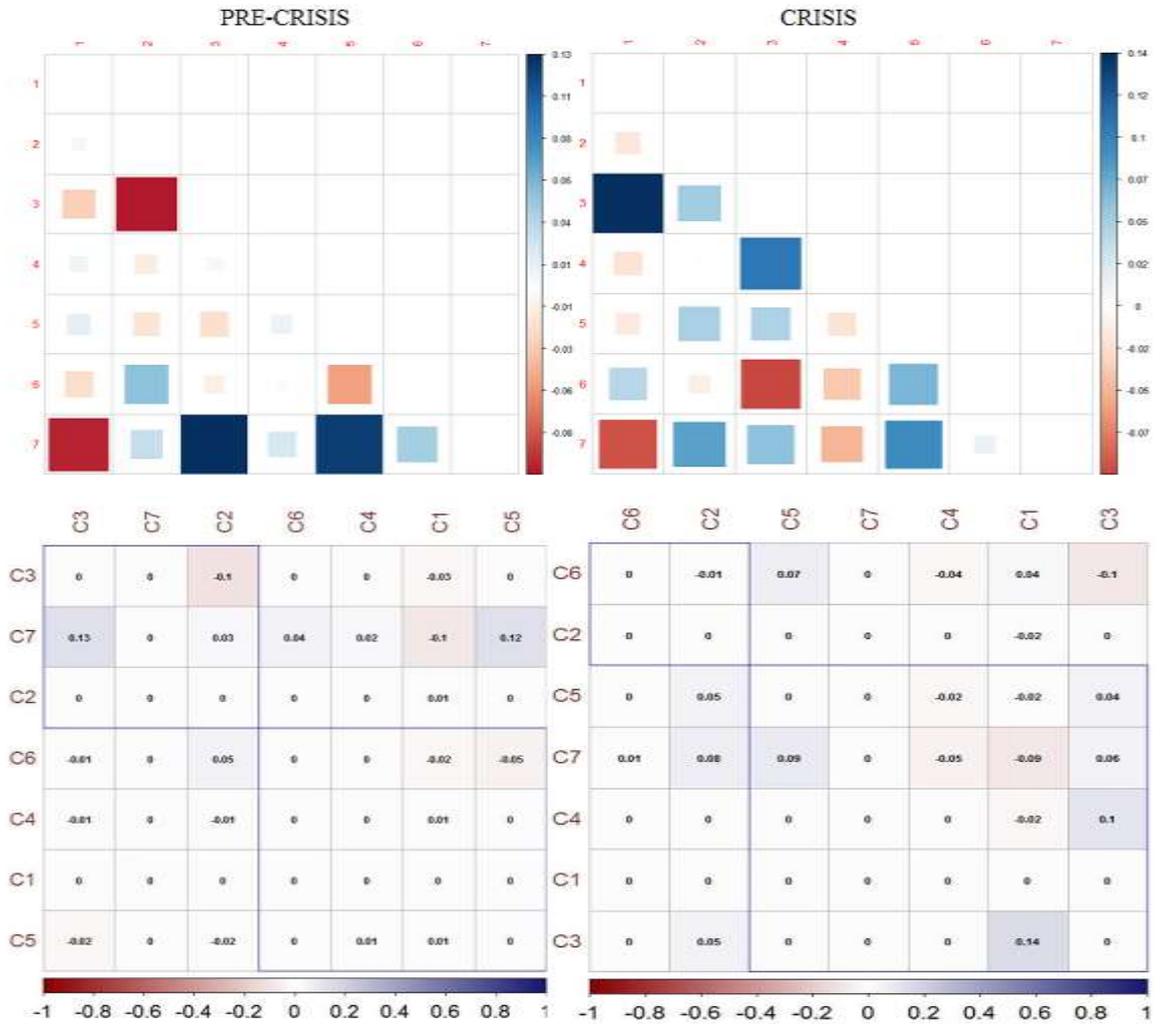

Figure 4.5 gives the long-run correlation matrix, displaying long-run correlation among seven equity markets of the U.S. (SP500), France (CAC40), Germany (DAX), Japan (NIKKEI), South Korea (KOSPI), Indonesia (JKSE) and India (BSE30). The upper panel of Figure 4.5 shows the fractal connectivity matrix whereas the lower panel gives the clustered version, using hierarchical clustering algorithm, of the long-run matrix of correlations. The left panel shows the correlation matrix of equity returns during the pre-subprime crisis period whereas the right panel gives the matrix of equity returns during the crisis period. The color coded legend, on the right side of fractal connectivity matrix and towards the bottom of the clustered matrix, helps in identifying the strength of long-range correlations. The strength of correlation rises as we move from red (low) to blue (high). The returns of seven aforementioned equity markets are labeled numerically from 1 to 7 in the upper panel and alphanumerically from C1 to C7 in the lower panel, where "C1, C2, C3, C4, C5, C6 and C7" correspond to SP500, CAC40, DAX, NIKKEI, KOSPI, JKSE and BSE30, respectively. It is evident from the long-run correlation matrix (upper panel) that long-range correlations significantly rises during the subprime crisis period, as indicated by larger number of elements in blue depicting positive correlations. The clustering of markets according to similar fractal structures is different during pre-crisis and crisis periods. Moreover, five markets (C1, C3, C4, C5 and C7) are clustered together during the subprime crisis period reflecting similar long memory behavior among these markets during crisis period. This is in line with the results from previous section where fractal structure of returns from SP500 (C1), DAX (C2), NIKKEI (C3), KOSPI (C4) and BSE30 (C7) behave similarly during the subprime period.

Figure 4.6 gives the long-run correlation matrix among seven Asian equity markets of South Korea (KOSPI), Malaysia (KLSE), Taiwan (TAIEX), China (SSE), Singapore (STI), Hong Kong (HSI) and India (BSE30).

Figure 4.6 Fractal connectivity matrix of select equity returns

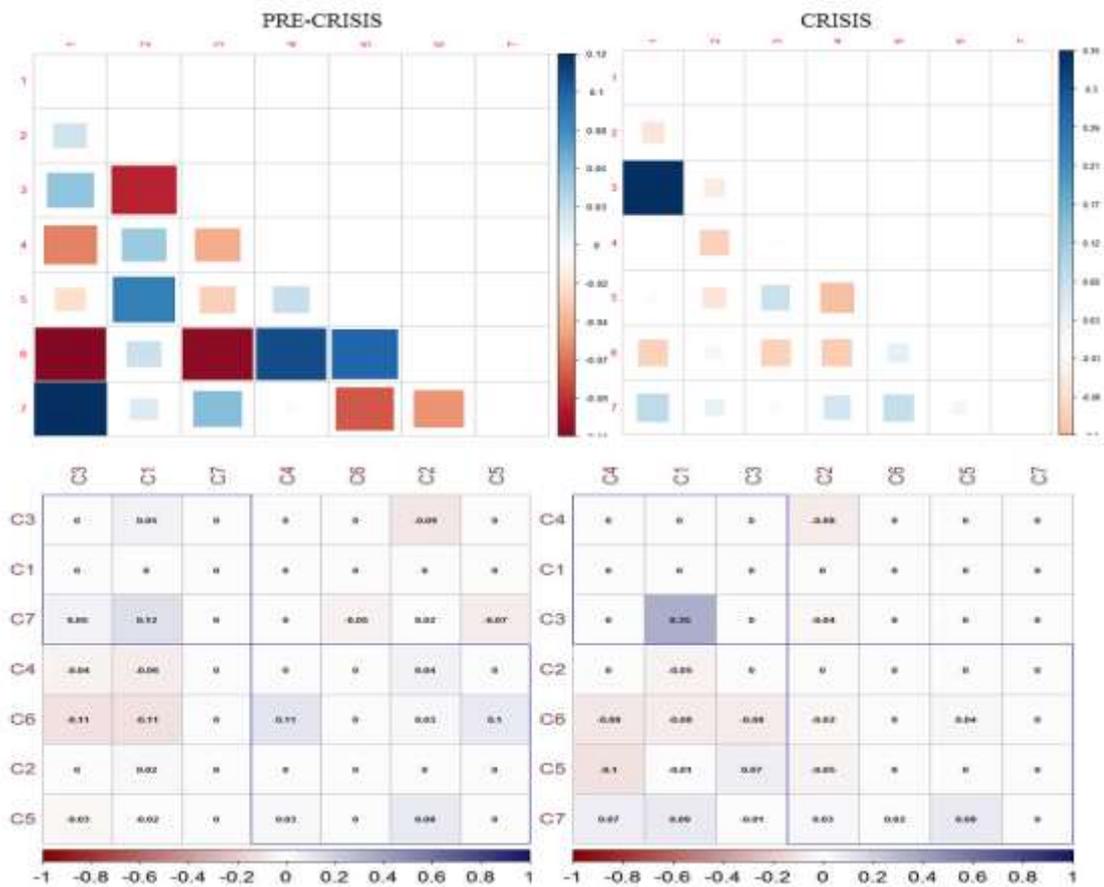

The fractal structures of Asian equity returns given in Figure 4.6 evidences the rise in long-range correlation between South Korean and Taiwanese equity returns during the subprime crisis. The long-range correlation among other Asian markets are very low during both crisis and non-crisis periods indicating dissimilar fractal structures. This is also evidenced from clustering of equity returns from these markets where markets forming clusters are almost similar during both crisis and non-crisis periods. This is in contrast with the developed western markets where fractal structures, based on long-range correlation coefficients and clustering of the same, are not similar during crisis and non-crisis periods. However, structure of fractality based on fractal connectivity matrix helps one in investigating the long memory properties of equity markets in greater detail. Nonetheless, multivariate wavelet estimator of long-range correlation and fractal similarities provide an efficient way of analyzing equity markets' correlation structure. However, analyses of global equity returns using the aforementioned method demands more thorough investigation as there are no studies in literature analyzing fractal connectivity of financial markets.

## 4.5 Conclusion

This chapter investigated the phenomenon of long memory among global equity returns using methods from both univariate and multivariate class of wavelet based long memory estimators. Some evidence of long-memory in equity returns of emerging markets of Malaysia, Taiwan, Pakistan, China and Indonesia are unearthed. However, the application of improved fractal estimators of Abry et al. (2003), aided by the logscale diagram of wavelet based scaling estimates, detected significant long memory in the emerging markets of India, China and Indonesia. On the other hand, equity returns of developed markets from Europe and the U.S. did not exhibit long-range dependence, thus validating results from existing studies that reject long memory in developed markets. Moreover, markets from developed economies are said to be more efficient where efficiency is inversely related to the persistence of returns and prices. Therefore, the presence of long memory in equity returns rubbishes the notion of market efficiency. However, equity markets are in a constant stage of development which can influence efficiency and predictability. Since dependence structure of equity returns over time can be time-varying, the analysis of long memory is extended to analyze time varying long memory behavior of equity returns. This helps in examining the evolution of long memory parameter over time, thereby allowing one to detect phases of market efficiency and inefficiency. Therefore, analysis of the evolutionary nature of long memory is captured using rolling windows estimation method where long memory of equity returns from both emerging and developed markets are investigated. The results indicate that the developed equity markets of Europe and the U.S. show relatively less degree of persistence. However, phases of long-memory, though smaller, are detected for some developed markets. In contrast, markets from emerging economies are found to have relatively more phases of inefficiency, indicating presence of arbitrage opportunities. Moreover, emerging markets' equity returns are found to move between phases of long memory and short memory. Furthermore, long memory is not evidenced during the subprime crisis period of 2008 for majority of markets which is in line with the wavelet based study of Tan et al. (2014), where faster information disseminated among investors during financial crises is said to curtail speculative behavior, thereby affecting predictability of markets. Likewise, the time-varying nature of long memory and varying phases and stages of market efficiency is consistent with the conception of adaptive markets, where market efficiency should be viewed from an evolutionary framework.

The results obtained from time varying long memory analysis reinforces the notion that markets are not always efficient. On the other hand, markets tend to traverse through different dynamics and are subjected to evolutionary patterns, where stages of both efficiency and inefficiency come into play. The explanation of market inefficiency be directly related to the presence of herding and investment cascades. Investment momentum is generated by feedback trading which in turn is a result of herding (Shiller 1989; DeLong et al. 1990; Barberis and Shleifer 2003), and cascades are a result of reinforcing trading activities (Mandelbrot et al. 1997; Calvet and Fisher 2002).

# Chapter 5

## Concluding Remarks

### 5.1 Introduction

This thesis seeks to investigate the relationship among global equity returns, with a specific focus on the Indian equity market, using a battery of methods from time-frequency analysis based wavelet techniques. More specifically, the structure and features of global equity returns are analysed using various wavelet methods, allowing information extraction from both domains of time and frequency, as opposed to the traditional time domain and spectral analyses where simultaneous information extraction from both time and frequency domain is not possible. The existence of multiple investment horizons, in financial markets where investors with certain time preference and investment horizon of interest operate independently, necessitates a careful and thorough analysis of each investment holding periods separately. This heterogeneity of investors, with respect to investment decisions based on their choice of time horizons, can be effectively captured via multiresolution analysis technique which encompasses the basic structure of almost all applied wavelet theory.

The presence of multiple time horizons, with varying levels of complexity, requires one to investigate financial time series from a heterogeneous market perspective where market players are said to operate at different investment horizons. Therefore, the theory of investment heterogeneity, as explored in Muller et al. (1997) where the theory of heterogeneous market hypothesis is expounded, can be suitably explored in the wavelet domain. In view of the behaviour of heterogeneous market participants, which essentially leads to the formation of multiple layers of investment horizons or investment holding periods, this thesis attempts to analyse the structure of global equity markets from a heterogonous market viewpoint.

In view of the above, multiscale correlation methods from wavelet domain are employed in the second chapter to investigate interdependence among global equity markets, thereby allowing one to adjudicate investment horizon-specific information on global equity market interdependence and market integration. Furthermore, the issue of portfolio diversification and its implication for Indian investors are analysed in detail. The issue of market interdependence is then extended, in the third chapter, to include an analysis of global equity market interdependence during times of financial crises. Therefore, the third chapter on contagion seeks to examine the effects of financial turbulence on the Indian equity market. Moreover, the implications of contagion for the Indian investors are addressed from a time-frequency and heterogeneous investors' perspective. Finally, the fourth chapter on long memory seeks to understand the long memory behaviour of global equity returns using novel methods from wavelet analysis. Moreover, long-run correlation structure among global equity returns are analysed within the framework of multivariate long memory methods based on wavelet analysis. More specifically, this thesis extends the application of wavelet based methods in the analysis of global equity markets with a special focus on the Indian market. The dearth of studies concerning wavelet based analysis of interdependence, contagion and long memory among global equity markets, particularly the Indian equity returns, necessitates an exploration based on multiscale methods, thereby seeking to understand the relationship among global equity markets with the help of novel time-frequency methods that this thesis attempts to analyse.

In this backdrop, this study tries to jointly address the issues of global equity market interdependence, financial contagion and the presence of fractality in equity returns. In doing so, the following objectives are addressed:

1. Should Indian investor invest in developed or emerging markets to gain benefits from international portfolio diversification?
2. How will international portfolio diversification change investor stock holding period?
3. To examine the contagious effects of financial crises on Indian equity market and access its implication for international portfolio diversification.
4. To investigate the efficiency (or inefficiency) and multifractality of global equity returns.

## 5.2 Summary of Findings

The first empirical chapter of this thesis investigates the relationship between global equity returns using multiscale methods with an attempt to gain insights on multi-horizon equity market behaviour from the standpoint of an international investor. An analysis of interdependence among global equity markets is first carried out using the classical wavelet correlation and cross-correlation methods given in Percival and Walden (2000) and Gencay et al. (2002). In the next stage, the analysis proceeds with the implementation of improved wavelet correlation methods as given in Fernandez-Macho (2012) and Polanco-Martinez and Fernandez-Macho (2014). Eight levels of wavelet decomposition is carried out to extract information from investment horizons corresponding to "1-2 days, 2-4 days, 4-8 days, 8-16 days, 16-32 days, 32-64 days, 64-128 days and 128-256 days". Daubechies least asymmetric LA (8) filter is selected for the multiresolution decomposition of global equity returns as use of this particular filter is widely supported in literature concerning financial time series (Percival and Walden, 2000). Risks accompanying assets at varying investment holding periods can be suitably identified by analysing correlation between markets at different time horizons. Significantly stronger correlations are observed at long-run timescales whereas correlations at short run time horizons are found to be weak. For e.g. correlation at daily timescale is very weak and becomes stronger beyond monthly time horizon. Hence, correlation among global equity returns are found to be scale dependent.

Overall, the correlations tend to increase as we move from shorter time horizons to long horizons, where the daily timescale (1-2 days) appear to have the lowest correlation. The multiscale correlations, across most of the timescales, are observed to be higher between equity markets from the developed European economies, indicating better market integration. More generally, multi-horizon correlation among equity markets from geographical region constituting Europe and the U.S. are demonstrated to be significantly stronger across different investment-horizons, indicating strong market integration among developed markets of Europe and the U.S. Similarly, correlations among equity markets from Asian economies are found to be strong, validating the work of Pretorius (2002) where regional proximity, and the associated trade and financial linkages that geographical proximity brings about, plays a pivotal role in determining equity market interdependence and integration.

However, weak multiscale correlations are reported between developed equity markets and India, indicating weak market integration. However, Indian market seems to be integrated with some markets from East Asian economies. Good market integration between India and markets from South Korea, Japan, Malaysia and Taiwan is observed at many investment horizons. Multiscale integration with the Chinese market is however not strong. Moreover, the lead-lag estimates obtained from wavelet cross-correlation reveal the leading behaviour of markets from developed economies. Strong market integration is observed between markets from Eurozone economies. Interestingly, Indian market is found to be integrated with the Austrian market, which is an exceptional case as integration with all other developed markets is weaker. The correlations and cross-correlation between Indian and global equity markets are all found to be timescale dependent. Information from time scale dependence and horizon specific linkages can help Indian investors in formulating better investment strategies. Based on the empirical results, the following scenarios are noted:

- Strong correlations at various investment horizons between Indian and East Asian markets are observed, reducing the benefits from international portfolio diversification.
- Indian Investors should be cautious while including stocks from Asian and East Asian Markets as interdependence exists from medium to long run time horizons.

- Diversification benefit exists for Indian investors when stocks from developed economies, with an exception of the Austrian market, are included in the portfolio.

The analysis of contagion, being the main focus of the second empirical chapter of the thesis, follows the study on interdependence, and primarily focuses on analysing comovements between Indian and global equity markets. Pure and fundamental based contagion cannot be effectively disentangled by time domain methods. Therefore, this study uses the wavelet coherence method to investigate the strength of comovements across time and frequencies. The comovements between India and the US is found to be stronger only in the long run time horizons during the neighbourhood of the global financial crisis. Comovements between India and developed European markets of France and Germany are also found to be stronger in the longer time horizons. No evidence of pure contagion from developed markets to India is recorded as long run comovements only imply market interdependence. However, strong comovements, even at finer scales, between Indian and East Asian markets like South Korea and Japan are recorded. The robustness of the results obtained from wavelet coherence is checked, for all pairs of countries, by obtaining a time-series of wavelet correlation using the rolling window wavelet correlation method. This is followed by a two sample t-test to check the significant difference in correlation before and after the crisis event. The results obtained support the findings from wavelet coherence based comovements analysis. Strong evidence of some contagion between developed markets is revealed. Comovements at all scales between Indian and some East Asian markets are found to be significant. Short term shocks due to excessive linkages are said to be generated by investors' behaviour (herding, panic etc.) during turbulent periods whereas long term shocks are ascribed to trade and financial market interdependence. Short term shocks are detected between Indian and some East Asian markets, indicating diversification risks for Indian investors during periods of financial turbulence. Since information about market efficiency and returns predictability might influence investment strategy, the study on interdependence and contagion would not be complete without investigating equity market efficiency.

Market efficiency is empirically tested for twenty four developed and emerging markets using wavelet based estimators of long memory. Wavelet estimates of Hurst parameter (H) computed for twenty four stock markets show less evidence of long memory among developed economies and weak long memory among indices of emerging markets. However, returns and volatility structure often changes due to market conditions, government policies and financial turbulence. Therefore, in order to capture the dynamic structure and evolution of equity returns, a rolling window estimation of the long memory parameter is carried out generating a time-series of Hurst parameter. Less evidence of dynamic long memory among all equity returns is observed. The strength of long memory is found to be very weak for Indian and other emerging markets during financial crisis, implying no arbitrage opportunities during crisis. Also, the high volatility state during turbulent periods reduces diversification opportunities.

However, Indian market is found to be shifting between periods of persistence and anti-persistence, indicating some arbitrage opportunity. This is also in tune with the adaptive market hypothesis as the obtained time varying long memory parameter series of Indian and other equity markets helps one to analyse markets from a dynamic and evolutionary perspective. Additionally, the results from the wavelet based multivariate estimates of fractal connectivity matrix reveals stronger correlations among developed markets of Europe, the U.S. and Asia during the subprime crisis, thereby revealing similar fractal structures among these markets. Moreover, the Indian equity market (BSE30) is clustered together, from a fractal perspective, with the markets of the U.S., Germany, Japan and South Korea.

The results obtained from the univariate, time-varying and multivariate estimates of long memory and other fractal parameters validates the standpoint of adaptive market hypothesis that markets are evolutionary in nature and are not always efficient. On the other hand, markets tend to traverse through different dynamics and are subjected to evolutionary patterns, where stages of both efficiency and inefficiency come into play.

## 5.3 Policy Implications

This study examines interdependence, contagion and fractal similarities among global equity markets from a time scale perspective aiding one in breaking down the global relationships at heterogeneous investment horizons. The section on interdependence particularly focuses on the multi scale correlation structure between Indian and select global markets to investigate diversification benefits for Indian investors. Since multi scale correlation generates information about market linkages at varying investment horizons, investors with multi scale information are better equipped to formulate investment decisions based on their investment holding period. Contrary to studies that examine portfolio diversification benefits using a homogenous time period where investment heterogeneity is not captured, wavelet based timescale study of interdependence aids in formulating decisions based on the requirements of heterogeneous market participants. In this regard, we argue that Indian investors operating at various time horizons will gain valuable information which might help in maximising the benefits of international portfolio diversification.

Information on correlation structure at varying time-horizons will aid investors in diversifying portfolios with global asset combinations, where portfolios diversified using international assets is empirically demonstrated in the literature to reduce portfolio risks (Grubel, 1968; Agmon, 1972; Dajcman, 2012 etc.). Furthermore, the information on correlation structures at different investment holding periods will provide additional inputs for investors whose risks might not be the same for all investment decisions that they undertake. Therefore, an analysis based on these lines aids investors in internationally diversifying their portfolios while incorporating different investment holding periods, or time-horizons, into their strategy.

Indian investors who invest in equity markets of the U.S. and developed European markets may benefit from reduced portfolio risks as correlation between the Indian stock returns and returns of these developed western markets, at almost all time-horizons, is very low. Additionally, Indian investors might also be well off if they invest in the Chinese stock market. However, Indian investors should be cautious if they include assets from Austrian, Brazilian and East Asian markets as multiscale correlation between BSE30 and markets from these regions, for a majority of investment holding periods, are very significant. Since heterogeneity of Investment horizons and corresponding information at multiple time scales allow heterogeneous Indian investors to carefully diversify their portfolio, the results obtained from this analysis might aid Indian investors in their investment decisions. Nonetheless, investors should consider their investment holding periods and the associated risks when they make risk management and portfolio allocation decisions.

Similarly, the exposure of Indian investors to financial contagion is explored and certain policy recommendations, primarily derived based on the results from interdependence and contagion among global equity markets, are presented. The separation of short-run and long-run shocks alongside their relative power, in a time-frequency framework, allows investors to clearly formulate optimal investment strategies based on risks involved at various investment horizons. With respect to portfolio diversification, Indian investors are required to exercise caution, especially in the short-run, while formulating portfolios comprising of stocks from some Asian economies. However, strategic investors might benefit in the long run by including stocks from some European and Asian markets. With respect to the mitigation of short-run contagious shocks, stabilisation policies aimed at the short-run can help infected markets to bypass speculative attacks emanating via investors' psychology and herding behaviour. Moreover, transparency in financial policies, proper disclosure of data, supervision and regulation of financial sectors etc. might help strengthen the country's financial system, thereby providing some immunity from contagious shocks.

Finally, the efficiency of Indian equity market is analysed to understand efficiency during both stable and turbulent periods. Evidence of heterogeneity in market efficiency during both stable and turbulent periods is demonstrated via several measure of market fractality. Therefore, investors should tactically formulate investment decisions based on the fractal structures of equity markets which is evolutionary and time varying. To summarize, the empirical findings from this study on interdependence, contagion and long memory among global equity markets have the following policy implications:

- Indian investors should be cautious while including stocks from Asian and East Asian markets as evidence of market interdependence is found for medium and long run investment horizons.
- Stocks from some developed economies, with an exception of the Austrian equities, can be included in portfolios of Indian investors to maximise the benefit from international portfolio diversification.
- Investors should be careful during periods of financial turbulence when dealing with equities from East Asian markets, as Indian market is found to be prone to contagious effects from these markets.
- Investors should also take cognizance of the dynamic and evolutionary nature of market efficiency when formulating investment strategies. The results obtained exposes the tendency of Indian market to shift between phases of efficiency and inefficiency.

The above mentioned results and inferences are obtained by the application of wavelet based multiscale methods to Indian and global equity markets. This study gives new insights regarding the optimization of financial investment strategies, with a special focus on Indian investors. Heterogeneity of Investment horizons and corresponding information at multiple time scales allow heterogeneous Indian investors to carefully diversify their portfolio. The results obtained might aid Indian investors in their investment decisions. Therefore, investors ought to tactically consider their investment horizon when they make decisions related to risk mitigation, allocation of portfolios and other investment related strategies.

## 5.4 Limitations and Scope for Further Research

The main drawback in the analysis of interdependence, contagion and long memory among global equity markets as presented in this thesis lies in the absence of concurrent use of wavelets with latest time series models. Moreover, the issue of some information loss while decomposing the time series using several wavelet filters demands simultaneous use of traditional models to check the significance of wavelet based results. Therefore, the use of several combinations of time domain models with wavelet analysis give rise to variety of pathways that can be adopted for the extension of this thesis for future research. Furthermore, analyses based on this thesis can be suitably extended to investigate the nature of high frequency equity market data which is not covered in this thesis. Moreover, the development of newer continuous wavelet methods and wavelet methods of multifractality opens up several dimensions in the analysis of financial markets in general.